\documentclass[review]{elsarticle}

\usepackage{lineno,hyperref,fullpage}
\usepackage{comment}
\usepackage{afterpage}
\usepackage{enumitem}
\modulolinenumbers[5]

\journal{Elsevier}

\usepackage{nomencl}
\usepackage{amsmath}
\usepackage{amsfonts}
\usepackage{multirow}
\usepackage{subcaption}

\usepackage{floatrow}
\makenomenclature
\usepackage{xpatch}
\xpatchcmd{\thenomenclature}{\section*{\nomname}
}{}{\typeout{Success}}{\typeout{Failure}}
\makenomenclature
\RequirePackage{ifthen}
\printnomenclature[0.8cm]

\usepackage{soul}
\usepackage{xcolor}
\usepackage{bm}
\usepackage[colorinlistoftodos,disable]{todonotes}

\usepackage{url}

\newcommand{\sx}{\mathsf{x}} %
\newcommand{\oy}{\mathsf{y}} %
\newcommand{\mf}{\mathsf{z}} %
\newcommand{\wv}{\boldsymbol{\omega}} %

\renewcommand\nomgroup[1]{%
  \item[\itshape%
  \ifstrequal{#1}{A}{Symbols}{%
  \ifstrequal{#1}{B}{Roman Letters}{%
  \ifstrequal{#1}{C}{Greek Letters}{%
  \ifstrequal{#1}{D}{Abbreviations}{}}}}%
]}

\let\oldequation\equation
\let\oldendequation\endequation
\renewenvironment{equation}
  {\linenomathNonumbers\oldequation}
  {\oldendequation\endlinenomath}

\begin{document}

\begin{frontmatter}
  \title{Regularized ensemble Kalman methods for inverse problems}

  \author[vt,ensam]{Xinlei Zhang}

  \author[vt]{Carlos Michel\'en-Str\"ofer}

  \author[vt]{Heng Xiao\corref{mycor}}
  \cortext[mycor]{Corresponding author}
  \ead{hengxiao@vt.edu}
  \ead[url]{https://www.aoe.vt.edu/people/faculty/xiaoheng.html}

  \address[vt]{Kevin T. Crofton Department of Aerospace and Ocean Engineering, Virginia Tech, Blacksburg, VA 24060, USA}
  \address[ensam]{Univ. Lille, CNRS, ONERA, Arts et Metiers Institute of Technology, Centrale Lille, UMR 9014 - LMFL - Laboratoire de Mécanique des fluides de Lille - Kamp\'e de F\'eriet, F-59000 Lille, France}

  \begin{abstract}
    Inverse problems are common and important in many applications in computational physics but are inherently ill-posed with many possible model parameters resulting in satisfactory results in the observation space.
    When solving the inverse problem with adjoint-based optimization, the problem can be regularized by adding additional constraints in the cost function.
    However, similar regularizations have not been used in ensemble-based methods, where the same optimization is done implicitly through the analysis step rather than through explicit minimization of the cost function.
    Ensemble-based methods, and in particular ensemble Kalman methods, have gained popularity in practice where physics models typically do not have readily available adjoint capabilities.
    While the model outputs can be improved by incorporating observations using these methods, the lack of regularization means the inference of the model parameters remains ill-posed.
    Here we propose a regularized ensemble Kalman method capable of enforcing regularization constraints.
    Specifically, we derive a modified analysis scheme that implicitly minimizes a cost function with generalized constraints.
    We demonstrate the method's ability to regularize the inverse problem with three cases of increasing complexity, starting with inferring scalar model parameters.
    As a final case, we utilize the proposed method to infer the closure field in the Reynolds-averaged Navier--Stokes equations, a problem of significant importance in fluid dynamics and many engineering applications.
  \end{abstract}

  \begin{keyword}
    Bayesian inference \sep Ensemble Kalman method \sep Regularization \sep Parameter estimation \sep Field inversion
  \end{keyword}
\end{frontmatter}

%\linenumbers

%
%

\section{Introduction}
\label{sec:intro}
  Inverse problems are frequently encountered in computational physics applications such as complex fluid flows where physical fields need to be inferred. A classic example of inverse problems is to estimate the stationary background flow velocity field from the concentration of passive scalars (e.g., pollutant or dye) that are advected by, and diffusing within, the fluid~\cite{borggaard2018bayesian}. The data that are available and used in such an inversion are often partial,  noisy observations of the concentration field. The inverse problem is motivated by the fact that concentrations are often easier to measure than velocities. The forward problem corresponding to the above-mentioned inverse problem is computing the concentration  field $\mathsf{z}(\bm{x})$, where $\bm{x}$ denotes spatial coordinates, by solving the steady-state advection--diffusion equation
  \begin{equation}
      \bm{u} \cdot \nabla \mathsf{z} - \kappa \nabla^2 \mathsf{z} = 0
      \label{eq:adv-diff}
  \end{equation}
  with a given background velocity field $\bm{u}(\bm{x})$, along with other auxiliary constraints such as boundary conditions and physical properties (e.g., diffusivity $\kappa$) of the passive scalar. Hence, the partial differential equation (PDE)-based forward model above implies a functional mapping $\mathcal{M}: \bm{u} \mapsto \mathsf{z}$, or more concisely, $\mathsf{z} = \mathcal{M}(\bm{u})$.
  Another example of solving inverse problems is the data assimilation used for weather forecasting, where partial, time-dependent observations of the atmosphere (e.g., wind speed, temperature, humidity) and numerical simulations are jointly used to infer the full initial state of the system.
  Inverse problems are typically many times more expensive to solve than the corresponding forward problems. This is not only due to the limited amount of observation data compounded by the uncertainties therein, but also due to the nonlinearity of the PDE-governed system and its high-dimensional state space that lead to the inverse problem being ill-posed.

  The example of inferring background velocities can be posed as an optimization problem, i.e., finding a velocity field $\bm{u}^\text{opt}$ that leads to a concentration field best matching the observations ($\mathsf{z}^\text{obs}$) at the measured locations. That is,
  \begin{align}
      \bm{u}^\text{opt}  = \mathop{\arg \min}_{\bm{u}} \lVert  \widetilde{\mathcal{H}}\left[\mathcal{M}(\bm{u})\right] - \mathsf{z}^\text{obs} \rVert^2 \text{,}
  \end{align}
  where $\mathcal{M}(\bm{u})$ involves solving the PDE for the concentration, $\widetilde{\mathcal{H}}$ is the observation operator (e.g., extracting values at the observed locations from the concentration field), and $\|\cdot\|$ denotes a norm in a Hilbert space (e.g., $L_2$ norm in Euclidean space or that weighted by the state covariance).
  In a terminology consistent with that used in the data assimilation community, the velocity field to be inferred is referred to as the state~($\sx$), and the measured concentrations called observations ($\oy$).
  We further define $\mathcal{H} \equiv \widetilde{\mathcal{H}} \circ \mathcal{M}$ as a composition of the model operator $\mathcal{M}$ and the observation operator $\widetilde{\mathcal{H}}$.
  The inverse problem above can thus be written as
  \begin{equation}
      \sx^\text{opt}  = \mathop{\arg \min}_{\sx}  J  \qquad
      \text{with} \quad  J(\sx) = \|\mathcal{H}[\sx] - \oy \|^2 \text{,}
  \label{eq:da-opt}
  \end{equation}
  where $J(\sx)$ is the cost function to be minimized, which corresponds to the discrepancy between the model outputs and the observations.

  \subsection{Adjoint- vs. ensemble-based inversion methods}
  \label{sec:intro_methods}
    In order to solve the optimization problem in the field inversion, the gradient descent method or one of its variants is often used, where the search for the optimal solution is guided by following the local gradient~$\partial J / \partial \sx$ of the cost function~$J$ with respect to the control parameter~$\sx$. Note that~$\sx$ usually resides in a space of very high dimensions (corresponding to the number of degrees of freedom of the discretized velocity field in the above example, which can be in the order of millions). A highly efficient way to compute such a derivative is the adjoint method~\cite{giles2000introduction}, where the derivative is obtained by solving an adjoint equation at a cost similar to solving the PDEs in the forward model (referred to as primal equation).
    Adjoint methods have been used for different fluid mechanics problems.
    Dow and Wang~\cite{dow2011quantification} proposed an adjoint-based Bayesian inversion method to quantify the structural uncertainty in Reynolds-averaged Navier--Stokes simulations.
    Gronskis et al.~\cite{gronskis2013inflow} adopted the variational method to infer the inflow and initial condition for a problem using direct numerical simulation (DNS) of the Navier--Stokes equations. Papadimitriou and Papadimitriou~\cite{papadimitriou2015bayesian} applied a Bayesian framework coupled with a high-order adjoint approach to quantify the uncertainty in the parameters in the Spalart--Allmaras turbulence model~\cite{spalart1992one-equation}. Singh and Duraisamy~\cite{singh2016using} proposed an approximate Bayesian inference framework based on the adjoint method to infer the multiplicative correction term in the Spalart--Allmaras model and the~$k$--$\omega$ model~\cite{wilcox2006turbulence}.
    Foures et al.~\cite{foures2014data} used the adjoint-based variational method and Lagrange multipliers to reconstruct the full velocity field from coarse-grid particle image velocimetry (PIV) measurements of velocity magnitude from only part of the domain. They imposed the Reynolds-averaged Navier--Stokes equations as a constraint in the minimization and used the Reynolds stress divergence as a control parameter.
    Recently, Beneddine et al.~\cite{beneddine2017unsteady} further extended this technique to the reconstruction of the unsteady behavior of a round jet at a Reynolds number of $Re = 3300$ from the mean flow field and unsteady measurements at a single point.
    Meldi and Poux~\cite{meldi2017reduced} integrated the Kalman filter into the structure of a segregated CFD solver and imposed the zero-divergence condition for the velocity field. They further proposed model reduction strategies to reduce the computational costs within the Kalman analysis. The framework has been used to reconstruct different flow configurations including three-dimensional unsteady flows~\cite{meldi2017reduced} with comprehensive sensitivity analysis performed~\cite{meldi2018augmented}.

    A major shortcoming of the adjoint method, however, is the effort required to develop the adjoint solver. For the discrete adjoint method, which is the most commonly used adjoint method in computational fluid dynamics (CFD) applications, this involves differentiating each operation (i.e., each line of code) in the primal solver~\cite{giles2003algorithm,nielsen2010discrete}. This is a laborious process and a daunting task for complex simulation codes such as CFD solvers. Taking CFD for example, while some codes intended for design and optimization have been developed with adjoint capability~\cite[e.g.,][]{economon2015su2, biedron2019fun3d}, many other popular solvers are not equipped with a native, production-level adjoint capability.
    Most notably, the CFD code OpenFOAM~\cite{opencfd2018openfoam} does not have any native discrete adjoint solver capabilities. Although there have been efforts to build one for OpenFOAM based on automatic differentiation~\cite{towara2013discrete}, it is not yet at production level at this time.

    The limited availability of physical simulation codes with adjoint capability has prompted the inverse modeling community to develop ensemble-based, derivative-free optimization methods.
    The iterative ensemble Kalman method proposed by Iglesias et al.~\cite{iglesias2013ensemble} is among such attempts for general inverse problems.
    In the data assimilation community, ensemble methods~\cite{evensen2003ensemble, evensen2009data, evensen2018analysis} have been developed to complement or replace the traditional variational (adjoint) methods (e.g., 3DVar, 4DVar)~\cite{gauthier2007extension, courtier1994strategy}.
    In ensemble methods, the covariance estimated from the ensemble is used in lieu of the derivatives to guide the optimization.
    A number of primal simulations with different samples of the system states are run, which is in contrast to solving adjoint equations along with the primal equations.
    A critical advantage of ensemble methods over adjoint methods is that they are non-intrusive, i.e., the forward model (primal solver) does not need to be changed, and adjoint solvers are not needed.
    Many works have used ensemble methods for inverse problems in fluid mechanics.
    For instance, Kato and Obayashi~\cite{kato2013} leveraged the ensemble Kalman filter to infer the value of empirical parameters in the Spalart--Allmaras turbulence model and demonstrated the effectiveness of ensemble Kalman methods for CFD problems.
    Mons et al.~\cite{mons2016reconstruction} applied different ensemble-based methods including the ensemble Kalman filter to infer the inlet and initial conditions for CFD simulations and thus reconstruct the unsteady flow around a cylinder.
    Xiao et al.~\cite{xiao2016quantifying} used an iterative ensemble Kalman method to infer the Reynolds stress discrepancy field and reconstruct the velocity field  accurately for flows over periodic hills as well as flows in a square duct.
    However, compared to adjoint methods, ensemble methods do not have the flexibility to introduce regularization to tackle ill-posed inverse problems. This shortcoming shall be examined in more detail below.

  \subsection{Ill-posedness and regularization of inverse problems}
  \label{sec:intro_illposed}
    We introduce the concept of ill-posedness by examining the operator~$\mathcal{H}$ in the optimization formulation of the inverse problem as in Eq.~\eqref{eq:da-opt}.
    As described above, computing the cost~$J(\sx)$ associated with the state~$\sx$ involves
    \begin{enumerate}[label=({\alph*})]
        \item solving the forward model (e.g., Eq.~\eqref{eq:adv-diff} with the given velocity field),
        \item mapping the results to observation space (e.g., sampling at specific locations), and
        \item comparing with the observations to find the discrepancy. 
    \end{enumerate}
    While the advection--diffusion equation happens to be linear, in many other problems (e.g., inferring the velocity field from partial observations of itself) the model~$\mathcal{M}$ is highly nonlinear. Moreover, the operator~$\widetilde{\mathcal{H}}$ typically maps a high-dimensional state space, where~$\mathcal{M}(\sx)$ is in, to a low-dimensional observation space, where~$\oy$ is in.  For example, the concentration field discretized with a mesh of millions of cells may be observed at only a few locations. Because of these two factors,~$\mathcal{H}$ typically results in a many-to-one mapping. In other words, many different velocity fields lead to the same agreement with the observations and thus the same cost~$J$. Consequently, the inverse problem formulated as the optimization in Eq.~\eqref{eq:da-opt} does not have a unique solution and is thus ill-posed.

    To tackle the ill-posedness, inverse problems can be regularized by introducing an additional term~$J_r$ into the cost function in Eq.~\eqref{eq:da-opt}, i.e.,
    \begin{equation}
        J = \|\mathcal{H}[\sx] - \oy \|^2 + J_r \text{.}
    \end{equation}
    The term~$J_r$ serves to differentiate among the states that previously led to identical costs.
    Desired properties of the states that are commonly used for regularization include:
    \begin{description}
        \item [Spatial smoothness] i.e., preferring smooth fields over non-smooth fields among the candidate states~\cite[see, e.g.,][]{dow2011quantification,osher2005iterative}.  The corresponding cost function in Eq.~\eqref{eq:da-opt} becomes $J = \|\mathcal{H}[\sx] - \oy \|^2 + w \| \nabla \sx\|^2$, where~$J_r = w \| \nabla \sx\|^2$ is the regularization term and $w$ is a positive algorithmic parameter controlling the strength of the regularization. In the derivation latter $w$ will be replaced by a weight matrix $\mathsf{W}$ embedded in the definition of the weighted norm.
        \item [Prior mean values] i.e., preferring candidate states closer to the prior mean $\sx_0$ over those further away~\cite[see, e.g.,][]{singh2016using}. The regularization is $J_r = w \|\sx - \sx_0 \|^2$.
        \item [Physical constraints] e.g., in the example above where the state is the background velocity, this can be preferring velocity fields that satisfy the mass conservation (divergence-free condition for incompressible flows)~\cite{he2016numerical}. The regularization is thus $J_r = w \|\nabla \cdot \sx \|^2$. Similarly, other physical constraints include positivity of turbulent kinetic energy or eddy viscosity and realizability of Reynolds stresses~\cite{xiao2016quantifying}.
    \end{description}
    There exist many more types of prior knowledge than those enumerated above. For example, one could use regularization to favor (or penalize) specific wave numbers or pattern in the field to be inferred, or to favor smaller (or larger) values in certain regions. Essentially, regularization utilizes prior knowledge on the state to be inferred to constrain the optimization process. Consequently, the regularization terms to be introduced are inevitably problem-specific and can have a wide range of forms in different applications.

    Implementation of such regularization in optimization schemes is much more challenging for ensemble methods than for adjoint methods.
    In the adjoint-based inversion, regularization involves modifying the cost function with an additional term, which in turn may necessitate modifying the adjoint solver. This is usually straightforward (albeit laborious) process. In contrast, it has been far from clear how to implement a generic regularization in ensemble-based inversion methods. So far, a general procedure to introduce prior knowledge-based regularization to ensemble methods is still lacking. The difficulty partly stems from the fact that ensemble methods do not directly optimize a cost function. Rather, they use an analysis scheme to optimize the cost function implicitly.
    Nevertheless, it is well known that the adjoint-based and ensemble-based Bayesian inverse modeling methods are equivalent under some mild assumptions (e.g., Gaussian priors on the states, normal distribution of observation uncertainties, linear model)~\cite{sakov2010asynchronous, evensen2018analysis}.
    Specifically, under these assumptions the maximum a posteriori (MAP) estimate from the Bayesian approach is equivalent to the minimization problem in adjoint-based methods.
    Therefore, one can naturally expect that the regularization methods reviewed above for adjoint methods can be equally introduced into ensemble methods for inverse modeling.

  \subsection{Related works and contributions of present work}
  \label{sec:intro_contrib}
	Enforcing constraints in both the standard and ensemble Kalman filtering has been an established topic of research~\cite{simon2010kalman,amor2018constrained}. Simon et al.~\cite{simon2002kalman} incorporated the constraint as an additional term in the cost function used to obtain the standard Kalman filter. They derived the update scheme based on the new cost function by solving the constrained optimization problem with the Lagrangian multiplier method. This approach of enforcing hard constraints is referred to as \emph{projection method} and has since been extended to nonlinear state constraint problems~\cite{julier2007kalman,yang2009kalman} and to ensemble Kalman filtering~\cite{wang2009state}.  On the other hand, Nachi~\cite{nachi2007kalman} enforced constraints by augmenting the observation data with fictitious observations generated from
	the constraint function. These fictitious observations have zero variance as they are sampled from hard constraints. It has been shown that the projection method and the observation data augmentation method are equivalent~\cite{simon2010kalman}. The observation augmentation method has been extended to account for soft constraints by adding artificial, zero-mean noises to the fictitious observations~\cite{julier2007kalman}. 
	Moreover, several other authors enforced constraints in ensemble Kalman filtering by solving constrained optimization problems, e.g., by truncating the probability distribution of the initial ensemble~\cite{prakash2008constrained}, by formulating the update as a quadratic programming problem~\cite{janjic2014conservation}, or by rejecting the samples that violate the constraints~\cite{wang2009state}. Recently, Albers et al.~\cite{albers2019ensemble} provided a unified, application-neutral framework for constraining ensemble Kalman methods with rigorous analysis based on previously proposed application-specific approaches~\cite{wang2009state,janjic2014conservation}.

    Most of the above-mentioned works focus on enforcing hard constraints. However, note that ill-conditioned inverse problems require regularization, which has subtle yet fundamental differences from constraints. Although constraints are a common approach to regularize ill--conditioned problems, many forms of prior knowledge (e.g., sparsity in the solution) cannot be formulated as constraints straightforwardly and thus cannot be solved with the approaches above. 
    Specifically in ensemble Kalman methods, while the constrained methods force the updated samples onto a specified manifold, regularization involves modifying the cost function to narrow down to a unique optimal solution (or a smaller set of solutions) by eliminating many otherwise equally optimal solutions under the original cost function (i.e., norm of discrepancies from the prior and the observation ). As is evident from the review above, most existing works belong to the category of constrained Kalman methods and mainly focus on enforcing hard constraints (e.g., mass conservation and non-negativity of physical quantities~\cite{janjic2014conservation}). A rare exception is the recent work of Wu et al.~\cite{wu2019adding}, who proposed a method to enforce soft constraints by re-weighting the samples in the updated ensemble. This re-weighting step is performed after the standard Kalman update.

    Most of the methods reviewed above (e.g., observation augmentation, accept/reject, re-weighting) incur significant computational overhead for imposing the constraints. For example, in the observation augmentation approach, the dimension of the observation matrix can be very high when the state space is high-dimensional and the number of constraints is large (see details in ~\ref{app:projection}).  The accept/reject method requires a large number of samples to estimate the distribution adequately, and it can be costly to find an acceptable state. 
    The re-sampling method, on the other hand, bears some similarities with particle filters and is thus potentially susceptible to sample degeneration~\cite{li2014fight}. That is, a few samples can have large weights and dominate all other samples, reducing the effective sample size and the computational efficiency. Moreover, it is not straightforward to enforce inequality constraints through re-weighting.

    In our work, we propose a new approach to incorporate various prior knowledge (e.g., hard or soft constraints, smoothness, sparsity) into the analysis scheme of ensemble Kalman methods, resulting in a \emph{regularized ensemble Kalman method}. This is in contrast to the projection method, which uses Lagrange multiplier to enforce hard constraints. Although the two schemes are formally similar they differ in terms of not only motivation and derivation procedure but also in the final update scheme (see~\ref{app:projection} for details). 
    The proposed method involves only a minor algorithmic modification to the standard ensemble Kalman methods.
    This modification leads to a derivative-free method that incorporates constraints or regularizations in a mathematically equivalent manner as the commonly used adjoint-based inversion methods, i.e., via implicit minimization of a regularized cost function. 

    Specifically, we propose a method to introduce general regularization terms (including but not limited to the types reviewed above) into the ensemble Kalman methods.
    This is achieved by deriving an analysis scheme starting from the modified cost function. The result is an analysis scheme with minor modifications to achieve the desired regularization.
    The derivation is valid for ensemble Kalman methods in general, including the iterative ensemble Kalman method in~\cite{iglesias2013ensemble} and the ensemble Kalman filter~\cite{evensen2003ensemble,evensen2009data}.
    Note that we aim to derive a scheme to impose general constraints through modification of the analysis schemes.  Applications to the specific type of constraints (e.g., smoothness and prior knowledge) as discussed above will be illustrated in further examples presented in follow-on works.

    The rest of the paper is organized as follows. Section~\ref{sec:method} presents the derivation of the regularized ensemble Kalman method for optimization and its implementation. Modification compared to its traditional counterpart is highlighted. Section~\ref{sec:results} evaluates the proposed method on three inverse modeling problems of increasing difficulty levels ranging from optimization of a nonlinear function of scalars to inferring the closure field in the Reynolds-averaged Navier--Stokes (RANS) equations. The RANS equation closure problem is of significant importance in fluid dynamics and engineering applications since the closure models are considered the main source of uncertainty in the predictions. Finally, Section~\ref{sec:conclude} concludes the paper.

\section{Methodology}
\label{sec:method}
  The two widely used methods for solving inverse problems, adjoint-based optimization approach and ensemble Kalman methods, are equivalent under mild assumptions. From a Bayesian perspective, both of these approaches find the maximum a posteriori (MAP) estimates. The former obtains the MAP estimate through explicit optimization, while the latter achieves it via ensemble-based iteration.
  The objective of this section is to bridge the gap between enforcing constraints (regularization) for the two approaches. Specifically, we show that a generic constraint introduced into the cost function for the optimization approach can be equivalently implemented as modifications to the analysis scheme of the ensemble Kalman methods.

  \subsection{Equivalence between optimization and maximum a posteriori approaches}
  \label{sec:method_equivalence}
    The optimization approach for solving the inverse problem is presented above in Eq.~\eqref{eq:da-opt}. In contrast, from the Bayesian perspective, solving the inverse problem amounts to finding the probabilistic distribution $P(\sx\mid \oy)$ of the state $\sx$ conditioned on observation $\oy$. Based on Bayes' theorem, this is
    \begin{equation}
        \label{eq:Bayes}
        P(\sx \mid \oy) \propto P(\sx) P(\oy\mid \sx) \text{,}
    \end{equation}
    where $P(\sx)$ is the prior distribution before incorporating the observation data and $P(\oy\mid \sx)$ is the likelihood indicating the probability of observing $\oy$ given state $\sx$.
    For the likelihood, the following relation is assumed between $\sx$ and $\oy$:
    \begin{equation}
      \label{eq:x-to-y}
      \oy = \mathcal{H}[\sx] + \epsilon \text{,}
    \end{equation}
    where $\epsilon$ is a stochastic observation error.
    Direct sampling of the full posterior distribution $P(\sx\mid \oy)$ (e.g., by using Markov Chain Monte Carlo sampling) can be prohibitively expensive, as it can require millions of evaluations of the forward model and often must resort to surrogate models~\cite{ray2016bayesian,edeling2014bayesian,edeling2014predictive,edeling2018data}. Therefore, in practical applications, one often finds the state $\sx$ that maximizes the posterior, which is referred to as MAP estimation~\cite{edeling2018bayesian}.
    The derivation assumes that both the prior and the observation noises are Gaussian processes~\cite{rasmussen2003gaussian}, i.e.,
    \begin{subequations}
    \label{eq:g-prior-likely}
    \begin{align}
    P(\sx) \quad &\propto \quad \exp\left(-\|\sx - {\sx}^\text{f} \|_{\mathsf{P}^{-1}}^2 \right),
    \label{eq:g-prior}
    \\
    P(\epsilon) \quad &\propto \quad \exp \left(-\|\epsilon \|_{\mathsf{R}^{-1}}^2 \right) =  \exp \left(-\|\mathcal{H}[\sx] - \oy \|_{\mathsf{R}^{-1}}^2 \right).
    \label{eq:g-likely}
    \end{align}
    \end{subequations}
    where ${\sx}^\text{f}$ is the prior mean, $\mathsf{P}$ and $\mathsf{R}$ are the covariance matrices of the state $\sx$ and the observation $\oy$, respectively, and the norm $\| \cdot \|^2_\mathsf{A}$ is defined as $\| v \|^2_\mathsf{A}=\bm{v}^\top \mathsf{A} \bm{v}$ for a vector $\bm{v}$ and weight matrix $\mathsf{A}$ (with $\mathsf{A}$ being $\mathsf{P}^{-1}$ or $\mathsf{R}^{-1}$ in Eq.~\eqref{eq:g-prior-likely} and $\mathsf{Q}^{-1}$ later on. %
    The posterior is thus
    \begin{equation}
    P(\sx \mid \oy) \propto \exp \left( - \|\sx - {\sx}^\text{f} \|_{\mathsf{P}^{-1}}^2 - \|\mathcal{H}[\sx] - \oy \|_{\mathsf{R}^{-1}}^2\right).
    \end{equation}
    Maximizing the posterior amounts to minimizing its negative logarithm, i.e.,
    \begin{equation}
    \sx^\text{opt}  = \mathop{\arg \min}_{\sx} \,  J  \quad
    \text{with} \quad  J(\sx) = \|\sx - {\sx}^\text{f} \|_{\mathsf{P}^{-1}}^2 + \|\mathcal{H}[\sx] - \oy \|_{\mathsf{R}^{-1}}^2 \text{,}
    \label{eq:g-posterior}
    \end{equation}
    which is equivalent to the optimization approach in Eq.~\eqref{eq:da-opt} with the prior based regularization presented in Section~\ref{sec:intro_illposed}.

    More general regularization terms can be introduced into the cost function.
    These are formulated as a norm of some differentiable function $\mathcal{G}[\sx]$ that needs to be minimized.
    The cost function is then
    \begin{equation}
        J(\sx) = \|\sx - {\sx}^\text{f} \|_{\mathsf{P}^{-1}}^2 + \|\mathcal{H}[\sx] - \oy \|_{\mathsf{R}^{-1}}^2 + \|\mathcal{G}[\sx]\|^2_{\mathsf{Q}^{-1}}
        \text{,}
        \label{eq:adjoint}
    \end{equation}
    where $\mathsf{Q}$ is the covariance matrix associated with the constraint or regularization term. As in the case of $\mathsf{R}$ for the observation error covariance, $\mathsf{Q}$ can be specified based on the confidence on the constraint and its correlation structure. Alternatively, its inverse $\mathsf{W} \equiv \mathsf{Q}^{-1}$ can be specified, which is often referred to as precision matrix in the statistics literature. In this work $\mathsf{W}$ is also called \emph{weight matrix} as it can be used to adjust the weights on the constraints, that is, how strongly to enforce each constraint or to penalize each term.
    For example, in order to promote spatial smoothness of the inferred field, the regularization term can be 
    \[
    \| \mathcal{G}[\sx] \|^2_{\mathsf{W}} 
    = \| \nabla \sx\|^2_{\mathsf{W}} \text{,}
    \] 
    with $\mathsf{W}$ chosen to be proportional to cell size in the discretization of the field. 
    The regularized cost function has a clear interpretation from a Bayesian perspective. Assuming the constraint satisfies a Gaussian distribution, the posterior conditioned by the observation and the constraint can be expressed as~\cite{wu2019adding}
    \begin{equation}
    	P(\mathsf{x} \mid \mathsf{y}, \mathcal{G}[\mathsf{x}]) \propto P(\mathsf{x}) \, P(\mathsf{y} \mid \mathsf{x})  \, P(\mathcal{G}[\mathsf{x}] \mid \mathsf{x} ) \text{,} \label{eq:bayes_extended}
    \end{equation}
    and consequently, Eq.~\eqref{eq:adjoint} corresponds to the MAP of the posterior above.
    Clearly, the covariance matrix~$\mathsf{Q}$ has a similar interpretation to those of~$\mathsf{P}$ and~$\mathsf{R}$ for the forecast and the observation data, respectively.
    Consequently, the proposed framework enforces soft constraints (i.e., prior knowledge with uncertainty, much like the observation data with errors) via regularizations rather than imposing hard constraints as in previous works~\cite[e.g.,][]{simon2002kalman,yang2009kalman,prakash2008constrained,janjic2014conservation}. This is consistent with the Bayesian interpretation of the Kalman filtering algorithm.
    In light of the equivalence between the two approaches, it can be shown that the analysis scheme in ensemble Kalman methods can be derived from the optimization formulation of the inverse problem~\cite[see, e.g.,][]{evensen2009data,evensen2018analysis}.
    We will reproduce and follow such derivations below and introduce the modification needed to incorporate the constraint $\mathcal{G}[\sx]$ along the way.

  \subsection{Derivation of the regularized ensemble Kalman method}
  \label{sec:method_derivation}
    We present the derivation of the regularized ensemble Kalman method. 
    Some algebra has been omitted for brevity and ease of understanding, but the full derivations are given in \ref{sec:derivation_REnKF}.
    In ensemble Kalman methods, the prior in Eq.~\eqref{eq:g-prior} and the likelihood in Eqs.~\eqref{eq:x-to-y} and~\eqref{eq:g-likely} are represented as ensembles $\{\sx^\text{f}_j\}$ and $\{\oy_j\}$, respectively, where $j = 1, \cdots, M$ with $M$ being the number of samples in the ensemble.  For each pair of ensemble member $\sx^\text{f}_j$ and observation $\oy_j$, the analysis scheme aims to find a posterior realization $\sx^\text{a}_j$ that minimizes the cost function $J(\sx_j)$, i.e.,
    \begin{equation}
    \sx_j^\text{a}  = \mathop{\arg \min}_{\sx} \,  J  \qquad
    \text{with} \qquad  J(\sx_j) = \|\sx_j - {\sx}^\text{f}_j \|_{\mathsf{P}^{-1}}^2 + \|\mathcal{H}[\sx_j] - \oy_j \|_{\mathsf{R}^{-1}}^2 \text{,}
    \label{eq:g-posterior-sample}
    \end{equation}
    which is the ensemble-based representation of the optimization formulation in Eq.~\eqref{eq:g-posterior}. If a regularization term is to be introduced to the cost function, the formulation in Eq.~\eqref{eq:g-posterior-sample} becomes
    \begin{equation}
    \qquad  J(\sx_j) = \|\sx_j - {\sx}^\text{f}_j \|_{\mathsf{P}^{-1}}^2 + \|\mathcal{H}[\sx_j] - \oy_j \|_{\mathsf{R}^{-1}}^2  +  \| \mathcal{G}[\sx_j] \|_{\mathsf{Q}^{-1}}^2 \text{.}
    \label{eq:reg-posterior-sample}
    \end{equation}
    Minimizing the cost function amounts to finding the $\sx^\text{a}_j$ that leads to $\partial J / \partial \sx_j = 0$.
    For Eq.~\eqref{eq:reg-posterior-sample} this becomes
    \begin{equation}
    \mathsf{P}^{-1}(\sx_j^\text{a} - \sx_j^\text{f})+(\mathcal{H'}[\sx_j^\text{a}])^\top \mathsf{R}^{-1}(\mathcal{H}[\sx_j^\text{a}]-\oy_j) + \mathcal{G'} [\sx_j^\text{a}] ^\top \mathsf{Q}^{-1} \mathcal{G} [\sx_j^\text{a}] = 0 \text{.}
    \label{eq:dJ}
    \end{equation}
    Assuming the observation operator $\mathcal{H}$ has only modest nonlinearity, one can introduce a linearization around~$\sx_j^\text{f}$ as %
    \begin{align*}
        &\mathcal{H}[\sx_j^\text{a}] \approx \mathcal{H}[\sx_j^\text{f}] + \mathcal{H'}[\sx_j^\text{f}](\sx_j^\text{a}-\sx_j^\text{f}) \text{,}\\
        \label{eq:ensemble_covariance}
        &\mathcal{H'}[\sx_j^\text{a}] \approx \mathcal{H'}[\sx_j^\text{f}] \text{,}
    \end{align*}
    where a prime ($'$) denotes derivative with respect to the state.
    Similarly, we introduce the following two assumptions for the regularization term:
    \begin{equation*}
    	\mathcal{G}[\sx^\text{f}] \approx \mathcal{G}[\sx^\text{a}] \qquad \text{and} \qquad \mathcal{G'}[\sx^\text{f}] \approx \mathcal{G'}[\sx^\text{a}] \text{.}
    \end{equation*}
    Different from~$\mathcal{H}[x]$, we assume a convergence condition for $\mathcal{G}[x]$ (i.e., the first derivative term is ignored) to simplify the derivation. 
    Furthermore, we introduce the tangent linear operator $\mathsf{H}$ for the observation operator~$\mathcal{H}$ so that  $\mathcal{H}[\sx]= \mathsf{H} \sx$ and $\mathcal{H}'[\sx] = \mathsf{H}$.
    Note that this linearization of the observation operator (which includes the forward model) is not done in the iterative ensemble Kalman filter, as will be detailed later.
    Equation~\eqref{eq:dJ} is then simplified to
    \begin{equation}
    \label{eq:dJ1}
        \mathsf{P}^{-1}(\sx_j^\text{a}-\sx_j^\text{f}) + \mathsf{H}^\top \mathsf{R}^{-1}(\mathsf{H}\sx_j^\text{f}-\oy_j + \mathsf{H}(\sx_j^\text{a}-\sx_j^\text{f})) +  \mathcal{G'} [\sx_j^\text{f}]^\top \mathsf{Q}^{-1} \mathcal{G} [\sx_j^\text{f}] = 0 \text{.}
    \end{equation}
    After some algebra (details in \ref{sec:derivation_REnKF}), this leads to the following analysis scheme:
    \begin{equation}
        \label{eq:dJ2}
    \sx_j^\text{a} = \sx_j^\text{f} +
    \underbrace{\mathsf{P}\mathsf{H}^\top ( \mathsf{R} + \mathsf{H} \mathsf{P} \mathsf{H}^\top)^{-1}(\oy_j - \mathsf{H} \sx_j^\text{f})}_{\text{Kalman correction}} \;
    \underbrace{- \; \mathsf{P}(I + \mathsf{H}^\top \mathsf{R}^{-1} \mathsf{H} \mathsf{P})^{-1} \;  \mathcal{G'}^\top \mathsf{Q}^{-1} \mathcal{G} }_{\text{regularization term}} \text{,}
    \end{equation}
    where the argument $\sx_j^\text{f}$ for the function $\mathcal{G}$ and its derivative $\mathcal{G}'$ are omitted for brevity of notation.
    This analysis scheme introduces two corrections to the prior realizations $\sx_j^\text{f}$.
    The first correction, the Kalman correction, comes from the classical ensemble Kalman methods and corresponds to the observation misfit term $\|\mathcal{H}[\sx_j] - \oy_j \|_\mathsf{R^{-1}}^2$ in the cost function in Eq.~\eqref{eq:reg-posterior-sample}, and the second correction corresponds to the regularization term~$ \| \mathcal{G}[\sx_j] \|_{\mathsf{Q}^{-1}}^2$.
    Note that multiple regularization terms can be added in Eq.~\eqref{eq:dJ2}, each with its own constraint function $\mathcal{G}_p$ and weight matrix $\mathsf{W}_p (\equiv \mathsf{Q}^{-1}_p$), where the subscript $p$ is an index denoting the different constraints.

\begin{figure}[!htb]
        \centering %
        \includegraphics[width=0.45\textwidth]{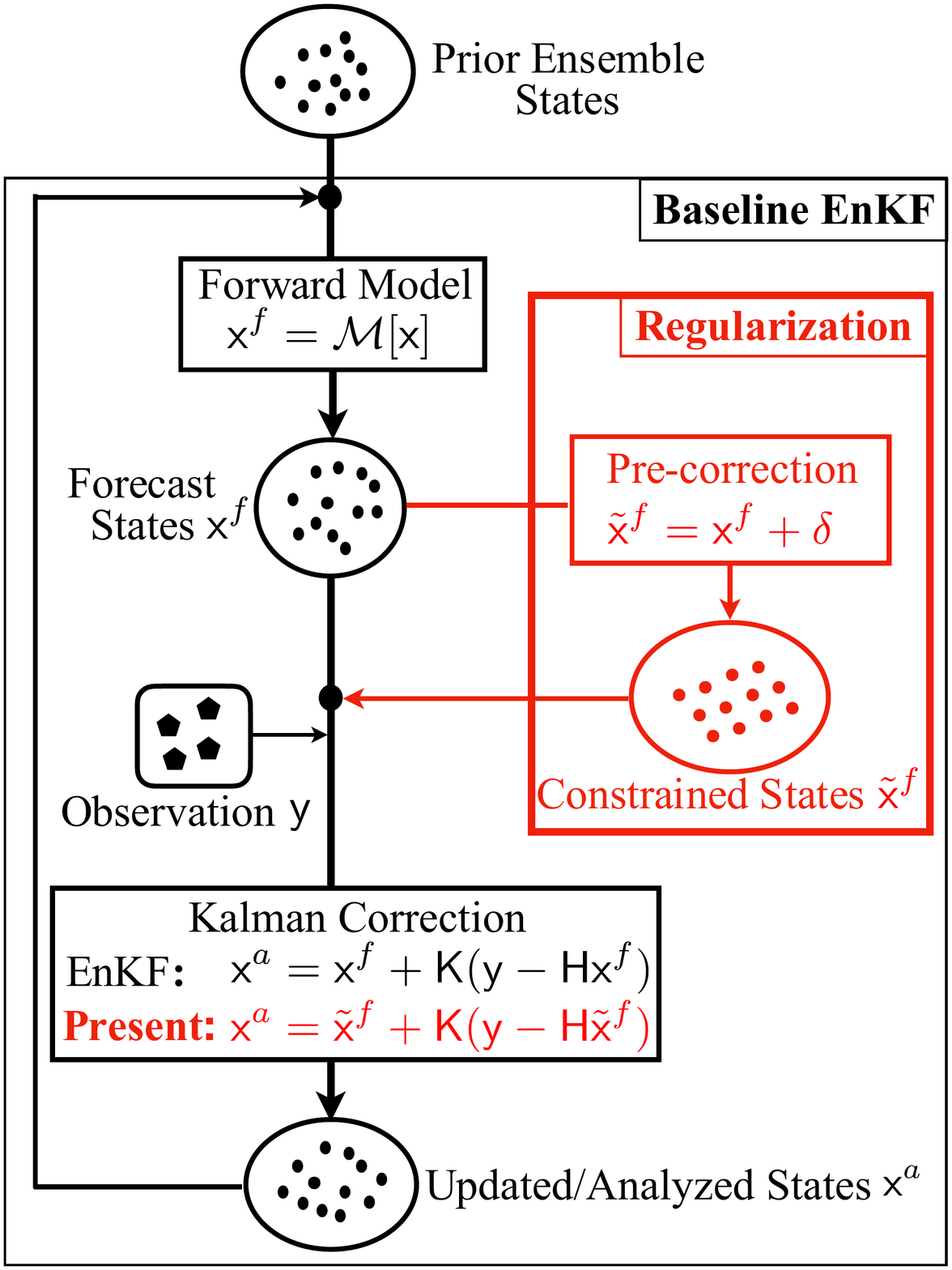}
        \caption{Schematic of ensemble Kalman methods by using ensemble Kalman filtering (EnKF) as example. The proposed regularization scheme consists of an additional correction $\delta$, defined in Eq.~\eqref{eq:delta-final}, to the forecast states $\sx^\text{f}$ before the Kalman correction. Such a correction enforces constraints and is equivalent to the penalty term $\|\mathcal{G}[\sx]\|^2_{\mathsf{Q}^{-1}}$ in adjoint methods as in Eq.~\eqref{eq:adjoint}.
        Our contribution that differentiates the present method from the baseline ensemble Kalman methods is highlighted in red/grey box. The Kalman correction in the regularized EnKF has the identical form as that in standard EnKF except that it acts on the pre-corrected states $\tilde{\sx}^\text{f}.$
        }
        \label{fig:renkf-overview}
    \end{figure}

    The analysis scheme in Eq.~\eqref{eq:dJ2} can be further simplified to facilitate interpretation and to gain insight into its relationship with that of classical Kalman update.
    First, we can expand the term $-\mathsf{P}(I + \mathsf{H}^\top \mathsf{R}^{-1} \mathsf{H} \mathsf{P})^{-1}$ in Eq.~(\ref{eq:dJ2}) to
    \[
    -\mathsf{P} + \mathsf{P}\mathsf{H}^\top ( \mathsf{R} + \mathsf{H} \mathsf{P} \mathsf{H}^\top)^{-1} \mathsf{H} \mathsf{P}
    \]
    by using the Woodbury formula~\cite{hager1989updating} (see details in \ref{sec:derivation_REnKF}).
    Following the convention in the data assimilation literature, we write the Kalman gain matrix
    $\mathsf{K} = \mathsf{P}\mathsf{H}^\top ( \mathsf{R} + \mathsf{H} \mathsf{P} \mathsf{H}^\top)^{-1}$. Consequently, the Kalman correction term and the regularization term become
    \begin{equation}
    \mathsf{K}(\oy_j - \mathsf{H} \sx_j^\text{f}) \quad \text{and} \quad
    - \mathsf{P} \mathcal{G'}^\top \mathsf{Q}^{-1} \mathcal{G} +
    \mathsf{K} \mathsf{H} \mathsf{P} \mathcal{G'}^\top \mathsf{Q}^{-1} \mathcal{G} \text{,}
    \label{eq:delta-final}
    \end{equation}
    respectively.
    We further denote
    \begin{equation}
        \delta =  - \mathsf{P} \mathcal{G'}^\top \mathsf{Q}^{-1} \mathcal{G}
        \quad \text{or equivalently} \quad
        \delta =  - \mathsf{P} \mathcal{G'}^\top \mathsf{W} \mathcal{G} 
        \text{,}
        \label{eq:delta}
    \end{equation}
    with which the analysis scheme Eq.~\eqref{eq:dJ2} then takes the following simplified form:
    \begin{equation}
         \sx_j^\text{a} = \sx_j^\text{f} + \delta + \mathsf{K} (\oy_j - \mathsf{H} (\sx_j^\text{f} + \delta)) \text{,}
         \label{eq:renkm_update_1}
    \end{equation}
    or alternatively written as a two-step scheme:
    \begin{subequations}
    \begin{align}
        \tilde{\sx}_j^\text{f} & = \sx_j^\text{f} + \delta , \label{eq:pre-correction} \\
        \sx_j^\text{a} & = \tilde{\sx}_j^\text{f} + \mathsf{K} (\oy_j - \mathsf{H} \tilde{\sx}_j^\text{f}) \text{.}
        \label{eq:renkm_update_2}
    \end{align}
    \end{subequations}
    Note that Eq.~\eqref{eq:renkm_update_2} has the same form as the analysis scheme of regular ensemble Kalman methods, i.e., $\sx_j^\text{a} = \sx_j^\text{f} + \mathsf{K} (\oy_j - \mathsf{H} \sx_j^\text{f})$. In other words, the regularized analysis scheme introduces a pre-correction $\delta$ to the state vector $\sx_j^\text{f}$ to obtain the constrained state $\tilde{\sx}_j^\text{f}$ (see Eq.~\eqref{eq:pre-correction}).
    This pre-correction is what enforces the desired constraints. This is then followed by the Kalman correction in Eq.~\eqref{eq:renkm_update_2} to assimilate the observations.
    To enforce multiple constraints simultaneously, the regularization term can be written as a sum as follows:
    \begin{equation}
    \delta =  -\sum_{p}  \mathsf{P} \mathcal{G'}_p^\top \mathsf{Q}^{-1}_p \mathcal{G}_p \; 
    \quad \text{or equivalently} \quad
    \delta =  -\sum_{p} \mathsf{P} \mathcal{G'}_p^\top \mathsf{W}_p \mathcal{G}_p \; 
    \text{,}
    \label{eq:delta-sum}
    \end{equation}
    where the subscript $p$ is an index denoting the different constraints.
    A case with multiple regularization terms is shown in Section~\ref{sec:parameter_estimation}.
   The proposed regularized ensemble Kalman method is schematically illustrated in Fig.~\ref{fig:renkf-overview} by using the ensemble Kalman filtering (EnKF) procedure as an example, where our modification to the baseline EnKF is highlighted. 

  \subsection{Implementing regularization procedure for an iterative ensemble Kalman method}
  \label{sec:method_implementation}
    As presented above the regularized Kalman update is general for the numerous ensemble Kalman methods, including the ensemble Kalman filter and the ensemble Kalman smoother.
    In the test cases in this paper, we use an iterative ensemble Kalman method to solve steady-state inverse problems iteratively.
    The analysis step is modified to incorporate the regularized update scheme derived above.
    The analysis step is further modified here to overcome the effects of sample collapse on the regularization term and to avoid the dominance of the regularization term during early iterations.
    The details of this regularized iterative ensemble Kalman method used are presented below. 
    The method described below differs from the iterative ensemble Kalman method for steady problems~\cite{iglesias2013ensemble} only in the pre-correction step in the analysis. 
    The proposed method requires only a small algorithmic modification. 
    The unmodified method is used as a baseline for the test cases in Section~\ref{sec:results}.

    The iterative ensemble Kalman filter~\cite{iglesias2013ensemble} recasts steady state inverse problems as a dynamic data assimilation problem. 
    The state vector is augmented to include the projection of the state onto observation space, that is $\sx^{(\text{aug})} = [\sx,~\mathcal{H}(\sx)]^\top$. 
    The dynamic model describing the artificial dynamics is taken as $\sx^{\text{(aug)}}_{i+1} = [\sx^{(\text{aug})}_i,~\mathcal{H}(\sx^{(\text{aug})}_i)]^\top$, and the linear observation operator is then given as $\mathsf{H}=[0, I]$. 
    With these definitions the problem is equivalent to the standard EnKF for dynamic systems, where a linear observation operator is used and a single update is done at each observation time. 
    In this case the observation times are pseudo-time where the same observations are used each time, and a stopping criteria is used for determining the end time. 
    In the rest of this section the state, forward model, and linear observation operator are taken to be the augmented version described here and the ``$(\text{aug})$'' superscript is dropped. 
    The following description of the method is therefore equally applicable to the standard EnKF. 
    It is noted that in practice the iterative EnKF is implemented in an equivalent but more efficient manner, but here it is still presented in this way for consistency with other ensemble Kalman methods and with the preceding derivation.

    Sample collapse is a common issue when using ensemble Kalman methods~\cite{evensen2009data}. %
    Moreover, for iterative methods on stationary systems, the observation data are used repeatedly, which further exacerbates the sample collapse problem.
    This issue can be partly addressed by perturbing the observations (based on the observation error) at each iteration in addition to perturbing them for each sample.
    Once the samples collapse, the covariance matrix $\mathsf{P}$ approaches zero, and the magnitude of the weight matrix $\mathsf{W}$ (that is, $\mathsf{Q}^{-1}$) has to be very large in order to keep the regularization effective (i.e., to keep the regularization term of a similar order of magnitude to the data discrepancy term).
    In light of this observation, we recast the pre-correction term $\delta$ in Eq.~\eqref{eq:delta} as follows:
    \begin{equation}
        \delta = -\frac{\chi}{\| \mathsf{P} \|_\text{F}} \, \mathsf{P} \mathcal{G'}^\top \overline{\mathsf{W}} \mathcal{G} \text{,}
        \label{eq:modified_delta}
    \end{equation}
    where $\| \mathsf{P} \|_\text{F}$ is the Frobenius norm of matrix $\mathsf{P}$ and the weight matrix can be written as 
    \begin{equation}
        \mathsf{W} \equiv \frac{\chi}{\|\mathsf{P}\|_\text{F}} \overline{\mathsf{W}} \text{,} 
        \qquad
        \label{eq:lambda_collapse}
    \end{equation}
    with $\overline{\mathsf{W}}$ normalized such that its largest diagonal element is $1$. That is, the magnitude of $\mathsf{W}$ is dynamically adjusted based on $\|\mathsf{P}\|_\text{F}$ with $\chi$ kept constant.
    In doing so, only the ``direction'' of the covariance matrix $\mathsf{P}$  (i.e., information on the correlation of the samples) is preserved, which overcomes the detrimental effects of sample collapse on the added constraint.
    This makes it more intuitive to choose the algorithmic constant $\chi$.

    During the first few iterations, a large penalty parameter can lead to the regularization term being dominant and consequently the observations being ignored.
    For this reason, the parameter~$\chi$ is further modeled using a ramp function as
    \begin{equation}
      \chi(i) = 0.5 \chi_0 \left(\tanh \left(\frac{i-S}{d}\right)+1 \right) \text{,}
      \label{eq:regularization_function}
    \end{equation}
    where $\chi_0$ is the maximum value of~$\chi$ and $i$ denotes the iteration step.
    The parameters~$S$ and~$d$ control the slope of the ramp curve and are chosen to be $5$ and $2$, respectively, for all test cases in this paper.

    Given the prior distribution of the state vector $P(\sx)$, observation values $\mathsf{y}$ and error covariance matrix $\mathsf{R}$, and the constraint function $\mathcal{G}$ with the weight matrix $\mathsf{W}$ (or equivalently the covariance matrix $\mathsf{Q}$ of the constraint), the following steps are taken:
    \begin{enumerate}
    \item \textbf{(Sampling step)}\\
    Generate initial ensemble of state vectors, consisting of $M$ samples ${\{\sx^{(0)}_{j}}\}_{j=1}^M$, from the prior distribution of the states. %
    \item \textbf{(Prediction step)}\\
    For each sample, run the forward model $\sx^{a,(i-1)}\mapsto\sx^{(i)}$.
    \item \textbf{(Analysis step)}
    \begin{enumerate}[label={\roman*})]
        \item Estimate the sample mean $\overline{\sx}^{(i)}$ and covariance $\mathsf{P}^{(i)}$ as
    \begin{align}
      \overline{\sx}^{(i)} & = \frac{1}{M}\sum_{j=1}^M \sx^{(i)}_{j} \text{,} \\
    	\mathsf{P}^{(i)} & = \frac{1}{M-1} \mathsf{X}^{(i)}(\mathsf{X}^{(i)})^\top \text{,}
    \end{align}
    where $\mathsf{X}^{(i)}$ denotes the matrix formed by stacking the mean-subtracted sample vectors, i.e., $\mathsf{X}^{(i)} = \left[(\sx^{(i)}_1 - \overline{\sx}^{(i)}), \ldots, (\sx^{(i)}_M - \overline{\sx}^{(i)}) \right]$.
    
    \item Compute the Kalman gain matrix %
    \begin{equation}
     \mathsf{K}^{(i)}  = \mathsf{P}^{(i)}\mathsf{H}^\top ( \mathsf{R} + \mathsf{H} \mathsf{P}^{(i)} \mathsf{H}^\top)^{-1}
    \end{equation}
    \item Generate an ensemble of observations $\{ \oy^{(i)}_j \}_{j=1}^M$ from the joint normal distribution $\mathcal{N}(\oy,\mathsf{R})$.

    \item For each sample, constrain the state $\sx$ with a pre-correction $\delta$ as
    \begin{subequations}
    \begin{align}
      \delta^{(i)}_j &=  - \mathsf{P}^{(i)} (\mathcal{G'}[\sx^{(i)}_j])^\top \mathsf{W} \mathcal{G}[\sx^{(i)}_j] ,\\
      \tilde{\sx}_j^{(i)} & = \sx_j^{(i)} + \delta^{(i)}_j \text{,}
    \end{align}
    \label{eq:implementation_precorrection}
    \end{subequations}
    with the 
    weight matrix $\mathsf{W}$ and parameter $\chi$ determined from Eqs.~\eqref{eq:lambda_collapse} and~\eqref{eq:regularization_function}.  %
    
    \item For each sample, update the constrained state $\tilde{\sx}$ as
    \begin{equation}
        \label{eq:analysis}
        \sx^\text{a,(i)}_j = \tilde{\sx}^{(i)}_j + \mathsf{K}^{(i)} \left(\oy^{(i)}_j - {\mathsf{H}} \tilde{\sx}^{(i)}_j \right) \text{.} %
    \end{equation}
    \end{enumerate}
    \item Return to step~2 until the ensemble is statistically converged.
    \end{enumerate}

    We emphasize that the procedure described above differs from the baseline ensemble Kalman methods \emph{only} in the additional pre-correction step in Eq.~\eqref{eq:implementation_precorrection}.
    That is, the proposed regularization only requires this small algorithmic modification to existing ensemble Kalman methods and is thus very straightforward to implement. 
    The Python code for the proposed method and the test case from Section~\ref{sec:parameter_estimation} are provided in a publicly available GitHub repository~\cite{renkf-git}. Moreover, the proposed method is implemented in our software suite DAFI for data assimilation and field inversion in science and engineering applications~\cite{dafi_github,dafi_pypi,dafi_rtd}.

  \subsection{Generality and possible extensions to other ensemble Kalman methods}
    In Section~\ref{sec:method_implementation}, we described how the proposed regularized Kalman update was implemented for the specific iterative method used in the test cases in this paper. 
    However, we emphasize that it can be implemented into other ensemble Kalman methods straightforwardly.
    Examples of other ensemble Kalman methods include the ensemble Kalman smoother and the EnKF with multiple data assimilation (EnKF-MDA)~\cite{emerick2012history}.
    The ensemble Kalman smoother uses time dependent observations to infer the initial conditions of the observed fields. 
    The EnKF-MDA divides a Kalman update into a series of multiple, appropriately weighted, smaller updates. 
    Because of these differences, the specific implementation for the each method would be different, but the modification to the analysis step will be the same in all cases.
    Additionally, it is noted that the specific methods for dealing with the problem of sample collapse will vary from study to study, and the choice of the parameter $\chi$ as described in Section~\ref{sec:method_implementation} is specific to this study.

\section{Results}
\label{sec:results}
  We use three test cases to showcase the use and performance of the proposed regularized ensemble Kalman method. 
  First, we use the proposed method for the parameter estimation problem used by Wu et al.~\cite{wu2019adding}, which consists of a global minimization problem and for which the true solution is known.
  Parameter estimation problems typically have more observations than inferred parameters, and the inferred parameters are discrete scalars. 
  For the case tested, however, the number of observations is smaller than but of the same order of magnitude as the number of inferred parameters, making it ill-posed. 
  For this case we test a number of different constraints and prior mean (initial guess) and show that the proposed regularized method is effective in removing the ill-posedness of the data assimilation problem, making a better inference on the parameters.
  Parameter estimation problems in fluid mechanics include, for instance, determining the values of empirical parameters in specific turbulence models.
  Second, we test a field inversion problem: the one-dimensional diffusion equation on a finite domain with homogeneous boundary conditions.
  The quantity to be inferred is a discretization of a continuous field, which is fundamentally different from the discrete scalars inferred in the first case.
  In this case, the number of inferred values is much larger than the number of observations.
  Finally, we test the proposed method for a more complex and relevant field inversion problem: the RANS closure problem.
  In this case, we infer the eddy viscosity field for a two-dimensional turbulent flow over periodic hills. %
  For all cases, we show the advantage of the proposed regularized method over the traditional Kalman method in inferring the correct parameters or field by overcoming the ill-posedness intrinsic to inversion problems.

  Both parameter estimation and field inversion problems have applications in computational fluid dynamics (CFD). 
  In the case of field inversion, an important application is inferring the correct Reynolds stress field, and this is showcased in the third test case in this section. 
  In the case of parameter estimation, one important application is inferring the empirical parameters for the constructive turbulence models.
  Typically, many of these empirical parameters have underlying constraints determined from their theoretical derivation or numerical tests. 
  For instance, Poroseva and Bezard~\cite{poroseva2001ability} recommend the relationship $\sigma_\varepsilon / \sigma_k =1.5 $ in the $k$--$\varepsilon$ model~\cite{jones1972prediction} for aerodynamic simulations. 
  Oliver and Moser~\cite{oliver2011bayesian} used a Bayesian approach to quantify the uncertainty of model parameters and indicated that the parameter $k$ and $c_{v1}$ in the Spalart--Allmaras model are linearly related. These are equality constraints.  As an example of an inequality constraint, it has been shown through numerical experiments by Ray et al.~\cite{ray2018learning} that the parameters in the $k$-$\varepsilon$ model have to satisfy $C_{\varepsilon 2} > C_{\varepsilon 1}$. The physical reason behind this delineation is that the ratio $C_{\varepsilon 2} / C_{\varepsilon 1}$  corresponds to the spreading rate of a free jet. A ratio of  $C_{\varepsilon 2} / C_{\varepsilon 1} < 1$ would lead to a contracting jet, which is non-physical~\cite{xiao2019quantification}.
  Nevertheless, most current works on ensemble-based parameter inferences neglect such underlying constraints, partly because of the difficulty in enforcing constraints in existing ensemble-based inversion methods.

  \subsection{Parameter estimation}
  \label{sec:parameter_estimation}
    The first test case is the parameter estimation problem used by Wu et al.~\cite{wu2019adding}.
    The observable quantity~$\bm{\mathsf{z}} \in \mathbb{R}^2$ is a vector related to the parameter vector $\wv=[\omega_1,\omega_2]^\top$ (the state to be inferred) by the forward model $F$ as follows:
    \begin{equation}
      \bm{\mathsf{z}} =
      \begin{bmatrix}
        \mf_1 \\
        \mf_2
      \end{bmatrix}
        = F[\wv] =
      \begin{bmatrix}
        \exp(-(\omega_1 + 1)^2 - (\omega_2 + 1)^2) \\
        \exp(-(\omega_1-1)^2-(\omega_2-1)^2)
      \end{bmatrix} \text{.}
    \end{equation}
    The observation map is given by
    \begin{equation}
      \oy = H \bm{\mathsf{z}} = HF[\wv],
    \end{equation}
    with $H=[-1.5, -1.0]$.
    Given the observation $\oy = -1.0005$, the inverse problem consists of inferring the parameters $\omega^\text{opt}$ that minimize the discrepancy between the observation $\oy$ and model output $F[\wv]$ (after the latter has been projected to the observation space). That is,
    \begin{align}
      \mathrm{J} [\wv] = \parallel \oy - HF[\wv] \parallel^2 \text{,} \\
      \wv^{\text{opt}} = \mathop{\arg\min}_{\wv} \mathrm{J}[\wv] \text{.}
    \end{align}

\begin{figure}[!htbp]
        \centering
        \includegraphics[width=0.6\linewidth]{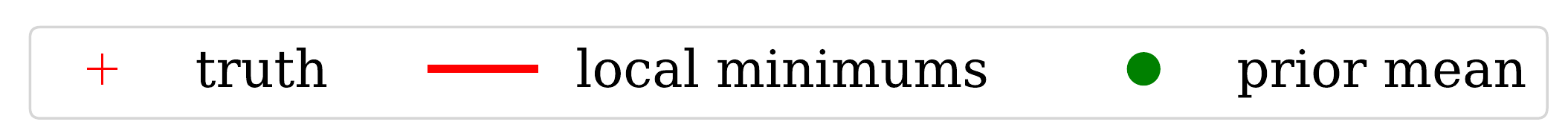}\\
        \includegraphics[width=0.5\linewidth]{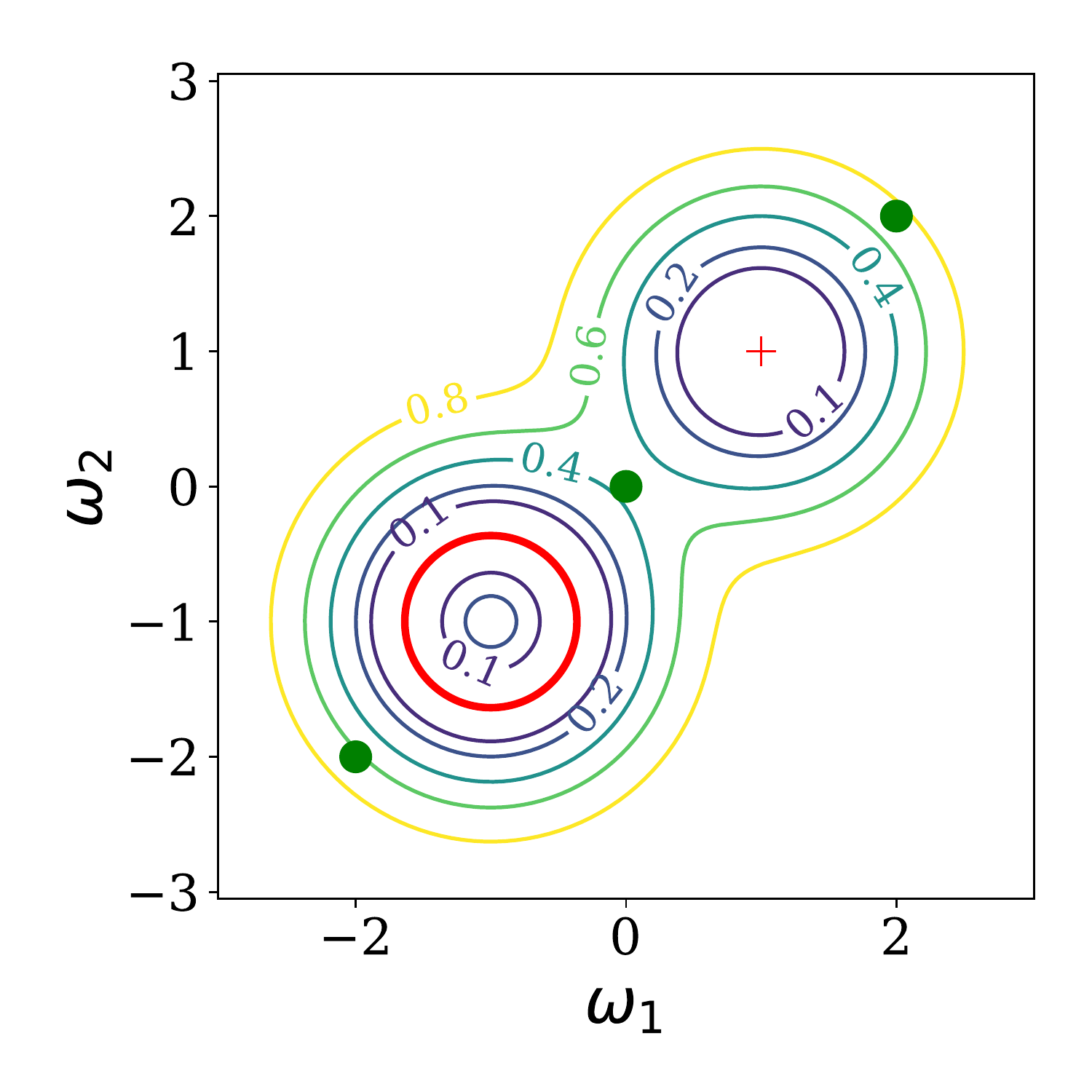}
        \caption{Contour plot of the discrepancy $\mathrm{J}[\wv]$. The two groups of local minima are indicated with the red/gray cross $`` + "$ (Group I) and the red/gray circle (Group II).}
        \label{fig:contour_I}
    \end{figure}

    A contour plot of $\mathrm{J}$ is shown in Figure~\ref{fig:contour_I}.
    This case has two groups of local minima: (Group I) the single point at $\wv=(1.0,1.0)$, and (Group II) the circle of points defined by
    \begin{equation}
        (\omega_1 + 1)^2 + (\omega_2 + 1)^2 = \log1.5 .
    \end{equation}
    Numerous local minima result in satisfactory agreement with the observation, which makes the inference of the true parameter $\wv$ challenging.
    Fundamentally, this results from insufficient information from the observations, and the goal of the proposed regularized method is to guide the inference to the true values of the parameters by incorporating additional sources of information.
    Here the robustness of the method is tested by using different constraints and three different prior means for the parameters, similar to Wu et al.~\cite{wu2019adding}.
    
    \subsubsection{Case details}
      The ensemble Kalman method is a Bayesian data assimilation framework and requires a prior distribution for the parameters.
      A Gaussian process is used with mean~$\wv_\text{p}$, equal standard deviation~$\sigma_\text{p}=0.1$ for both parameters, and zero covariance.
      Three different prior means $\wv_\text{p}$ are tested:  $(-1, -1)$, $(0,0)$, and $(2,2)$.
      For the observation, the true value of the parameters is taken to be~$\wv = (1.0, 1.0)$, and the observation to be~$\oy=-1.0005$ with standard deviation~$\sigma_\text{y}=0.01$.
      Three different sets of constraints are enforced: an equality constraint, an inequality constraint, and multiple inequality constraints.
      Combined with the three prior distributions, a total of nine constrained cases were tested in addition to three baseline cases with no constraints. 

      We consider three different sources of information on the quantity~$\omega_1 + \omega_2$: equal to~$2$, greater than~$1$, and less than~$3$, with corresponding constraint equations:
      \begin{align}
        h_{\text{eq}}[\wv] &= \omega_1 + \omega_2 - 2 = 0 \text{,} \\
        h_{\text{in1}}[\wv] &= -\omega_1 - \omega_2 + 1 < 0  \text{,} \\
        h_{\text{in2}}[\wv] &= \omega_1 + \omega_2 - 3 < 0 \text{.} 
      \end{align}
      For inequalities of the form~$h_\text{in}[\wv] < 0$ a penalty function~$\mathcal{G}[\wv]$ of the form
      \begin{align}
       \mathcal{G}[\wv] =  \phi[h_{in}[\wv]]=
        \left \{
        \begin{array}
        {ll}{0} & {\text { for } h_\text{in}[\wv]<0} \\
        {h_\text{in}[\wv]^{2}} & {\text { for } h_\text{in}[\wv]] \geq 0}
        \end{array}
        \right.
        \label{eq:inequality_constraint}
      \end{align}
      is used. 
      This means that for the inequality constraints, the penalty function is only active when the constraint is violated.
      The derivative can be obtained using the chain-rule as
      \begin{align}
       \mathcal{G}'[\wv] = \phi'[h_{\text{in}}[\wv]]=
        \left \{
        \begin{array}
        {ll}{ (0, 0)} & {\text { for } h_{\text{in}}[\wv]<0} \\
        {(-2 h_\text{in}[\wv], -2 h_\text{in}[\wv])} & {\text { for } h_\text{in}[\wv] \geq 0}
        \end{array}
        \text{.}
        \right.
        \label{eq:dev_inequality_constraint}
      \end{align}

      The three constraints used as regularization are summarized in Table~\ref{tab:summary_penalty_scalar}.
      The last case consists of multiple inequalities and serves as an illustration of combining multiple sources of information into the framework.
      The penalties in Table~\ref{tab:summary_penalty_scalar} are implemented as in Eq.~\eqref{eq:modified_delta}, with covariance set to the identity matrix $\overline{\mathsf{W}}=I$.
      A regularization parameter of $\chi_0=0.1$ is used. %

      \begin{table}[!htb]
        \caption{Summary of the constraints used in the parameter estimation problem.}
        \label{tab:summary_penalty_scalar}
        \begin{tabular}{l l l}
        \hline
        \hline
        case & constraint type & penalty function \\
        \hline
        C1 & equality & $\mathcal{G}(\wv) = h_\text{eq}[\wv]$ \\
        C2 & inequality & $\mathcal{G}(\wv) = \phi[h_\text{in1}[\wv]]$ \\
        C3 & multiple & $\mathcal{G}(\wv) = \phi[h_\text{in1}[\wv]] + \phi[h_\text{in2}[\wv]]$ \\
        \hline
        \end{tabular}
      \end{table}

    \subsubsection{Results}
\begin{figure}[!htbp]
          \centering
         \includegraphics[width=\linewidth]{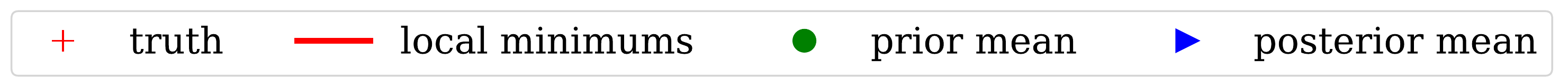}

          \begin{subfigure}[b]{0.49\linewidth}
              \centering            \includegraphics[width=\textwidth]{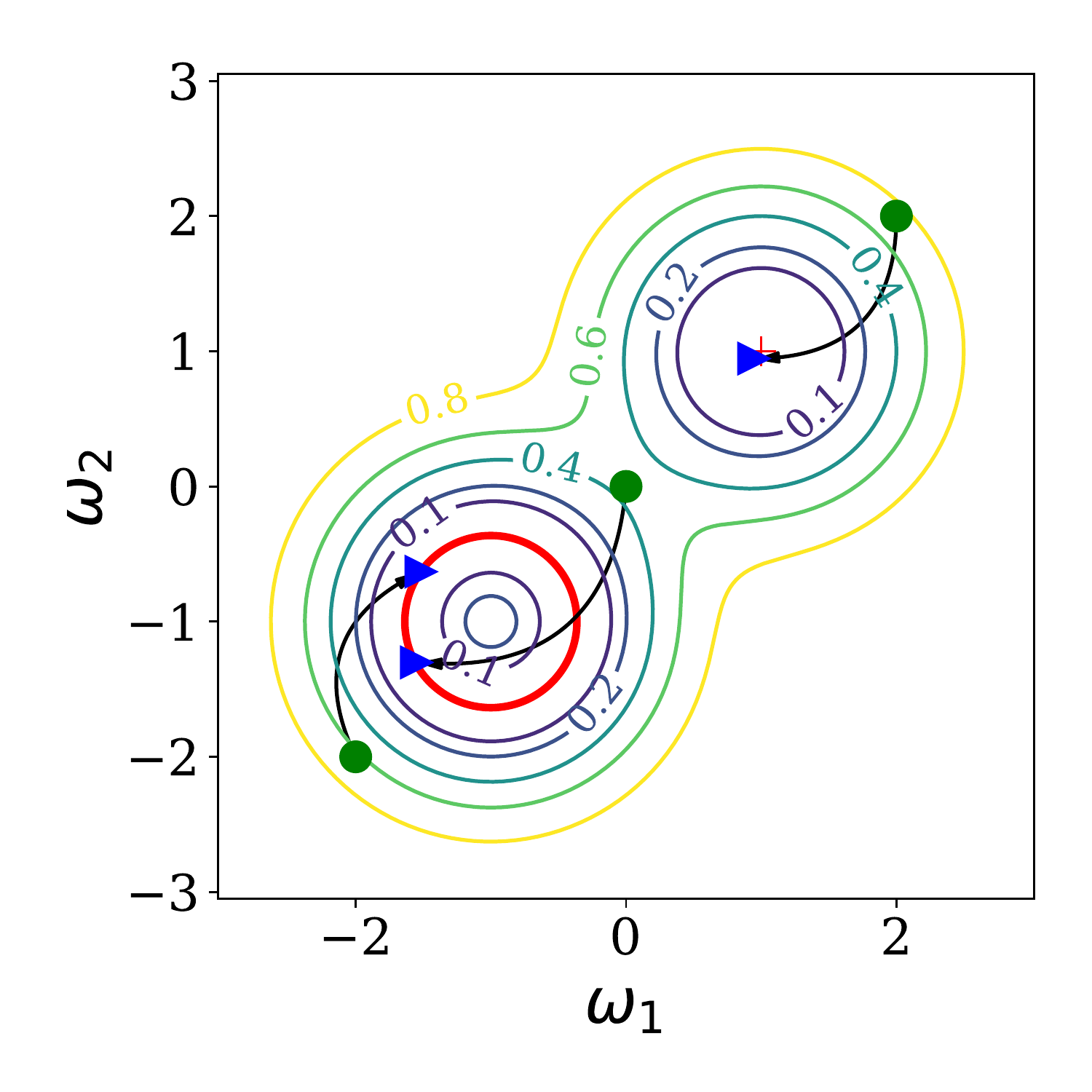}
              \caption{Baseline (no constraint)}
              \label{fig:case1_results_enkf}
          \end{subfigure}
          \begin{subfigure}[b]{0.49\linewidth}
              \centering
              \includegraphics[width=\textwidth]{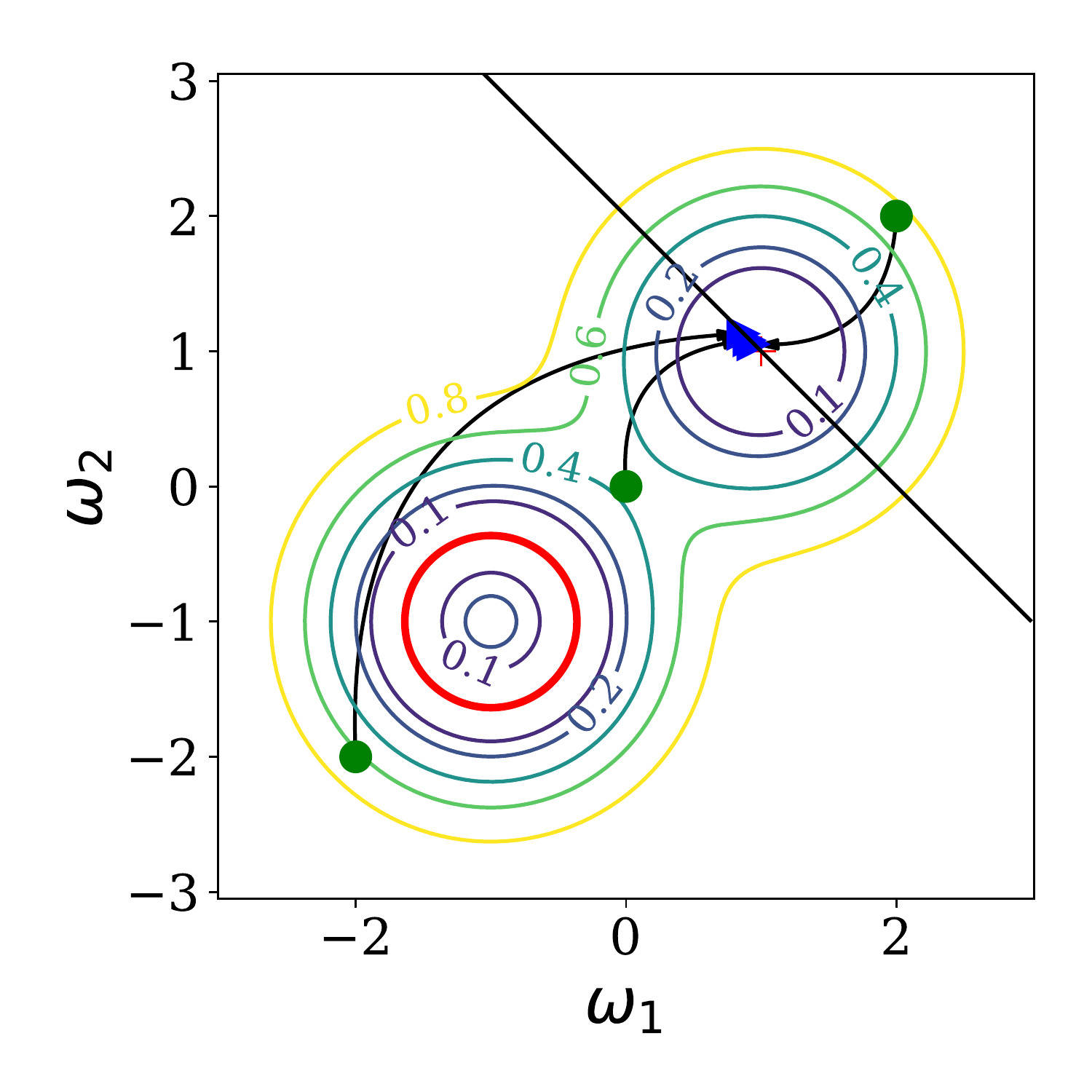}
              \caption{Regularized, Case C1}
              \label{fig:case1_results_equality}
          \end{subfigure}
          \\
          \begin{subfigure}[b]{0.49\linewidth}
              \centering
              \includegraphics[width=\textwidth]{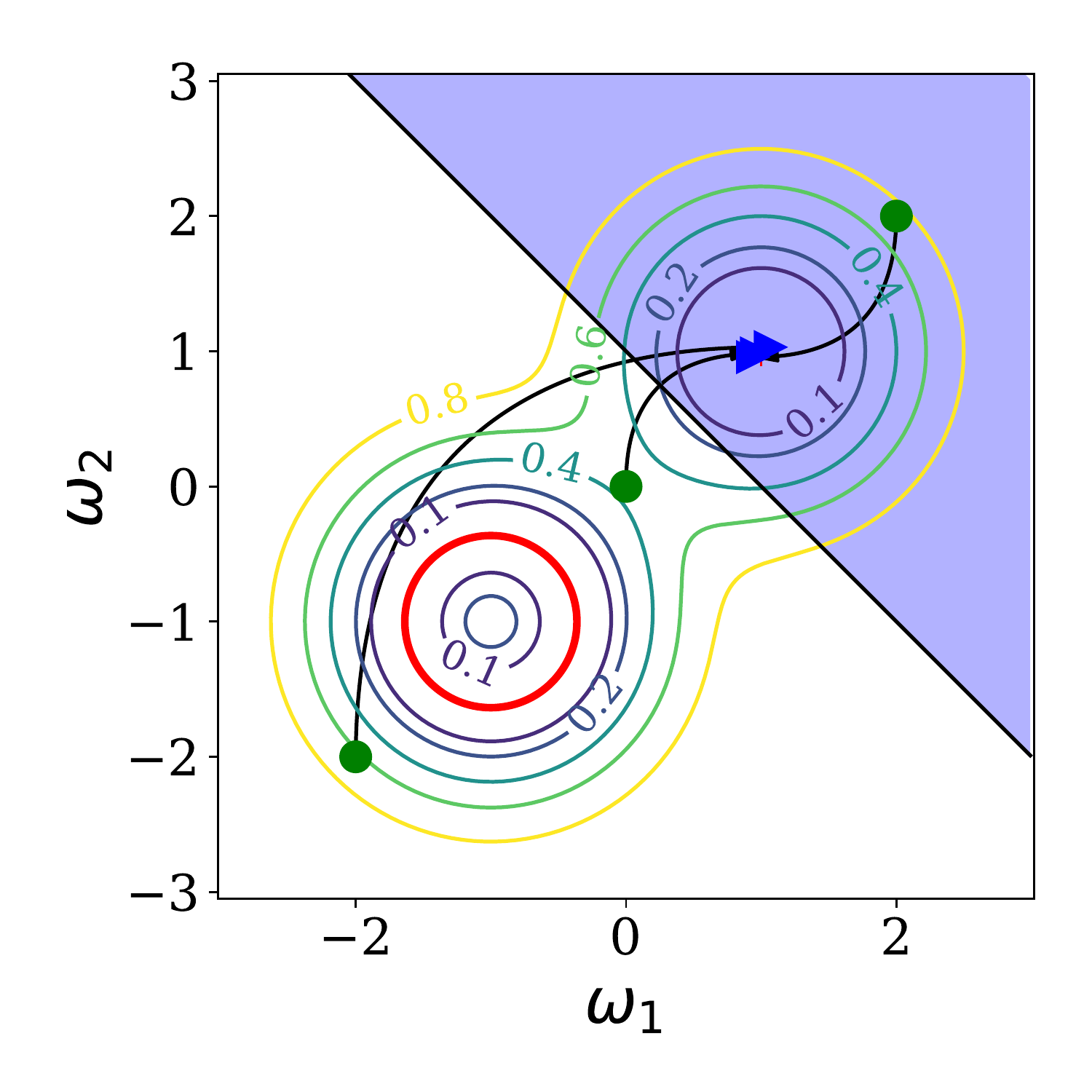}
              \caption{Regularized, Case C2}
              \label{fig:case1_results_inequality}
          \end{subfigure}
              \begin{subfigure}[b]{0.49\linewidth}
              \centering
              \includegraphics[width=\textwidth]{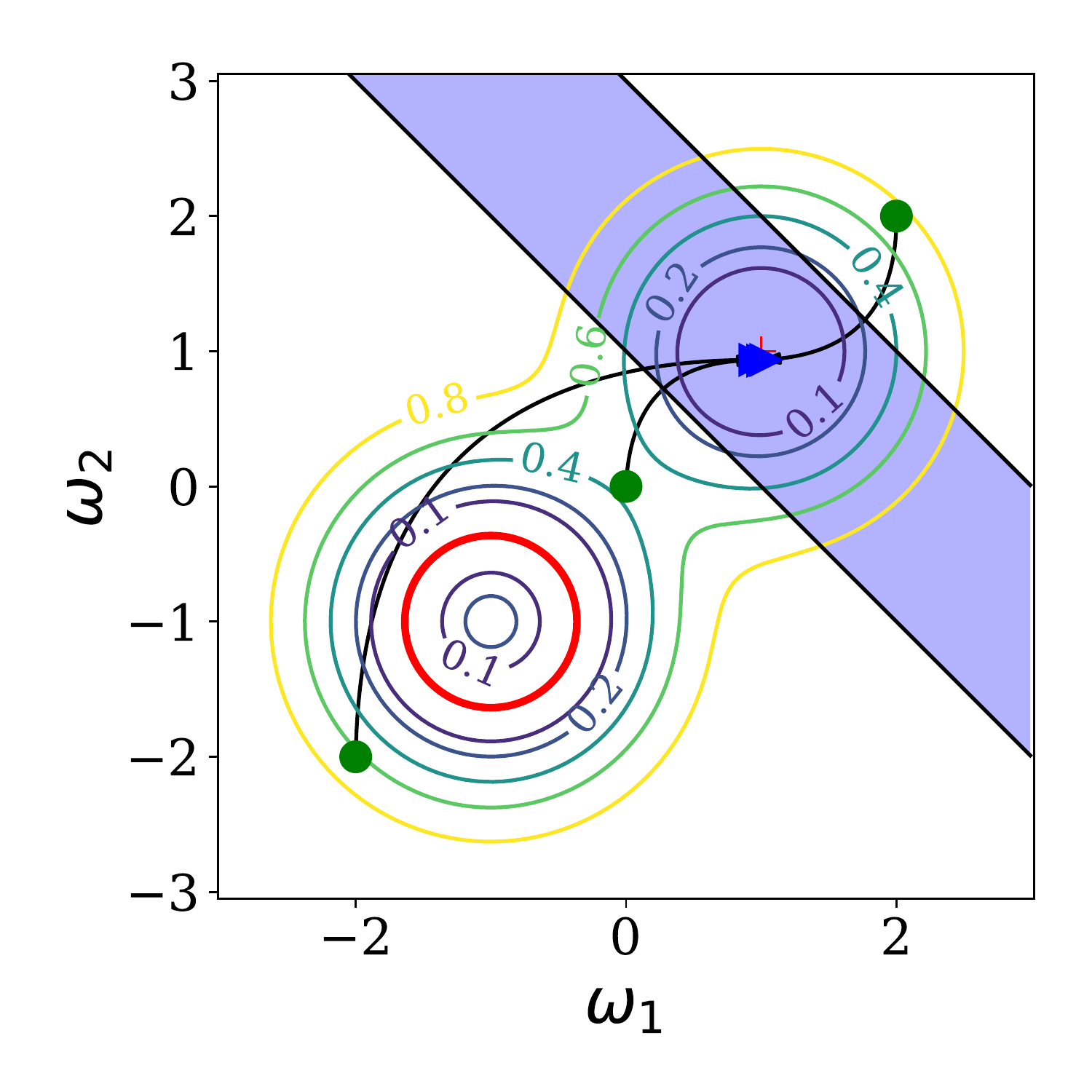}
              \caption{Regularized, Case C3}
              \label{fig:case1_results_multiple}
          \end{subfigure}
          \caption{Results of the parameter estimation problem using the baseline and proposed regularized methods with different constraints. For all methods three different prior means (green/gray dots) are considered. (a) Baseline case; (b) Case C1: proposed method with $G[\omega]=h_1[\omega]$ where the penalty function is indicated by the black straight line ; (c) Case C2: proposed method with $G[\wv] = \phi[h_\text{in1}[\wv]]$ where the blue/gray region indicates where the constraint is inactive; (d) Case C3: proposed method with $G[\wv] = \phi[h_\text{in1}[\wv]] + \phi[h_\text{in2}[\wv]]$ where the blue/gray region indicates  where the constraint is inactive. With the baseline method, different priors converge to different local minima. With the proposed regularized method all priors can converge to the true local minimum $(1, 1)$, indicated by a blue/gray triangle.}
          \label{fig:case1_results}
      \end{figure}

      As a baseline, the ensemble Kalman method is used without any regularization (constraints) for each of the three prior distributions considered.
      The results are shown in Fig.~\ref{fig:case1_results_enkf} and Table~\ref{tab:case1_results}.
      It is noticeable that for different priors the inference will converge to a different local minimum, with the priors with mean of~$(-2,-2)$ and~$(0,0)$ converging to local minima belonging to Group II.
      Next, the proposed regularized method is tested using the equality constraint (case C1).
      The results are shown in Fig.~\ref{fig:case1_results_equality} and Table~\ref{tab:case1_results}.
      Using the equality constraint the inference converges around the truth for all three priors considered.

      Similarly, the inequality constraint (case C2) is able to make the inference converge around the truth for all three priors considered, completely avoiding the Group II local minima.
      These results are shown in Fig.~\ref{fig:case1_results_inequality} and Table~\ref{tab:case1_results}.
      It should be noted that the penalty term in this case is only active when the constraint is violated.
      This results in that while this constraint can avoid inference dropping into the local minima in Group II it cannot further enhance the optimization result, as in the case with the equality constraint.
      Finally, the method is tested with multiple inequality constraints (case C3) in order to showcase how to incorporate multiple sources of information.
      Once again, the inference converges around the truth for all three priors considered, and the results are shown in Fig.~\ref{fig:case1_results_multiple} and Table~\ref{tab:case1_results}.

      \begin{table}
        \caption{Results of the baseline and regularized inference with different constraints.}
        \label{tab:case1_results}
        \begin{tabular}{c c | c r c c}
          \hline
          \hline
          method & initial $\omega$ & inferred $\omega$ & $HF[\omega]$ & error ($\omega$) & error ($HF[\omega])$ \\
          \hline
          truth/observation & --- & $(1.0, 1.0)$ & $-1.0005$ & --- & --- \\
          \hline
          Baseline & $(-2, -2)$ & $(-1.52,-0.63)$ & $-1.0010$ & $(252\%,163\%)$ & $0.05\%$ \\
           & $(0, 0)$ & $(-1.55,-1.30)$ & $-1.0108$ & $(255\%,230\%)$ & $1.03\%$ \\
           & $(2, 2)$ & $(0.94,0.95))$ & $-0.9947$ & $(6\%,5\%)$ & $0.58\%$ \\
          \hline
          C1 & $(-2, -2)$ & $(1.06, 0.93))$ & $-0.9921$ & $(6\%,7\%)$ & $0.84\%$ \\
           & $(0, 0)$ & $(1.06, 0.93)$ & $-0.9921$ & $(6\%,7\%)$ & $0.84\%$ \\
           & $(2, 2)$ & $(1.02, 0.98)$ & $-0.9997$ & $(2\%,2\%)$ & $0.08\%$ \\
          \hline
          C2 & $(-2, -2)$ & $(1.07, 1.03)$ & $-0.9946$ & $(7\%,3\%)$ & $0.59\%$ \\
           & $(0, 0)$ & $(0.96, 0.98)$ & $-0.9986$ & $(4\%,2\%)$ & $0.19\%$ \\
           & $(2, 2)$ & $(0.94, 0.96)$ & $-0.9956$ & $(6\%,4\%)$ & $0.49\%$ \\
          \hline
          C3 & $(-2, -2)$ & $(1.03, 0.94)$ & $-0.9961$ & $(3\%,6\%)$ & $0.44\%$ \\
           & $(0, 0)$ & $(1.01, 0.93)$ & $-0.9956$ & $(1\%,7\%)$ & $0.48\%$ \\
           & $(2, 2)$ & $(0.95, 0.94)$ & $-0.9947$ & $(5\%,6\%)$ & $0.58\%$ \\
          \hline
        \end{tabular}
      \end{table}

      The errors in the inferred quantities are quantified based on the agreement with their corresponding true values.
      The error on quantity~$q$ is defined as
      \begin{equation}
          \text{error} = \frac{\|q_{\text{truth}} - q_{\text{estimate}}\|}{\|q_{\text{truth}}\|}.
          \label{eq:error_def}
      \end{equation}
      The errors on the parameters $\wv$ and the observed point are shown in Table~\ref{tab:case1_results}.
      For all constraints and prior means considered, the proposed regularized method can infer the parameters accurately, comparable to the baseline case with prior mean of $(-2,-2)$. 
      For the estimated observation error, all cases, including the unregularized baseline cases, can give a satisfactory estimation in the observation space.

      The convergence criteria in this work is based on the discrepancy principle~\cite{schillings2018convergence}. That is,
      \begin{equation}
          \parallel \mathsf{Hx}-\mathsf{y} \parallel \le \tau \parallel \epsilon \parallel \text{,}
      \end{equation}
      where $\tau \ge 1$ and~$\epsilon$ is the observation error, the norm of which is estimated with~$\parallel \epsilon \parallel = \sqrt{\operatorname{trace}(R)}$. We chose $\tau=2$ in this work. For the cases with constraints, convergence is assumed when the observation and the constraint function meet the discrepancy principle simultaneously. 
      The number of iterations required for convergence for each case are provided in Table.~\ref{tab:summary_converge_scalar}.
      \begin{table}[!htb]
        \caption{Summary of the number of iterations at which the convergence is reached in the parameter estimation problem.}
        \label{tab:summary_converge_scalar}
        \begin{tabular}{l | c c c}
        \hline
        \hline
        case & prior $(-2, -2)$ & prior $(0, 0)$ & prior $(2, 2)$ \\
        \hline
        baseline & $5$ & $3$ & $32$ \\
        \hline
        C1 & $188$ & $371$ & $9$\\
        \hline
        C2 & $279$ & $57$ & $30$\\
        \hline
        C3 & $95$ & $75$ & $29$\\
        \hline
        \end{tabular}
      \end{table}
      It can be seen that for the priors $(-2, -2)$ and $(0, 0)$ with which the baseline case converges to local minima, enforcing constraints with the proposed method can lead to the truth but requires more iterations. That is because in the early iterations the state first reaches other local minimum and at the same time, the variance of samples is significantly reduced. After that, the constraint forces the state to jump out of local minima but will approach the truth very slowly due to the narrow searching space spanned by the samples.
      Also, the baseline case with prior $(2, 2)$ takes more iteration than that with prior $(0, 0)$ and $(-2, -2)$. That is likely due to the forward model having larger gradient at prior $(0,0)$ and $(-2,-2)$, and it can therefor reach the nearest minimum faster.
      For the prior $(2, 2)$, the inequality constraint (case C2) is inactive, and accordingly, the convergence speed is similar to that of the baseline, while enforcing the equality constraint (case C1) can speed up the convergence.
    
      It should be noted that the regularization parameter $\chi$ in the penalty term in the cost function is inflated as in Eq.~\eqref{eq:regularization_function} to ensure the robustness of the analysis step. 
      The hyper-parameters in the ramp function may affect the inference performance. 
      Concretely, the parameter~$\chi_0$ has to be inflated sufficiently to regularize the inference but not so much as to ignore the observations. 
      If the penalty term is too small it cannot drag the inference away from the erroneous local minima. 
      The hyper-parameters $\chi_0$, $S$, and $d$ in Eq.~\eqref{eq:regularization_function} were chosen based on a parameter study. 
      The parameter study suggests that the equality constraint is robust with a large range of parameters leading to correct inference. 
      However, the inequality constraint was found to be more sensitive to these parameters. This is due to the nature of such constraints and not caused by the intrinsic limitations of the proposed method.
      The equality constraint embeds more information about the truth, which can further enforce the inference to the expected point. 
      In contrast, inequality constraints can only drag the inferred parameters out of the region where the constraint is violated but cannot further inform the inference process as the equality constraint does. 
      Consequently, too large a penalty term may result in over-correction and lead to inference divergence, while too small a penalty term may not be sufficient to force the parameters out of the constraint-violating region and away from the undesired local minima. Detailed results of the parameter study are presented in \ref{sec:tunable_parameter}.

  \subsection{Field inversion}
  \label{sec:diffusion}
    The second test case is a field inversion case, in which observations of a field described by a partial differential equation (PDE) are used to infer a latent field in the PDE. 
    Specifically, we infer the diffusivity field in the one-dimensional diffusion equation by observing the output field (e.g., temperature) at a few locations.
    As is the case in general for field inversion problems, the number of observations is much smaller than the dimensions of the discretized domain.
    This increases the ill-posedness of the problem and makes it challenging to infer the true latent field.
    We apply the proposed method to regularize the problem and demonstrate its ability to infer the correct field by incorporating additional knowledge into the inversion scheme.

    The diffusion equation is given by
    \begin{equation}
         -\frac{d}{dx} \left(\mu[x] \frac{du}{dx}\right) = f[x] \text{,}
         \label{eq:diffusion_def}
    \end{equation}
    where~$x$ is the one-dimensional spatial coordinate,~$u$ is the quantity being diffused which is considered the output observable field,~$f[x]$ is a source term in units of~$u$ per time, and~$\mu[x]$ is the diffusivity field which is regarded as the latent field to be inferred. 
    Here we consider the diffusion of a non-dimensional quantity~$u$ (e.g., normalized by a reference value), but the equation can be used for many different applications. 
    For instance, it could be used for heat distribution along a rod, where~$u$ is temperature,~$f$ is distribution of heat sources, and~$\mu$ is thermal diffusivity of the material. 
    Another common application is pollutant concentration in a fluid, where~$u$ is concentration density,~$f$ is distribution of pollutant sources, and~$\mu$ is mass diffusivity of the pollutant in that medium. 
    We consider a domain of length~$L_x$, a source term~$f[x] = 100\sin(2\pi x/L_x)$, and homogeneous boundary conditions~$u|_{x=0}=u|_{x=L_x}=0$.
    The domain is discretized into~$50$ equal length cells, and the equation is discretized using the central difference scheme.
    The output field~$u$ is observed at nine equally spaced locations~$x/L_x=0.1,0.2, \cdots, 0.9$, and the goal is then to infer the value of the discretized diffusivity field at each of the~$50$ cells.

    \subsubsection{Case details}
\begin{figure}
      \centering
      \includegraphics[width=0.6\linewidth]{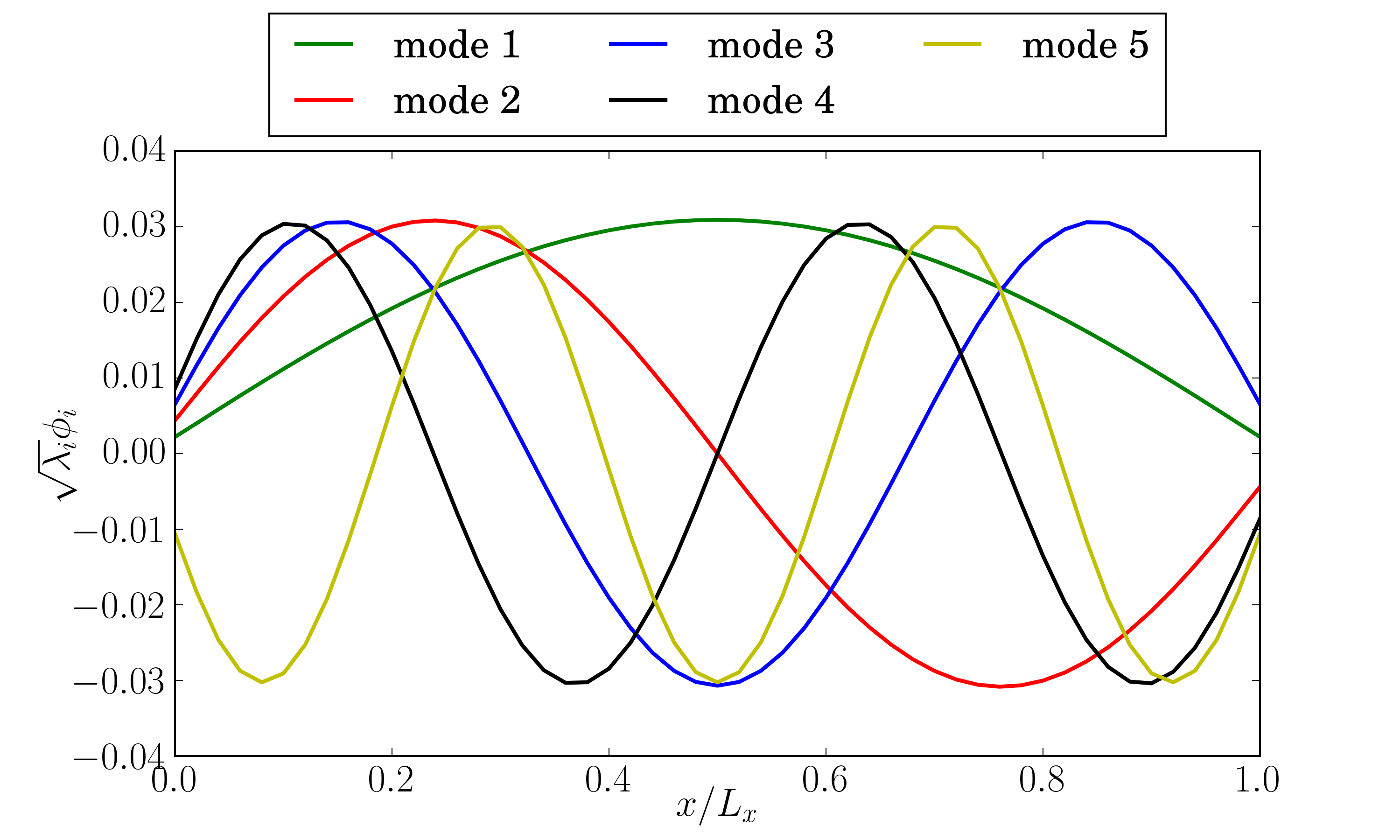}
      \caption{First 5 KL modes in the diffusion case scaled by their corresponding eigenvalues.} 
      \label{fig:diffusion_modes}
    \end{figure}

    The values of the discretized diffusivity field are not independent, and some sort of spatial correlation needs to be enforced.
    Furthermore, diffusivity is a field with physical meaning and subject to the physical constraint that it must be non-negative.
    To ensure positivity, the logarithm of diffusivity~$\log [\mu/\mu_0]$ is inferred, where~$\mu_0$ is a reference diffusivity value.
    To enforce spatial correlation and smoothness, the field~$\log [\mu/\mu_0]$ is assumed to be a sample of a Gaussian process~$\log[\mu/\mu_0]= \mathcal{GP}(0, \mathcal{K})$ with covariance kernel~$\mathcal{K}$.
    Using Kahunen-Lo\`{e}ve (KL) decomposition the field can be written as
    \begin{equation}
        \log[\mu/\mu_0] = \sum_{i=1}^n \omega_i \sqrt{\lambda_i} \phi_i,
        \label{eq:kl}
    \end{equation}
    where~$\lambda_i$ and~$\phi_i$ are the eigenvalues and unit eigenvectors of the kernel~$\mathcal{K}$ arranged in decreasing magnitude of the eigenvalues, i.e., $\lambda_1 \ge \lambda_2 \ge \ldots \lambda_n$, and~$\omega_i$ is the coefficient for mode $i$.
    While~$n$ is theoretically equal to the discretization size, it is common to set it to a much smaller value due to the usually rapid decrease of the magnitude of the eigenvalues. 
    This also results in dimensionality reduction, which can be beneficial in large~$2$- or~$3$-dimensional problems discretized on large meshes.
    The problem now consists of inferring the coefficients~$\omega_i$ rather than the discretized~$\log[\mu/\mu_0]$ field directly.
    We use the square exponential kernel with standard deviation~$\sigma_\text{p}$ and length scale~$l$, which for two points~$x$ and~$x'$ is given by
    \begin{equation}
        \mathcal{K}(x, x') = \sigma_\text{p}^2\exp \left(-\frac{\| x-x' \|^2}{l^2}\right).
        \label{eq:kernel}
    \end{equation}
    A standard deviation of~$\sigma_\text{p}=1.0$ is used, and the length scale is chosen as~$l=0.02L_x$, a relatively small value to allow for noisy inferred fields, making the problem artificially more difficult.
    The first five modes scaled by their respective eigenvalues are shown in Fig.~\ref{fig:diffusion_modes}.
    It can be seen that higher modes correspond to higher frequencies and that the magnitudes of the modes decrease slowly, due to the small correlation length in the covariance kernel.

\begin{figure}[!htbp]
      \centering
      \includegraphics[width=0.7\textwidth]{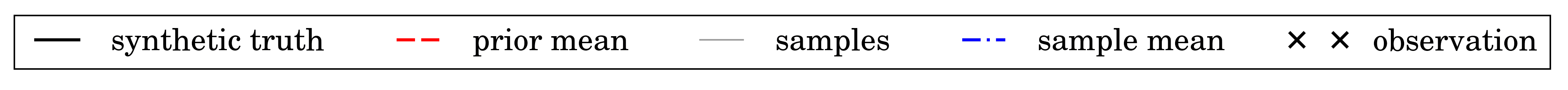}\\
      \begin{subfigure}[b]{0.32\linewidth}
          \includegraphics[width=\textwidth]{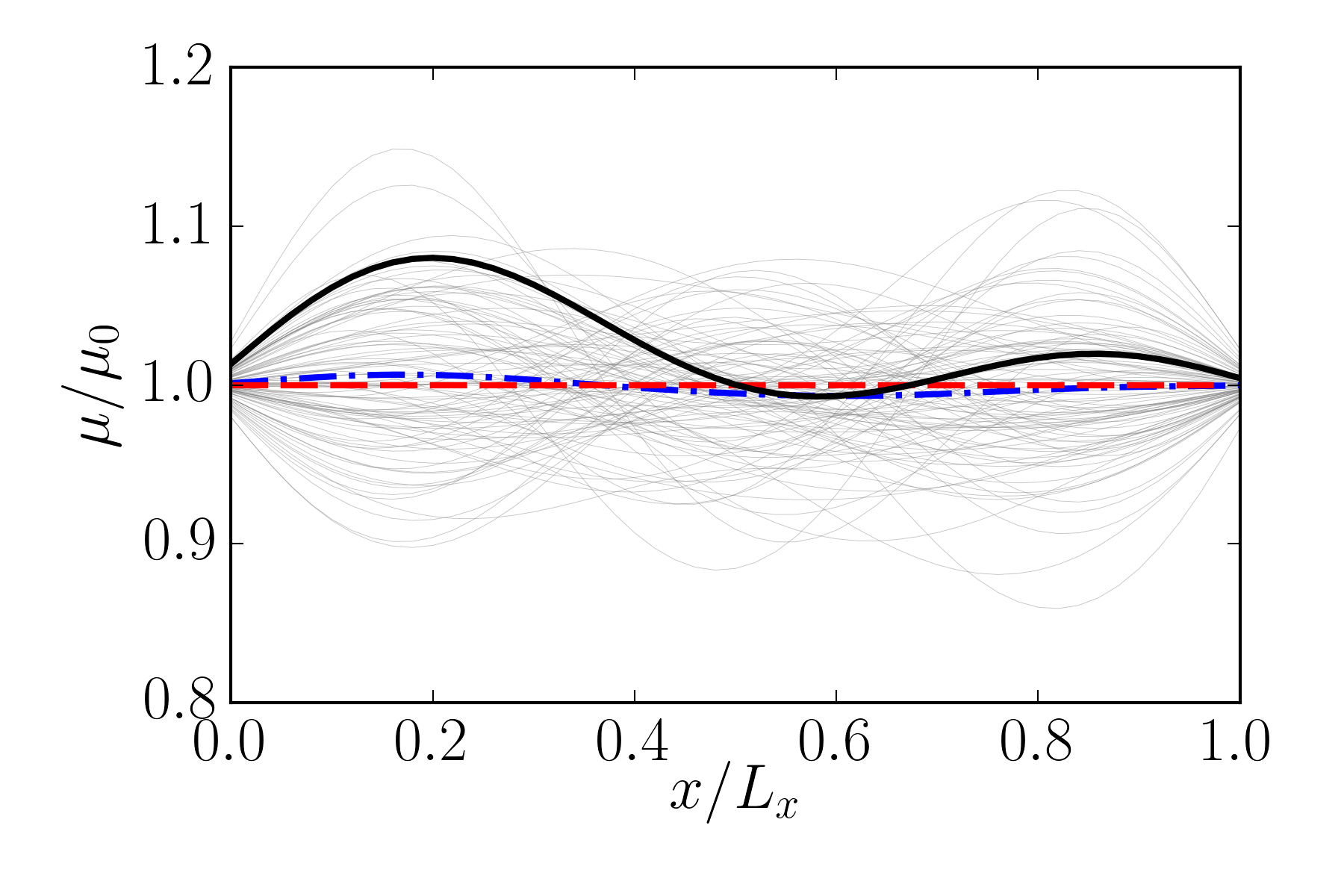}
          \caption{$\mu$, 3 modes}
      \end{subfigure}
      \begin{subfigure}[b]{0.32\linewidth}
          \includegraphics[width=\textwidth]{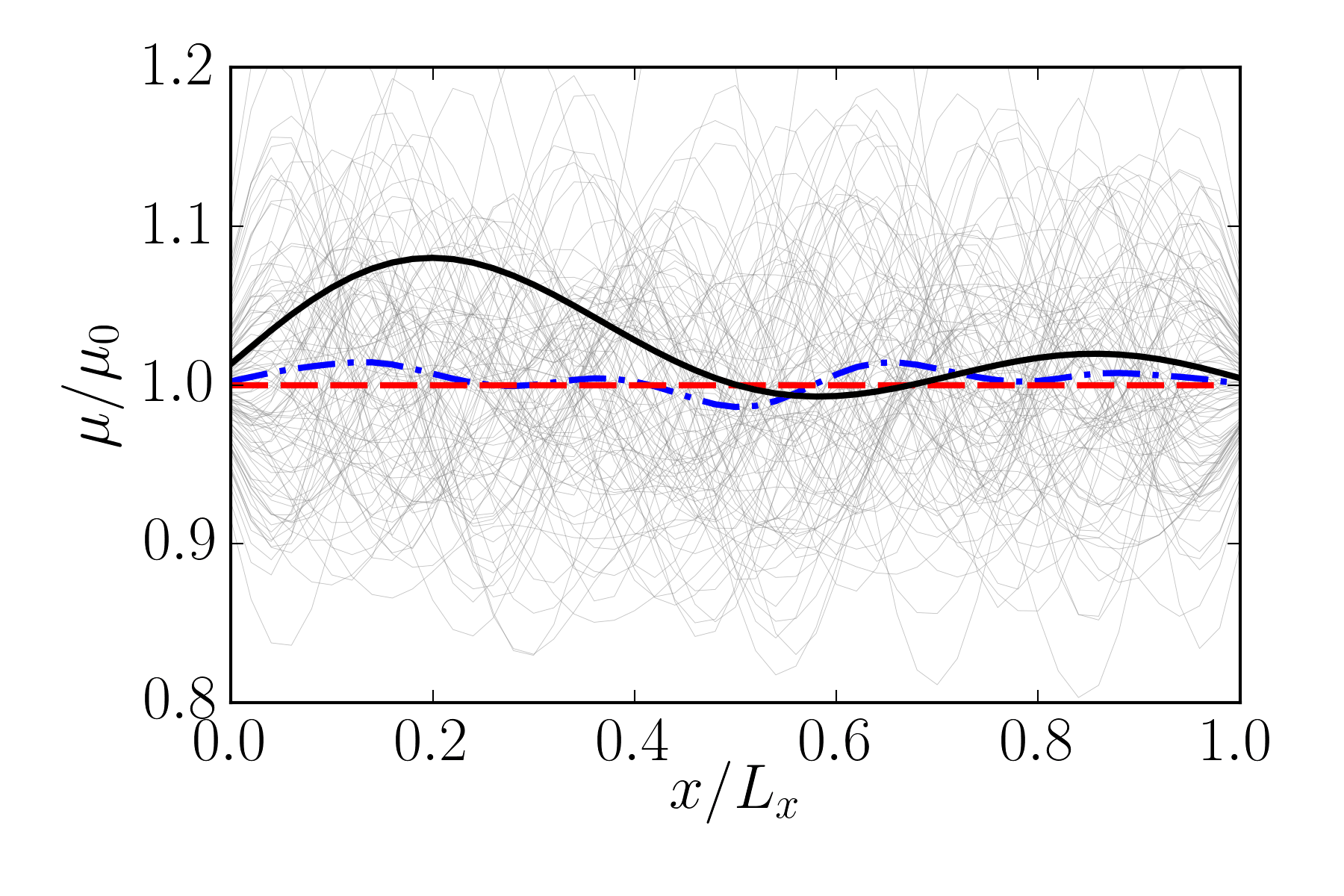}
          \caption{$\mu$, 10 modes}
      \end{subfigure}
      \begin{subfigure}[b]{0.32\linewidth}
          \includegraphics[width=\textwidth]{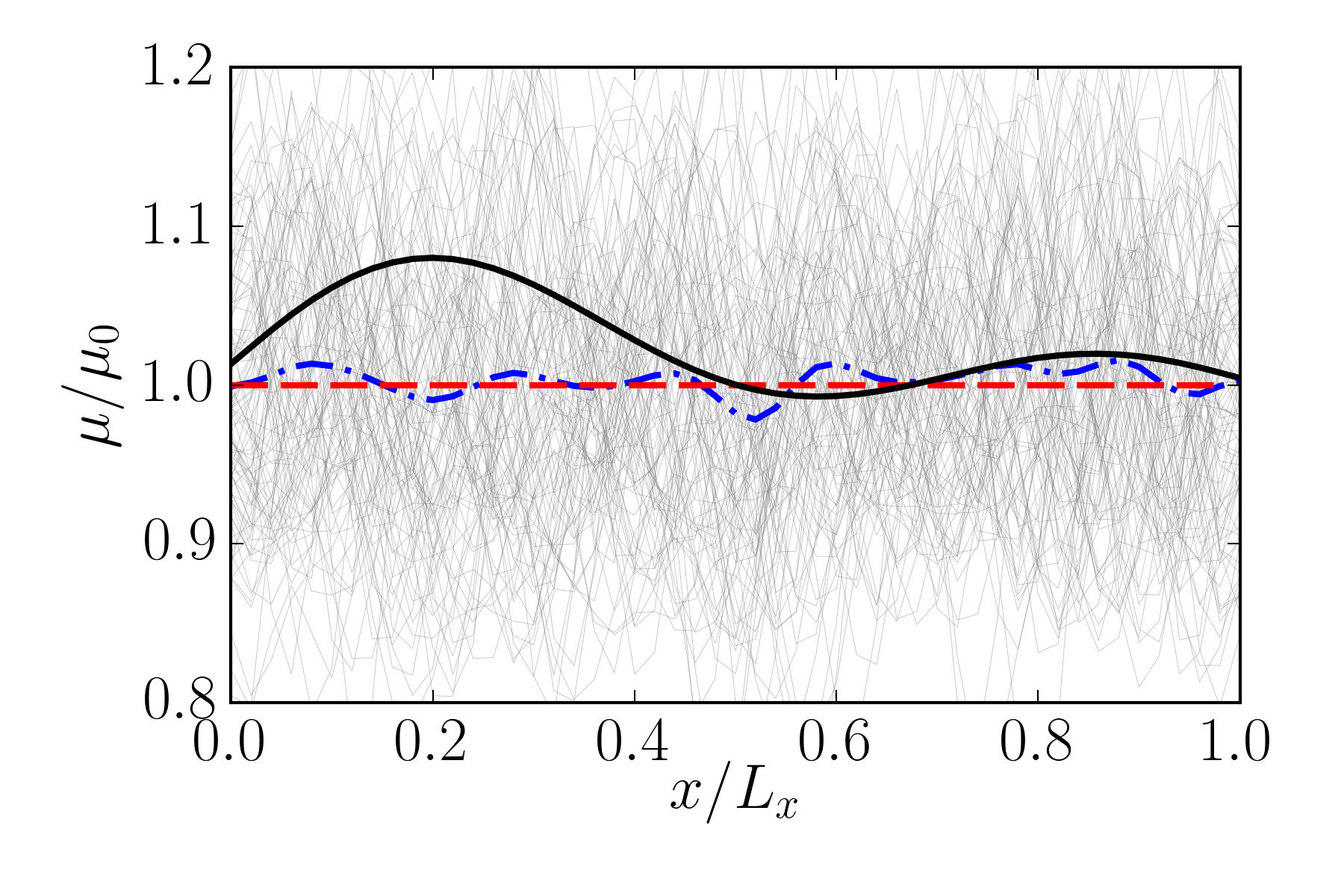}
          \caption{$\mu$, 20 modes}
      \end{subfigure}
      \\
      \begin{subfigure}[b]{0.32\linewidth}
          \includegraphics[width=\textwidth]{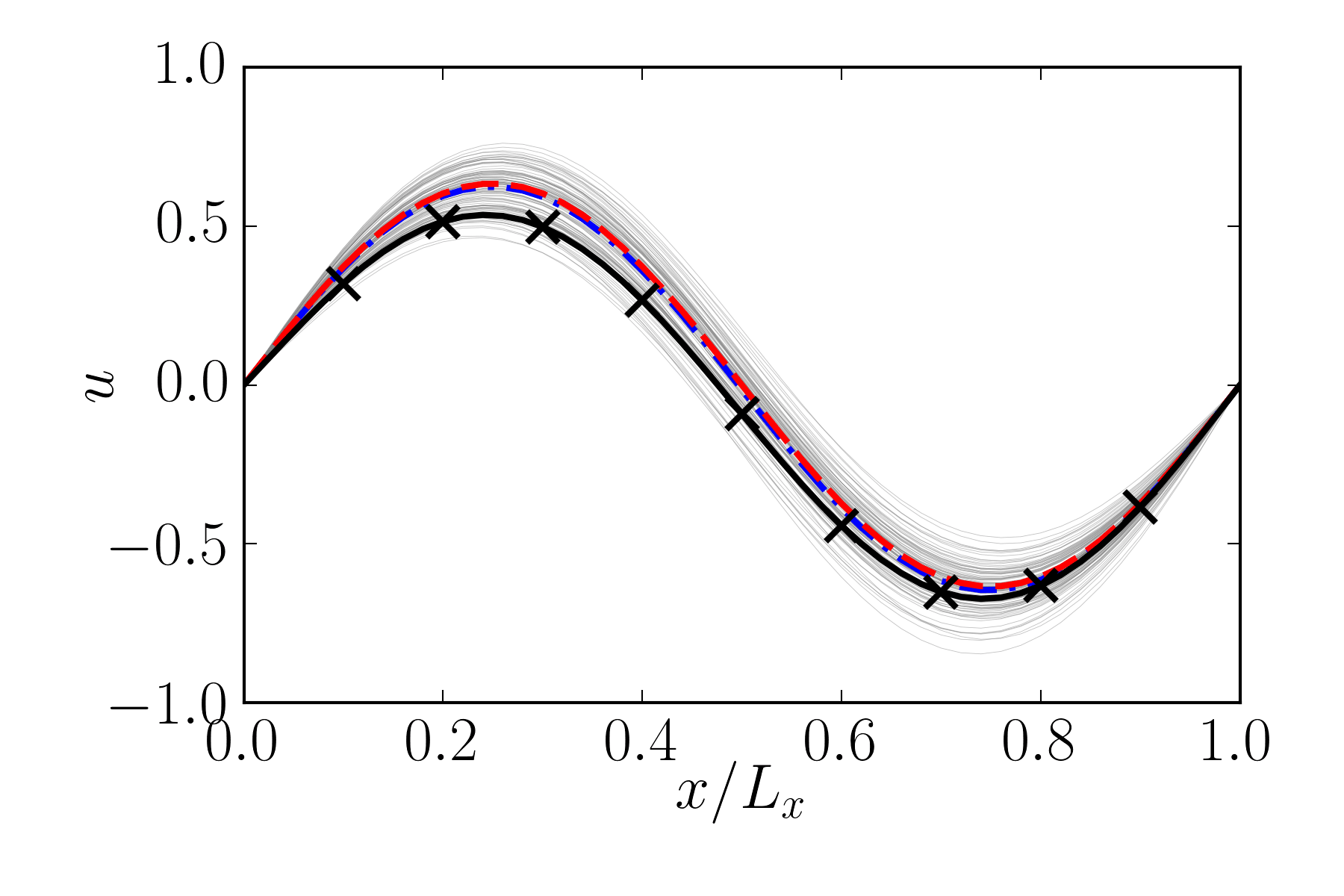}
          \caption{$u$, 3 modes}
      \end{subfigure}
      \begin{subfigure}[b]{0.32\linewidth}
          \includegraphics[width=\textwidth]{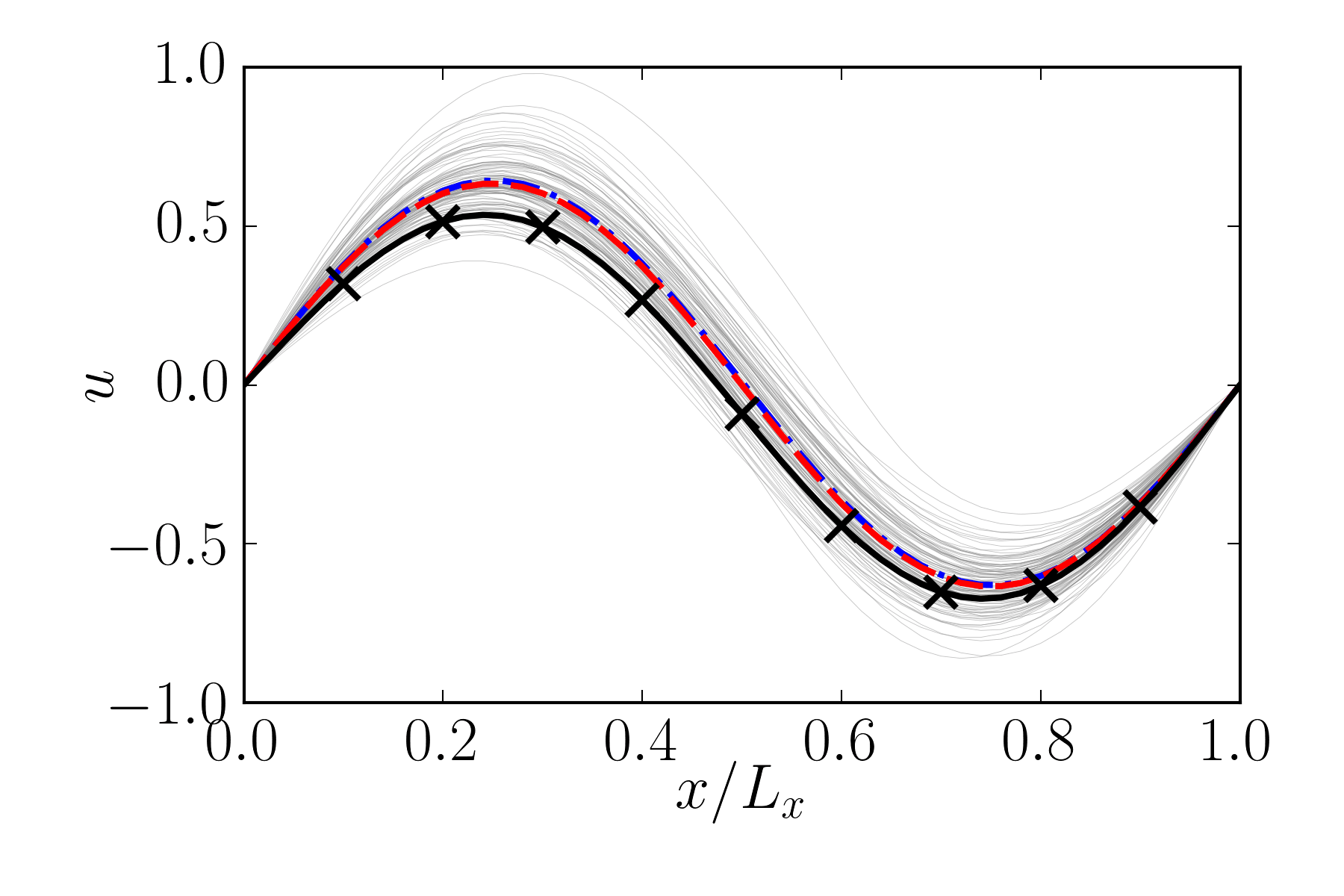}
          \caption{$u$, 10 modes}
      \end{subfigure}
      \begin{subfigure}[b]{0.32\linewidth}
          \includegraphics[width=\textwidth]{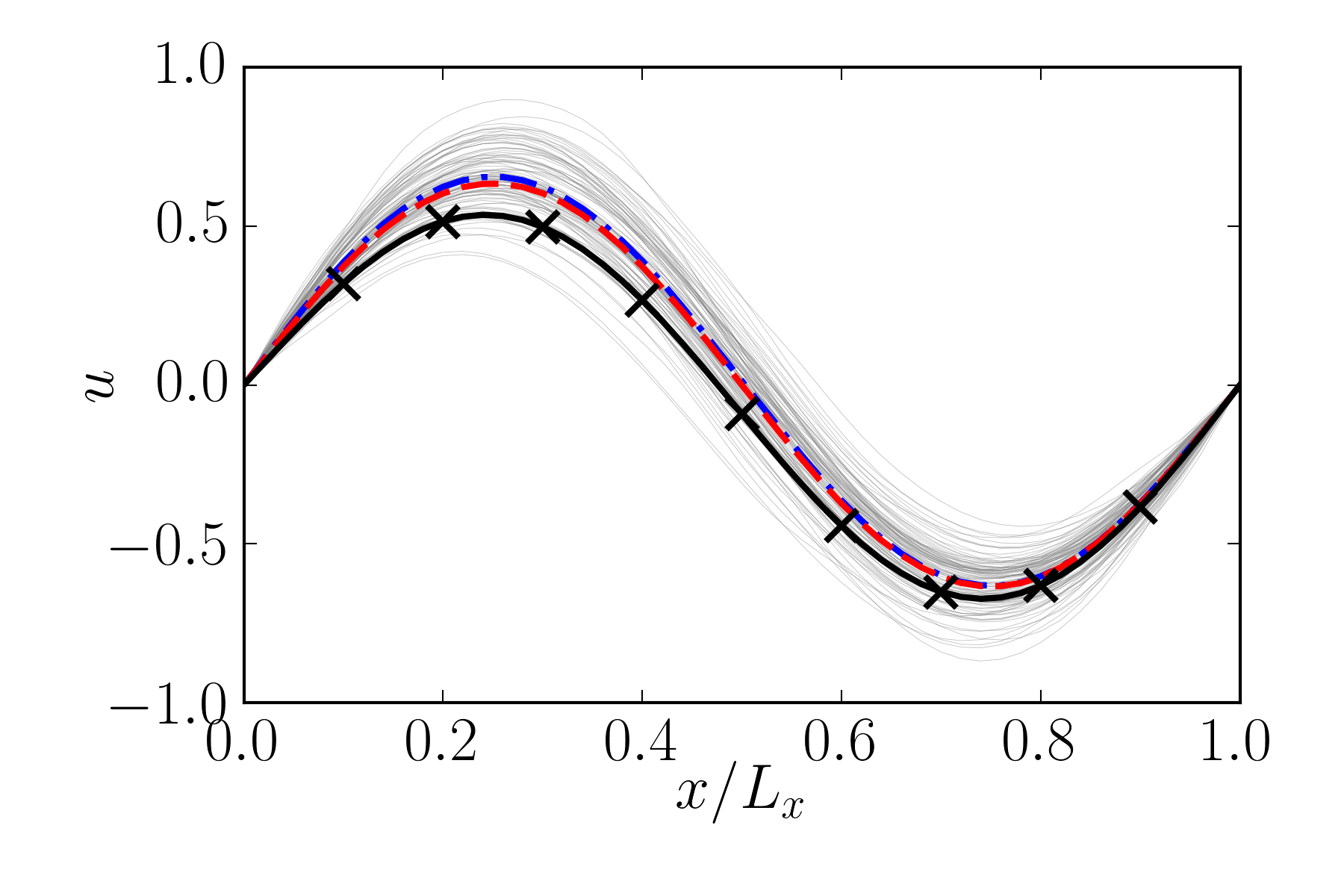}
          \caption{$u$, 20 modes}
      \end{subfigure}
      \caption{\textbf{Prior samples} of diffusivity $\mu$ (top) and corresponding output fields $u$ (bottom) using different number of modes.}
      \label{fig:case2_prior}
    \end{figure}

    For the Bayesian inversion scheme, the prior distribution of~$\log[\mu/\mu_0]$ is considered to be the Gaussian process described earlier with uniform mean~$\mu_\text{p}[x]=\mu_0$.
    A total of $80$ samples are used, created using the KL decomposition in Eq.~\eqref{eq:kl} with random coefficients with independent standard normal distributions, i.e.~$\omega_i \sim \mathcal{N}(0,1)$.
    The truth is constructed using the same decomposition in Eq.~\eqref{eq:kl} with only the first three modes, each with coefficient equal to~$1$, i.e.,~$\omega_1=\omega_2=\omega_3=1.0$ and~$\omega_j=0$ for~$j>3$.
    The observations in~$u$ are obtained by propagating this true diffusivity field through the diffusion equation and using an observation standard deviation of~$\sigma_\text{y}=0.0001$.
    Fig.~\ref{fig:case2_prior} shows the prior samples for the diffusivity field as well as the propagated output field using different number of modes for representing the field, as will be discussed later.

    The synthetic truth is constructed with only~$3$ modes, and the magnitude of the eigenvalues of the kernel decreases slowly due to the small length scale used in the correlation kernel.
    Thanks to these two facts, we can control the dimension of the inference space and the level of ill-posedness of the problem by setting the number of modes~$n$ used in the representation of the field (Eq.~\ref{eq:kl}).
    Specifically, if a large number of modes is used, many different diffusivity fields with increasingly different qualitative shapes can result in matching the observations in the output space.
    We consider as an additional source of knowledge that the lower modes are more important, and use REnKF to embed this information into the data assimilation process.
    To embed this information, we use a penalty function of the form
    \begin{equation}
      G[\bm{\omega}]=\bm{\omega} ,
    \end{equation}
    with a weight matrix
    \begin{equation}
      \overline{\mathsf{W}} = \text{diag} \left( \frac{1}{n},\frac{2}{n}, \frac{3}{n},\ldots,\frac{n-1}{n},1 \right) 
      \quad \text{or equivalently} \quad  \overline{\mathsf{Q}} = \text{diag} \left( n,\frac{n}{2}, \frac{n}{3},\ldots,\frac{n}{n-1},1 \right).
      \label{eq:constraint_cov_diffusion}
    \end{equation}
    where the higher modes are increasingly penalized.
    We use the ramp function in Eq.~\eqref{eq:regularization_function} with~$\chi_0=10$.
    With this constraint, we prefer the lower modes to the higher modes.
    It is noted that this is a soft constraint, which still allows for higher modes if they contribute to improving the agreement with observations.

    To test the performance of the proposed method, we perform the field inversion with both the baseline and regularized ensemble Kalman methods using different number of modes to represent the field. 
    Here the results with~$3$, $10$, and $20$ modes are presented.
    Figure~\ref{fig:case2_prior} shows the prior distributions (samples) using different number of modes.
    Note that with more modes there are much higher oscillations in the prior diffusivity fields, leading to samples that look very noisy.
    Nonetheless, even with the high noise all cases have similar distributions in the output field.
    This clearly shows the ill-posedness of this field inversion problem.
    Diffusivity fields that are qualitatively very different still result in very similar output fields, where the observation is made.
    The traditional ensemble Kalman method has no way to prefer one of these over the others as long as they match well with the observations in the output space. 
    This is true even though the traditional ensemble Kalman method has an embedded regularization based on the prior distribution. 
    This can be clearly seen by considering that most of the samples in Fig.~\ref{fig:case2_prior} have similar probability of coming from the prior distribution. 

    \subsubsection{Results}
\begin{figure}[!htbp]
        \centering
        \includegraphics[width=0.7\textwidth]{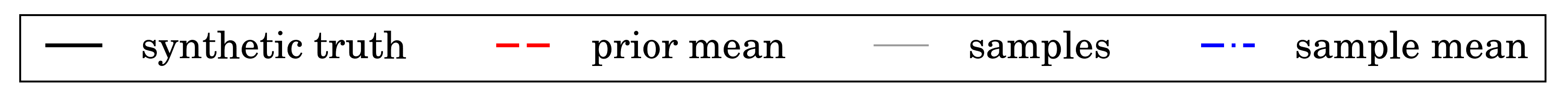}\\
        \begin{subfigure}[b]{0.45\linewidth}
           \includegraphics[width=\textwidth]{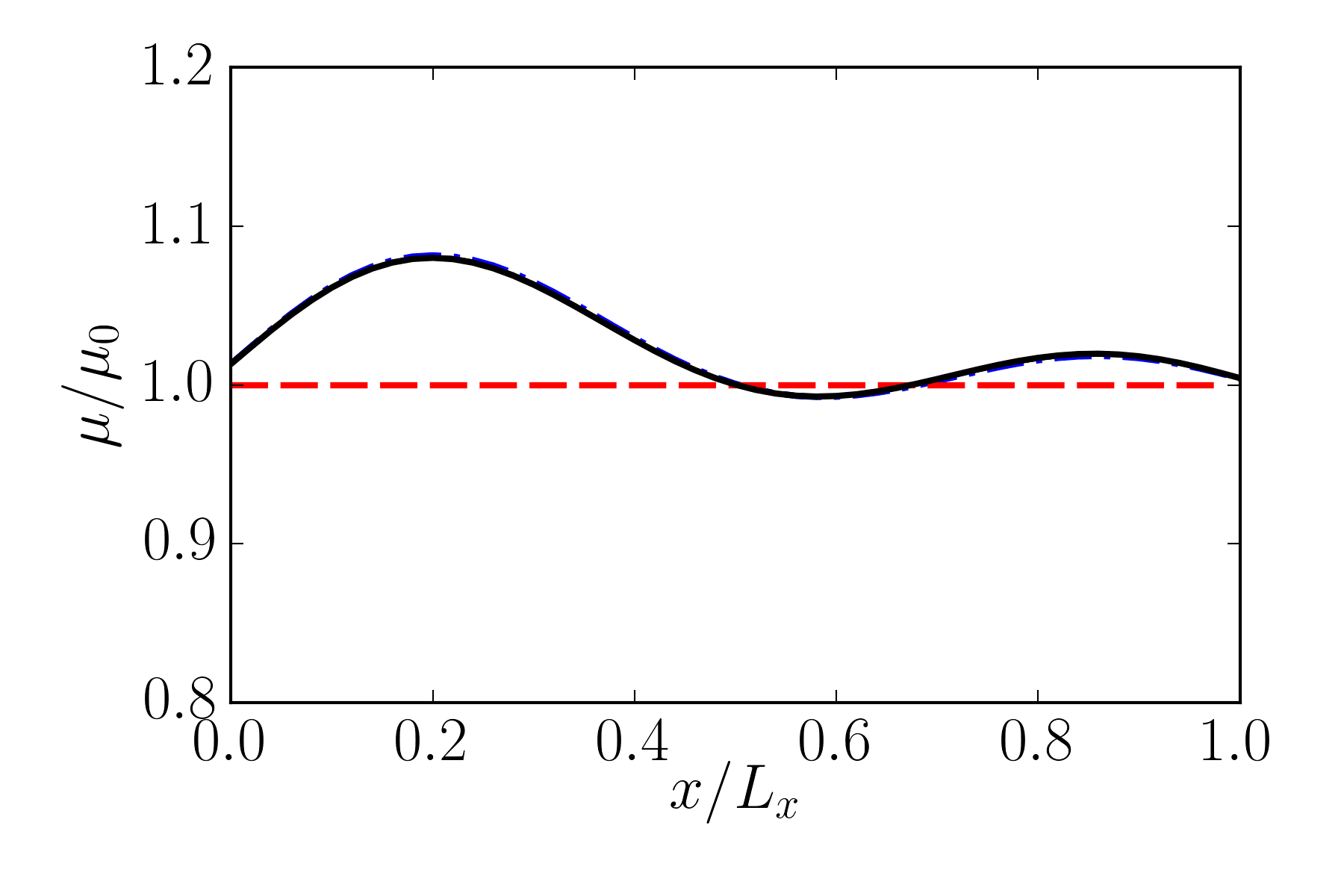}
           \caption{Baseline, 3 modes}
        \end{subfigure}
        \begin{subfigure}[b]{0.45\linewidth}
           \includegraphics[width=\textwidth]{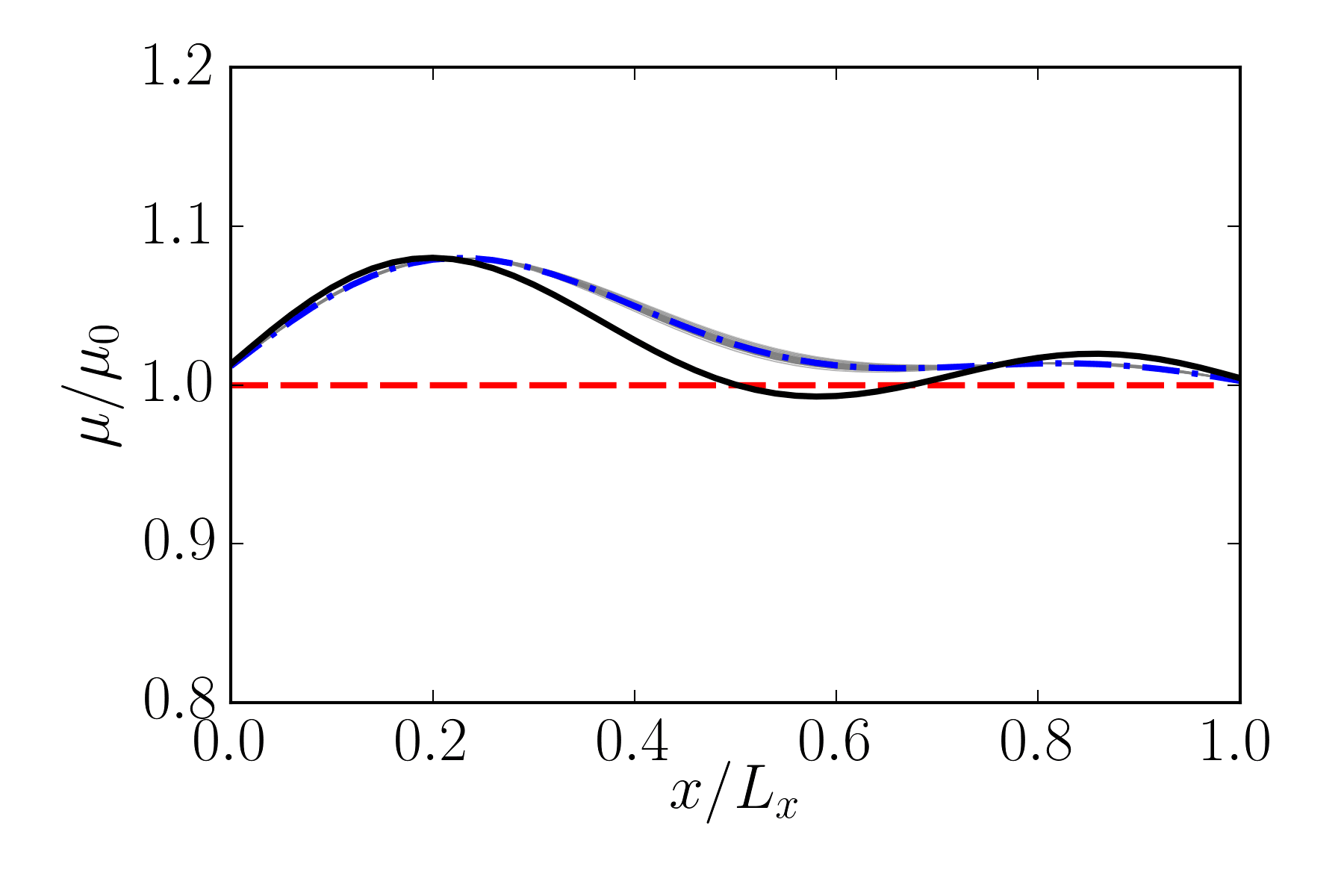}
           \caption{Regularized, 3 modes}
        \end{subfigure}
        \\
        \begin{subfigure}[b]{0.45\linewidth}
           \includegraphics[width=\textwidth]{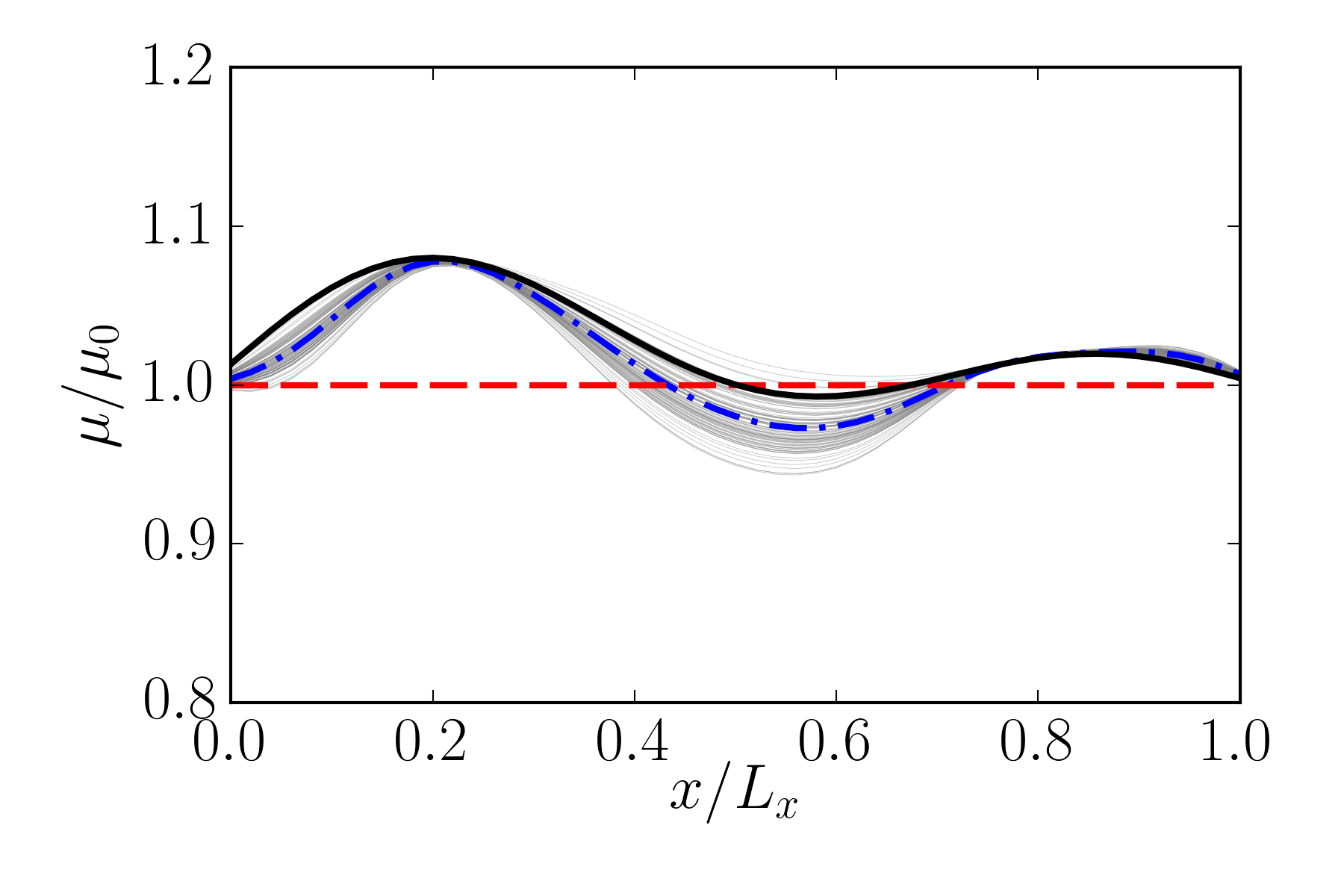}
           \caption{Baseline, 10 modes}
        \end{subfigure}
        \begin{subfigure}[b]{0.45\linewidth}
            \includegraphics[width=\textwidth]{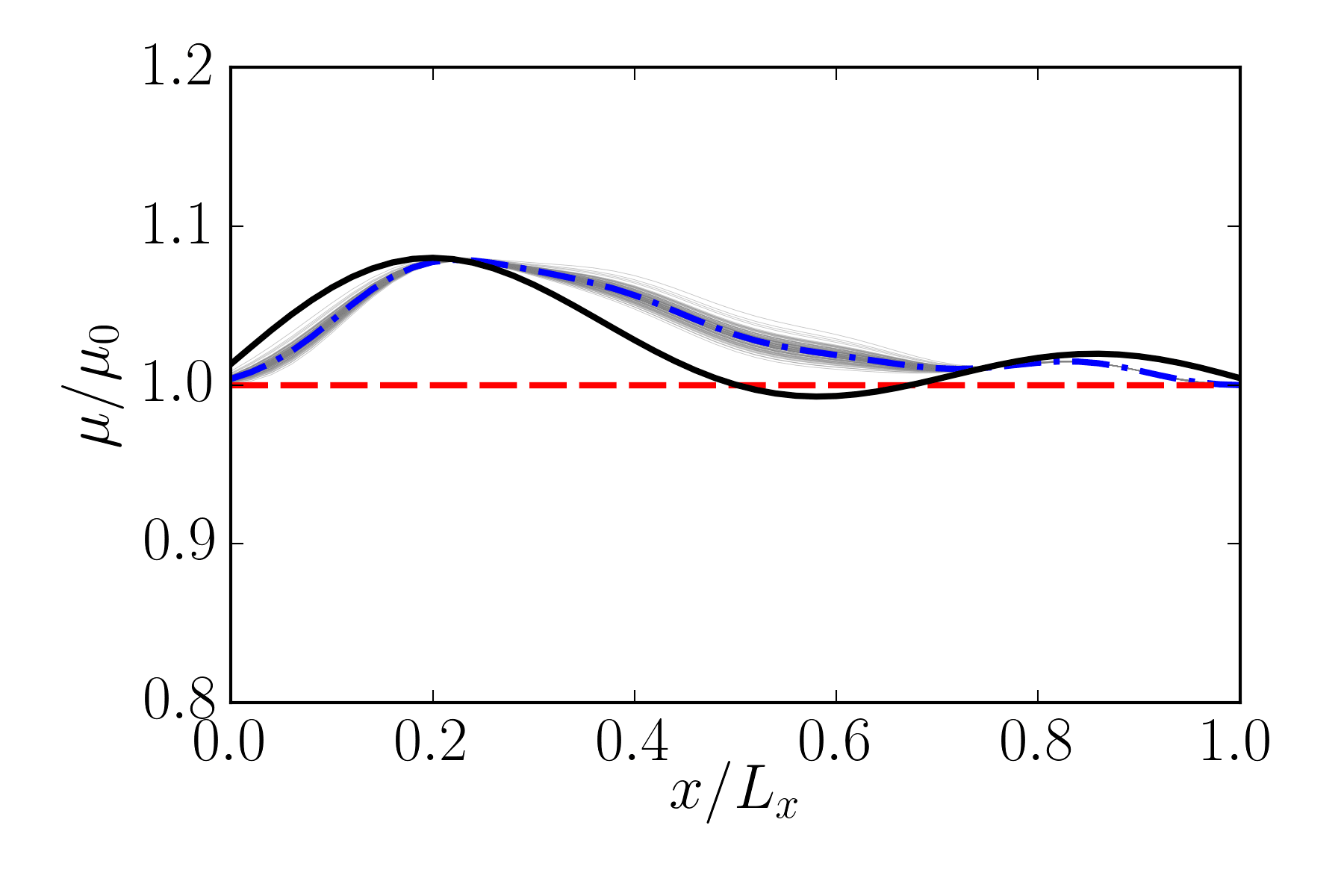}
            \caption{Regularized, 10 modes}
        \end{subfigure}
        \\
         \begin{subfigure}[b]{0.45\linewidth}
           \includegraphics[width=\textwidth]{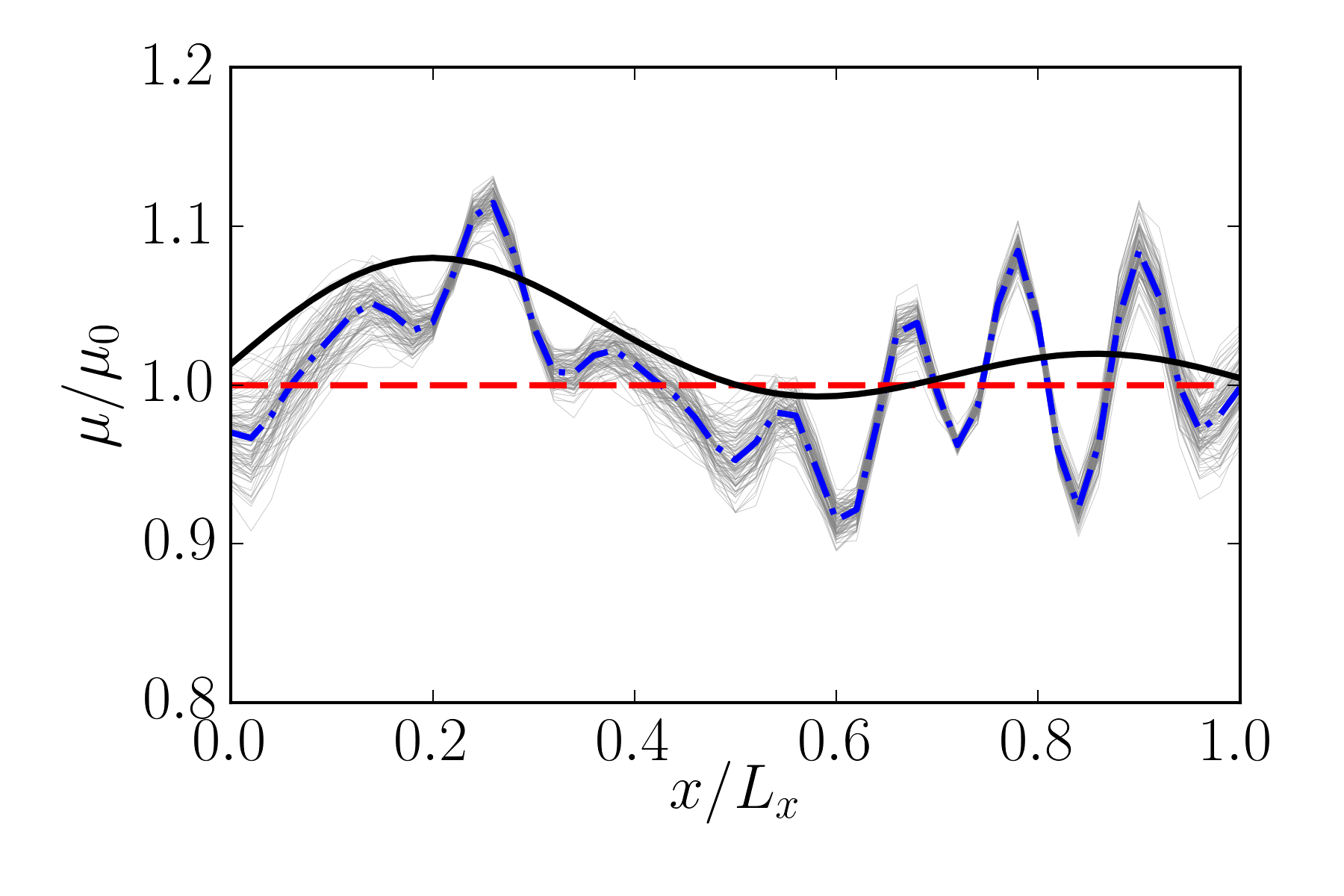}
           \caption{Baseline, 20 modes}
        \end{subfigure}
        \begin{subfigure}[b]{0.45\linewidth}
            \includegraphics[width=\textwidth]{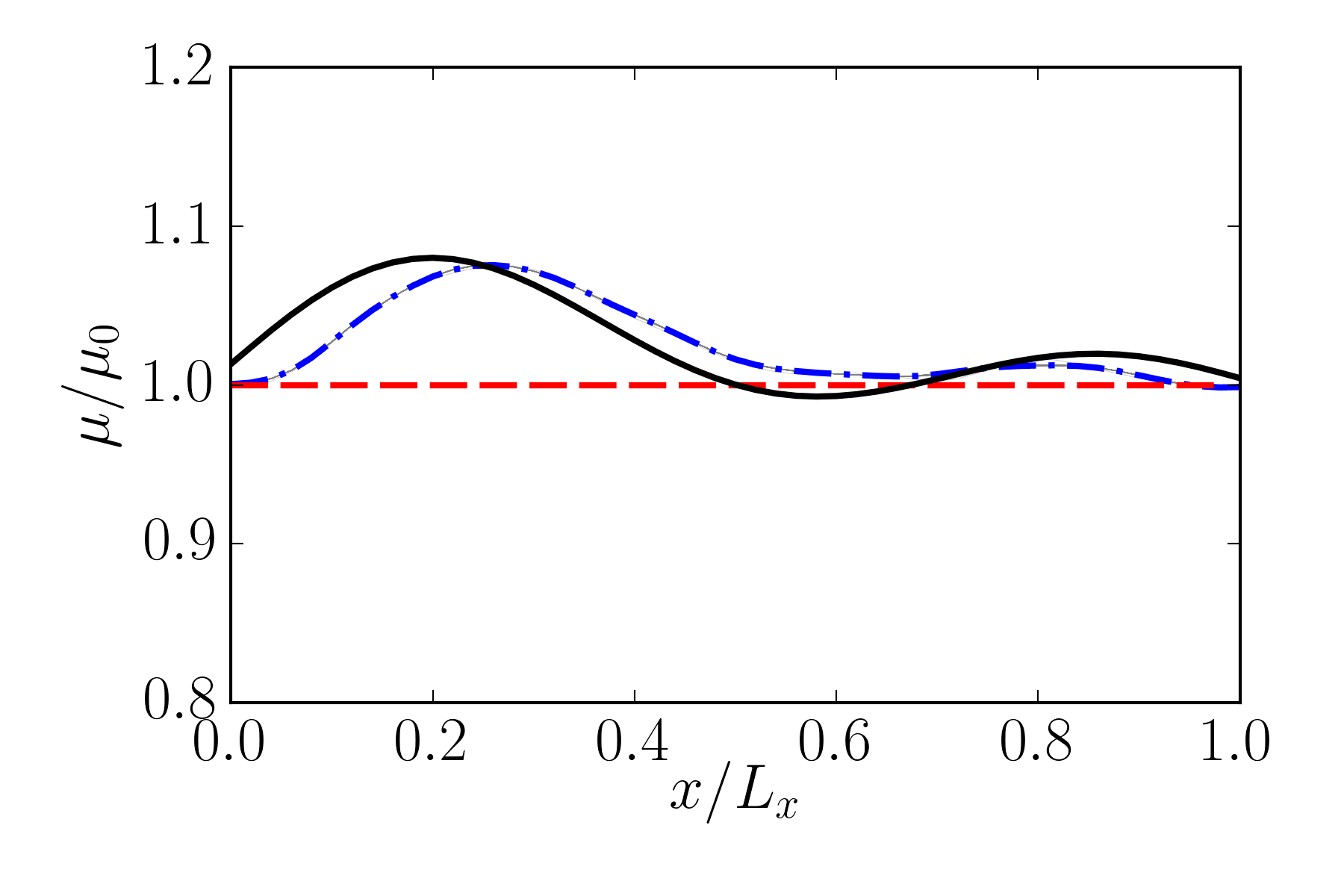}
            \caption{Regularized, 20 modes}
        \end{subfigure}
        \caption{Inferred diffusivity by using the \textbf{baseline method} (left column; panels a, c, and e) and the \textbf{proposed regularized method} (right column; panels b, d and f) using different number of modes.}
        \label{fig:case2_results}
      \end{figure}

      The results of the inferred field $\mu$ are shown in Figure~\ref{fig:case2_results}.
      Regardless of the number of modes used in representing the field, both EnKF and REnKF are able to give a satisfactory agreement in the observed field, and these results are omitted since they are visually indistinguishable.
      The difference between the results from the different methods lies in their ability to infer the correct latent diffusivity field.
      The baseline method with only three modes results in the correct diffusivity field as expected, since this problem is not ill-posed.
      On the other hand, in the case with only three modes adding a regularization term to the cost function results in worst agreement. This is expected since the problem is not ill-posed and therefor improvement to the regularization term necessarily means worsening of the original cost, and the minimization must balance both terms.
      However, when there is more freedom with increased number of modes used in the representation and hence increased dimensionality of the space of possible latent fields, the baseline method infers increasingly more qualitatively wrong diffusivity fields while still matching the observations and true output field.
      Incorporating the additional knowledge through the proposed regularized method results in a much improved diffusivity field being inferred  particularly in the cases with a large number of modes. 
      A clear improvement in the qualitative shape can be seen, with the inferred field exhibiting the correct, larger, correlation length.
      This is clearly seen in Fig.~\ref{fig:case2_results}. 

\begin{figure}[!htb]
        \centering
        \includegraphics[width=0.5\linewidth]{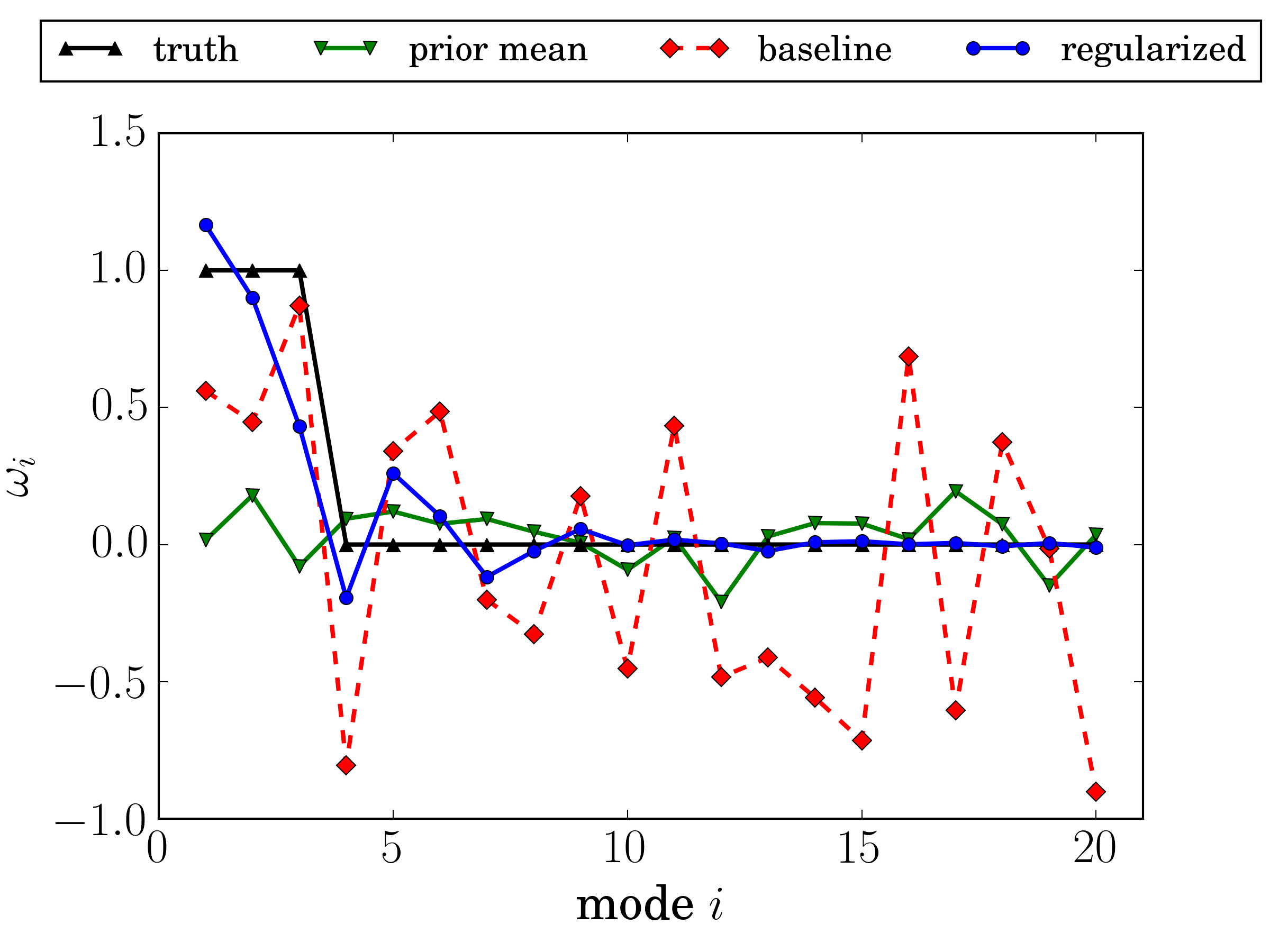}
        \caption{Comparison of inferred KL coefficients for \textbf{the diffusion case} by using the baseline method and the regularized method using $20$ modes.}
        \label{fig:case2_inferred_omega}
      \end{figure}

\begin{figure}[!htb]
        \centering
        \includegraphics[width=0.5\linewidth]{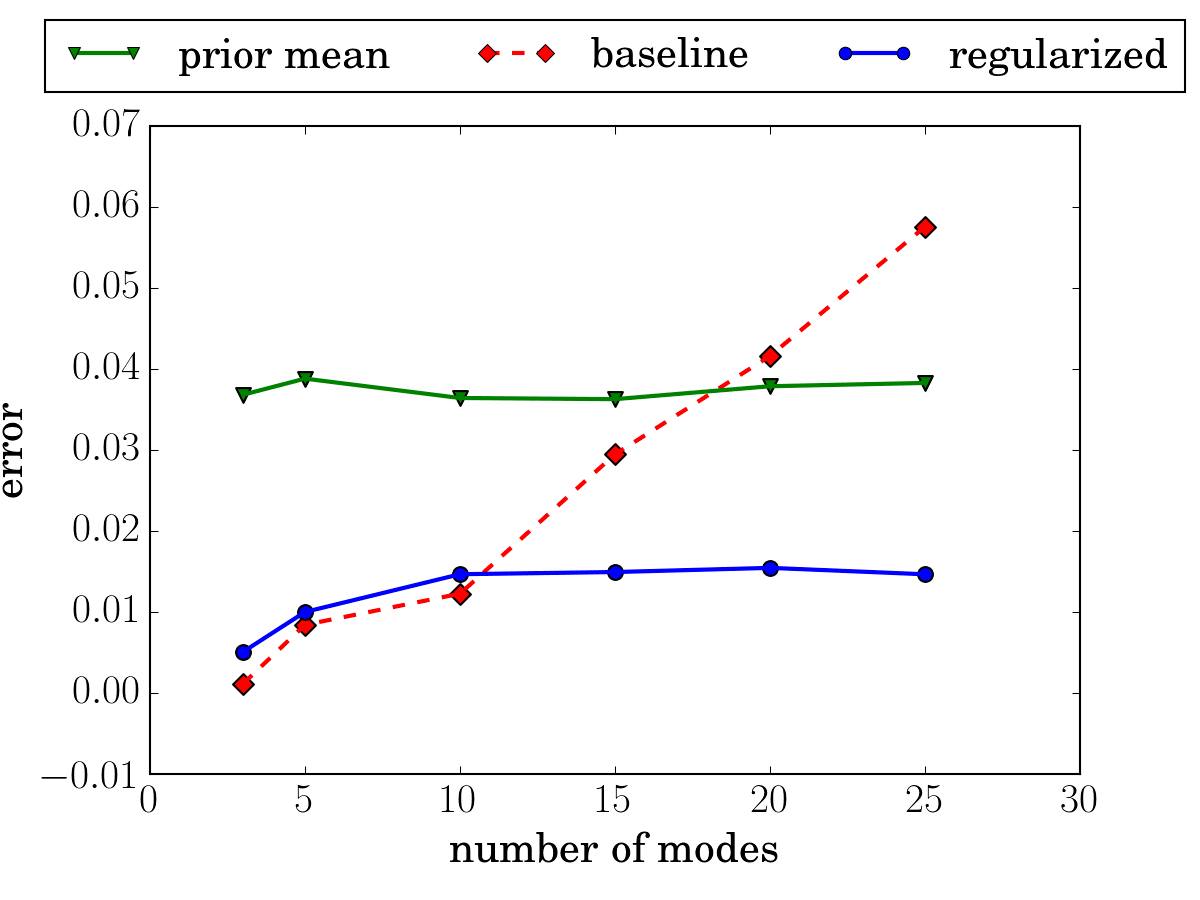}
        \caption{Errors in the inferred diffusivity in \textbf{the diffusion case} for the baseline method and the regularized method as a function of number of modes used in the field representation.}
        \label{fig:case2_error}
      \end{figure}

      The inferred KL coefficients for the baseline and regularized methods using 20 modes are shown in Fig.~\ref{fig:case2_inferred_omega}. 
      It is noticeable that the baseline method uses all the available modes without preferences, while the regularized method suppresses the higher modes.
      Moreover, the inferred coefficients with the regularized method follow a similar trend as the synthetic truth values. 
      The errors in the inferred diffusivity for the different methods are shown in Fig.~\ref{fig:case2_error} as a function of the number of modes used in the representation. 
      It can be seen that with increasing number of modes, the baseline method gives increasingly worse inference on the diffusivity, while with the regularized method the error remains relatively constant.  
      The regularized method can provide satisfactory inference regardless of the number of modes used in the representation.
      
  \subsection{RANS equations closure}
    As a final case we test the proposed regularized method for a field inversion problem of practical interest in fluid mechanics: closure of the Reynolds-averaged Navier--Stokes (RANS) equations. 
    The RANS equations describe the mean flow of fluids accurately; however, they are unclosed.
    The Reynolds stress term requires a turbulence model, and no universally good model exists.
    In practice, this means that complex flows cannot be confidently predicted in regions with separation or high pressure gradients.
    It is therefore of tremendous interest to infer the Reynolds stress in regions where the flow is too complex to be captured by current turbulence models.
    This can be achieved by incorporating sparse observations using inversion schemes such as the ensemble Kalman methods.
    Here we consider the steady two-dimensional incompressible RANS equations with linear eddy viscosity assumption.
    This means that a single scalar field, the eddy viscosity field, needs to be inferred rather than the full Reynolds stress tensor field.
    The RANS equations can then be written as
    \begin{subequations}
      \begin{align}
        \quad \frac{\partial U_i}{\partial x_i} & = 0  \\
        U_j \frac{\partial U_i}{\partial x_j} & =  -\frac{\partial {p}}{\partial x_i} + \frac{\partial}{\partial x_i} \left[ \left(\nu + \nu_\text{t} \right) \left( \frac{\partial U_i}{\partial x_j} + \frac{\partial U_j}{\partial x_i} \right) \right],
      \label{eq:rans-momentum}
      \end{align}
    \end{subequations}
    using Einstein summation notation, where $i\in{1,2}$ denotes spatial direction, $U$ is velocity, $x$ is spatial coordinate, $p$ is a pseudo pressure term, $\nu$ is the fluid viscosity, and $\nu_\text{t}$ is the eddy viscosity field to be inferred.

    For this test case, we use the canonical flow over periodic hills ~\cite{breuer2009flow} which has been extensively used for the investigation of numerical methods in CFD~\cite{gritskevich2012development}.
    A single hill is modeled with periodic boundary conditions.
    The domain is discretized with~$50$ cells in the stream-wise direction~$x_1$ and~$30$ cells in the wall-normal direction~$x_2$.
    The dimensionless wall distance~$y^+$ of the first cell is small enough to lie in the viscosity layer, and no wall model is used.
    All spatial coordinates are normalized by hill height~$H$ and all velocities by the bulk velocity~$U_\text{b}$ at the hill crest. 
    The Reynolds number based on~$H$ and~$U_\text{b}$ is~$2800$.
    
    In this case, we use OpenFOAM, an open-source CFD platform based on finite volume discretization, to simulate the incompressible, steady-state turbulent flows. %
    The SIMPLE (Semi-Implicit Method for Pressure Linked Equations) algorithm is used to solve the RANS equations. 
    Second-order spatial discretization schemes are applied to discretize the equations on an unstructured mesh. 
    The prior mean and synthetic truth are both created from RANS simulations using the built-in simpleFOAM solver but with different turbulence models.
    The synthetic truth is obtained using the $k$--$\varepsilon$ model~\cite{jones1972prediction} and the prior mean using the Spalart--Allmaras model~\cite{spalart1992one-equation}. 
    To propagate eddy viscosity to the velocity field, a modified solver was created that uses a constant (i.e., over iterations) specified eddy viscosity field rather than using a turbulence model. 
    This modified solver is the forward model which gives the output fields (velocities and pressure) given an input field (eddy viscosity).

    \subsubsection{Case details}
\begin{figure}[!htb]
          \centering
          \includegraphics[width=0.3\textwidth]{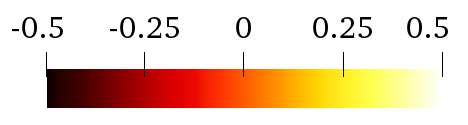}\\
          \begin{subfigure}[b]{0.23\linewidth}
              \includegraphics[width=\linewidth]{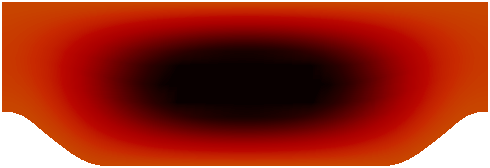}
              \caption{mode 1}
          \end{subfigure}
          \begin{subfigure}[b]{0.23\linewidth}
              \includegraphics[width=\linewidth]{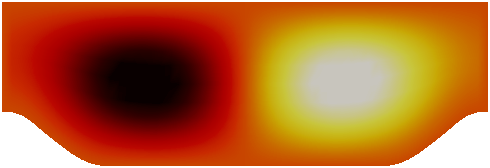}
              \caption{mode 2}
          \end{subfigure}
          \begin{subfigure}[b]{0.23\linewidth}
              \includegraphics[width=\linewidth]{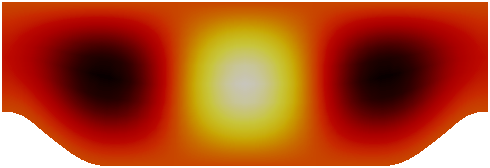}
              \caption{mode 3}
          \end{subfigure}
          \begin{subfigure}[b]{0.23\linewidth}
              \includegraphics[width=\linewidth]{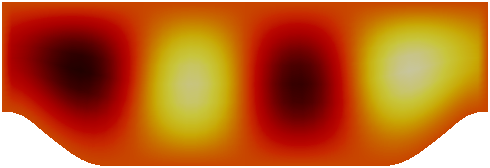}
              \caption{mode 4}
          \end{subfigure}
          \begin{subfigure}[b]{0.23\linewidth}
              \includegraphics[width=\linewidth]{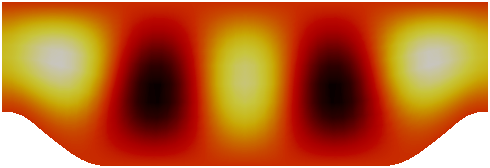}
              \caption{mode 5}
          \end{subfigure}
          \begin{subfigure}[b]{0.23\linewidth}
              \includegraphics[width=\linewidth]{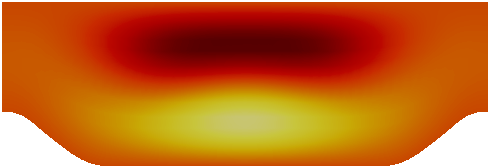}
              \caption{mode 6}
          \end{subfigure}
          \begin{subfigure}[b]{0.23\linewidth}
              \includegraphics[width=\linewidth]{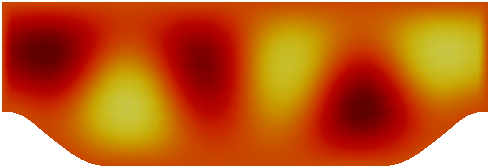}
              \caption{mode 7}
          \end{subfigure}
          \begin{subfigure}[b]{0.23\linewidth}
              \includegraphics[width=\linewidth]{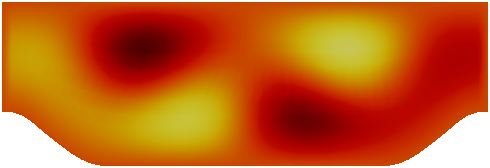}
              \caption{mode 8}
          \end{subfigure}
          \caption{The first $8$ modes from KL decomposition for the periodic hills case. The modes are scaled by their corresponding eigenvalues.}
          \label{fig:pehills_mode}
      \end{figure}

\begin{figure}[!htb]
          \centering
          \includegraphics[width=0.9\textwidth]{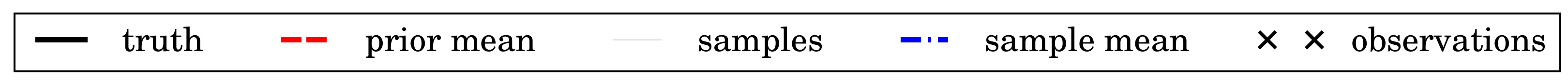}\\
          \begin{subfigure}[b]{0.48\linewidth}
              \includegraphics[width=\linewidth]{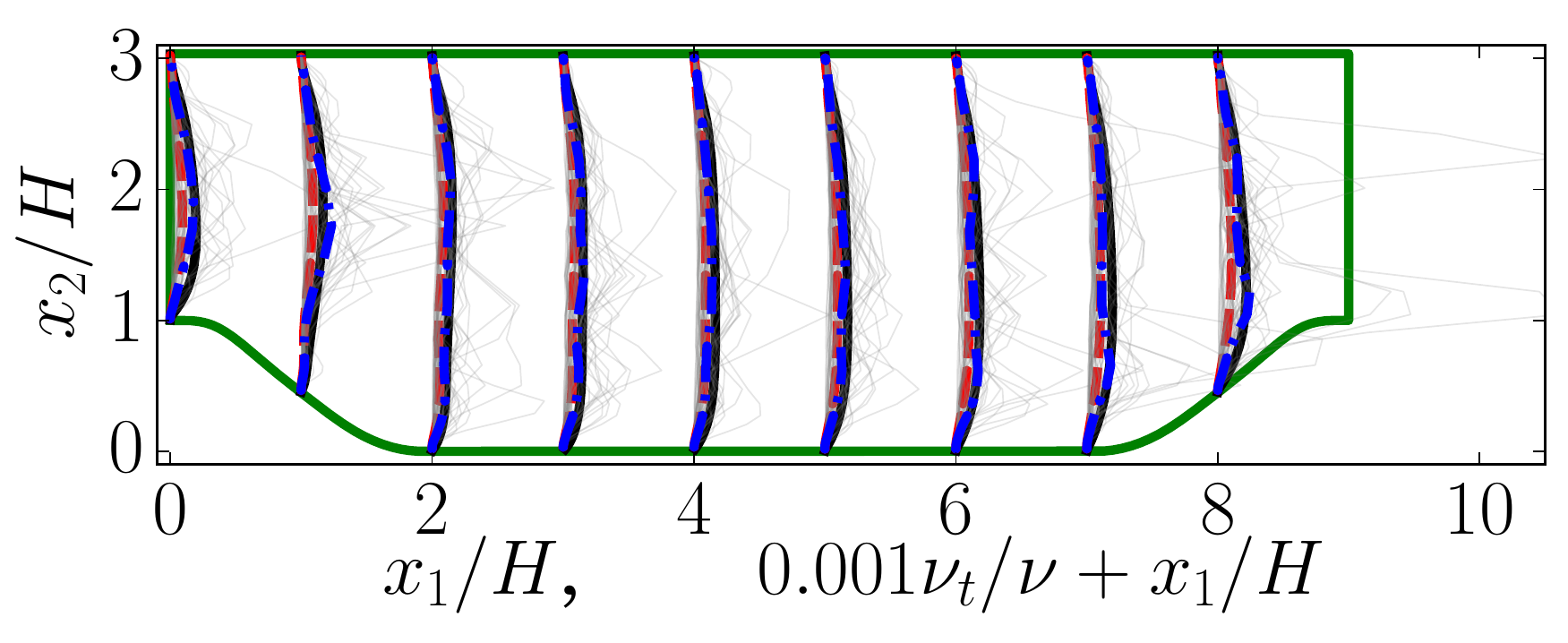}
              \caption{eddy viscosity}
              \label{fig:case3_prior_nut}
          \end{subfigure}
          \begin{subfigure}[b]{0.48\linewidth}
              \includegraphics[width=\linewidth]{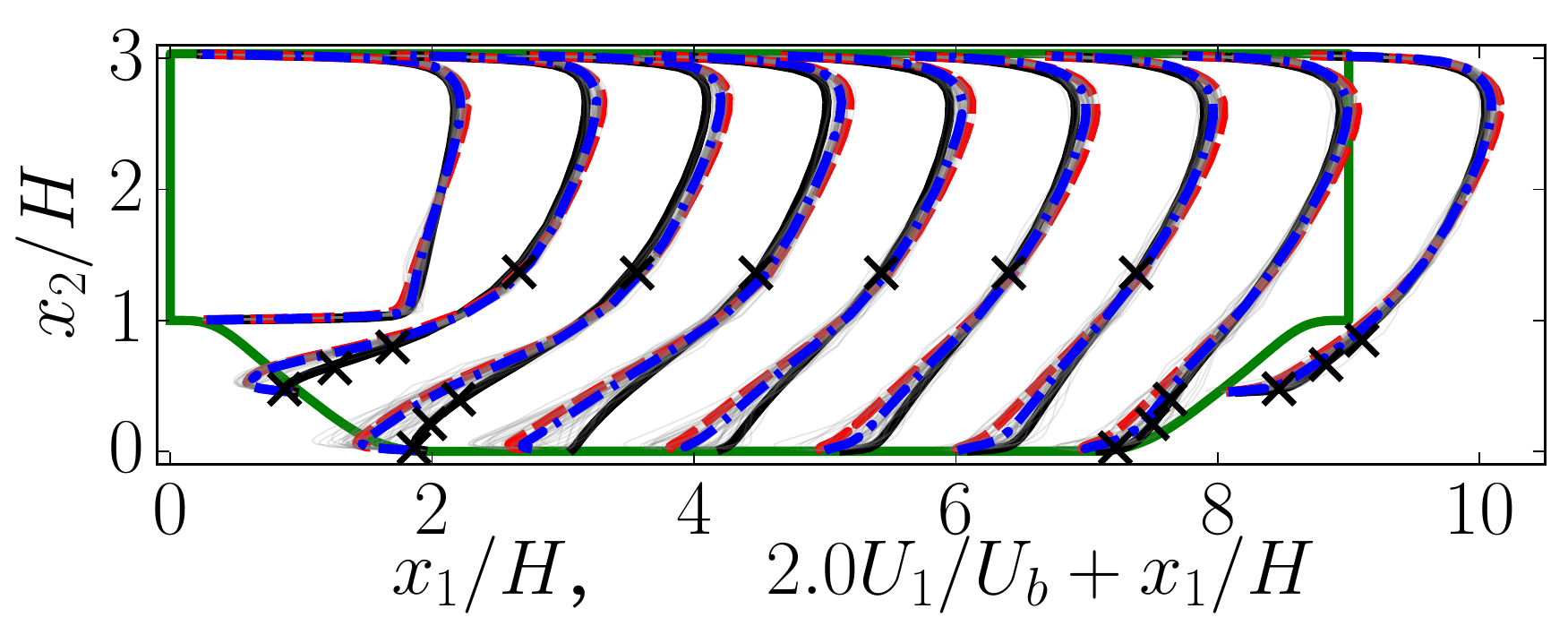}
              \caption{streamwise velocity}
              \label{fig:case3_prior_U1}
          \end{subfigure}
          \caption{Prior realizations of eddy viscosity $\nu_\text{t}$ and propagated streamwise velocity $U_1$ for the periodic hills case. The locations of the observations are indicated by crosses ($\times$).}
          \label{fig:case3_prior}
      \end{figure}

      The latent field to be inferred is the eddy viscosity field~$\nu_\text{t}$.
      Like the diffusivity field in the former case, the eddy viscosity is non-negative, and the same representation is used for it as for~$\mu$ in Eq.~\eqref{eq:kl}, inferring the logarithm of the field and using KL decomposition.
      The prior distribution is then $\log[\nu_\text{t} / \nu_{\text{t0}}] = \mathcal{GP}(0,\mathcal{K})$, where $\nu_{\text{t0}}$ is a reference eddy viscosity value.
      Again the square exponential covariance in Eq.~\eqref{eq:kernel} is used, with length scale $l=0.25H$ and variance $\sigma^2=1.0$.
      Moreover, we enforce the boundary conditions by incorporating them into the Gaussian processes of the prior. The readers are referred to~\cite{strofer2019enforcing} for further details.
      The first eight modes of the decomposition are shown in Fig.~\ref{fig:pehills_mode}.
      The lower modes represent the larger scale characteristics of the constructed field, while the higher modes have more oscillations.
      For the prior mean we use the results from a RANS simulation with the Spalart--Allmaras turbulence model.
      These results are projected into the KL modes to get the prior coefficients~$\omega_i$. The prior distribution is represented as an ensemble by using~$100$ samples.
      The prior distribution of eddy viscosity and the propagated streamwise velocity are shown in Fig.~\ref{fig:case3_prior}.
      Note the high oscillations in the prior eddy viscosity and the relatively smooth propagated streamwise velocities, which highlights the ill-posedness of the problem.
      The results from a RANS simulation with the~$k$-$\varepsilon$ turbulence model are taken as the truth which is used to create synthetic observations.
      The observations consist of streamwise velocity~$U_1$ at~$18$ points, shown in Fig.~\ref{fig:case3_prior_U1}, with observation error~$\sigma_\text{y}=0.001$.

      As a baseline, the inverse problem is solved using the traditional ensemble Kalman method.
      As before, different number of modes are used to study cases which are progressively more ill-posed. 
      The results for all cases are summarized but we choose to highlight the results for the case with~$200$ modes in more detail.
      As can be seen from the prior samples in Fig.~\ref{fig:case3_prior_nut}, the eddy viscosity field can have a qualitatively very different shape from the truth and still result in satisfactory results in the observation space.
      This problem can be exacerbated in the inference where the inferred values of the coefficients~$\omega_i$ are not restrained unlike in the prior where they are specified to have a standard normal distribution.
      This means that the inferred coefficients for higher modes can be very large.
      However, the modes from the KL decomposition have intrinsic importance embedded in them, indicated by the magnitude of their corresponding eigenvalues, and while this information is used in constructing the prior samples, it is ignored in the inference step. 
      We use this relative importance of the modes as an additional source of information to create a regularization constraint.
      Among equally fit candidate solutions, we will prefer the simplest one, i.e., the one that uses the fewest modes (e.g., low pass filter). 
      We use this preference as the regularization and use the relative importance of the modes to embed this preference into the inversion through the proposed method. 
      To achieve this, a penalty function
      \begin{equation}
        G[\bm{\omega}] = \bm{\omega} 
      \end{equation}
      is used with the weight matrix $\overline{\mathsf{W}}$ constructed from the inverse of the eigenvalues as 
      \begin{equation}
        \overline{\mathsf{W}} = 
        \text{diag} \left( \frac{1}{{\lambda}_1}, \dots, \frac{1}{{\lambda}_{n-1}}, \frac{1}{{\lambda}_n} \right) 
        \quad \text{or equivalently} \quad  
        \overline{\mathsf{Q}} =
        \text{diag} \left( \lambda_1, \dots, \lambda_{n-1}, \lambda_n \right) 
        \text{.}
          \label{eq:cov_RANS}
      \end{equation}
      A value of~$\chi_0=6$ is used for the ramp function in Eq.~\eqref{eq:regularization_function}. 

    \subsubsection{Results}
\begin{figure}[!htb]
          \centering
          \includegraphics[width=1.0\textwidth]{figure10-legend.pdf}\\
          \begin{subfigure}[b]{0.48\linewidth}
              \includegraphics[width=\linewidth]{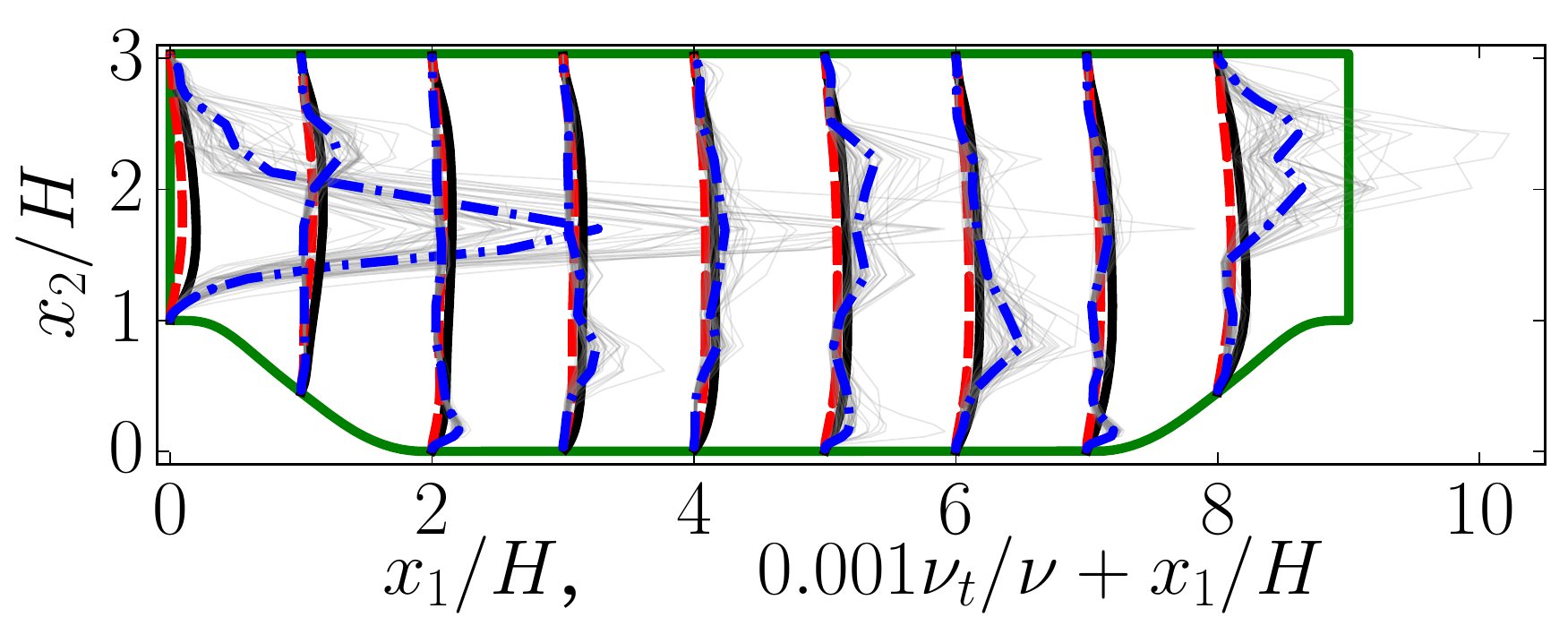}
              \caption{baseline, eddy viscosity}
              \label{fig:EnKF_nut}
          \end{subfigure}
          \begin{subfigure}[b]{0.48\linewidth}
              \includegraphics[width=\textwidth]{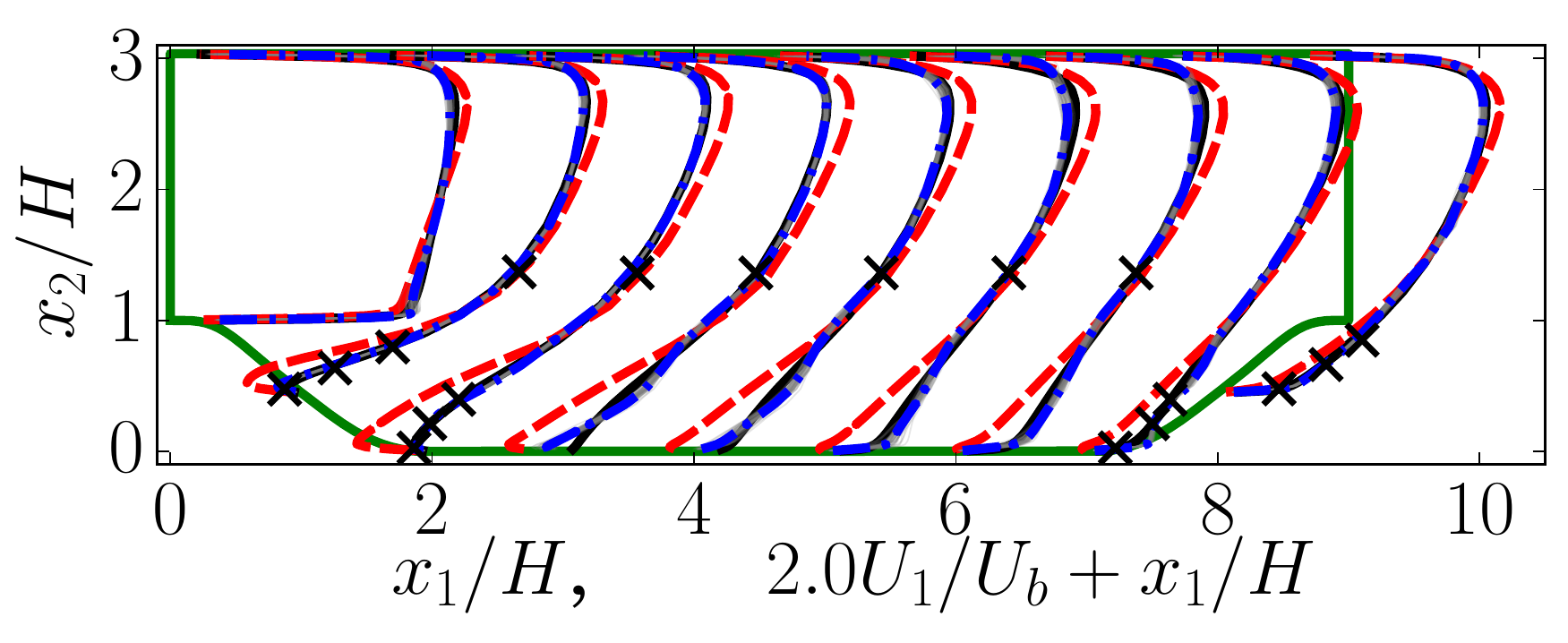}
              \caption{baseline, streamwise velocity}
              \label{fig:REnKF_nut}
          \end{subfigure}
          \\
          \begin{subfigure}[b]{0.48\linewidth}
              \includegraphics[width=\linewidth]{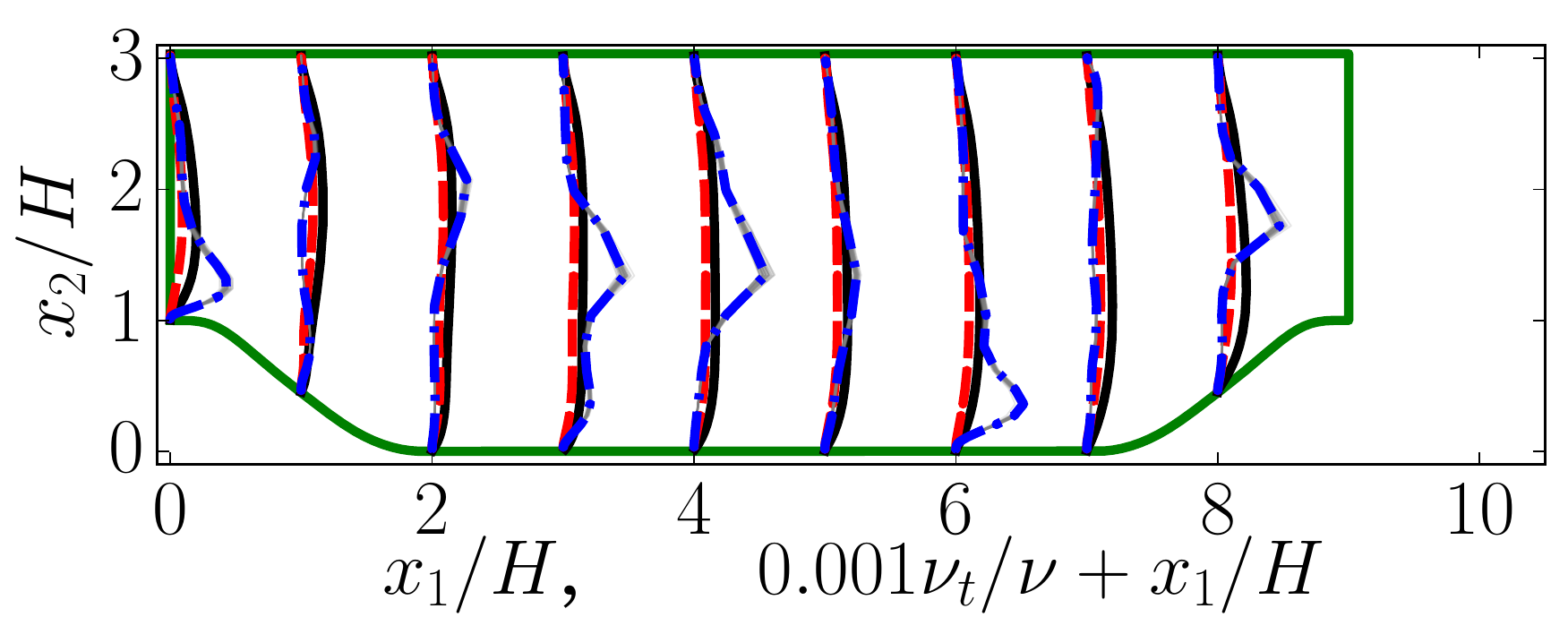}
              \caption{regularized, eddy viscosity}
          \end{subfigure}
          \begin{subfigure}[b]{0.48\linewidth}
              \includegraphics[width=\linewidth]{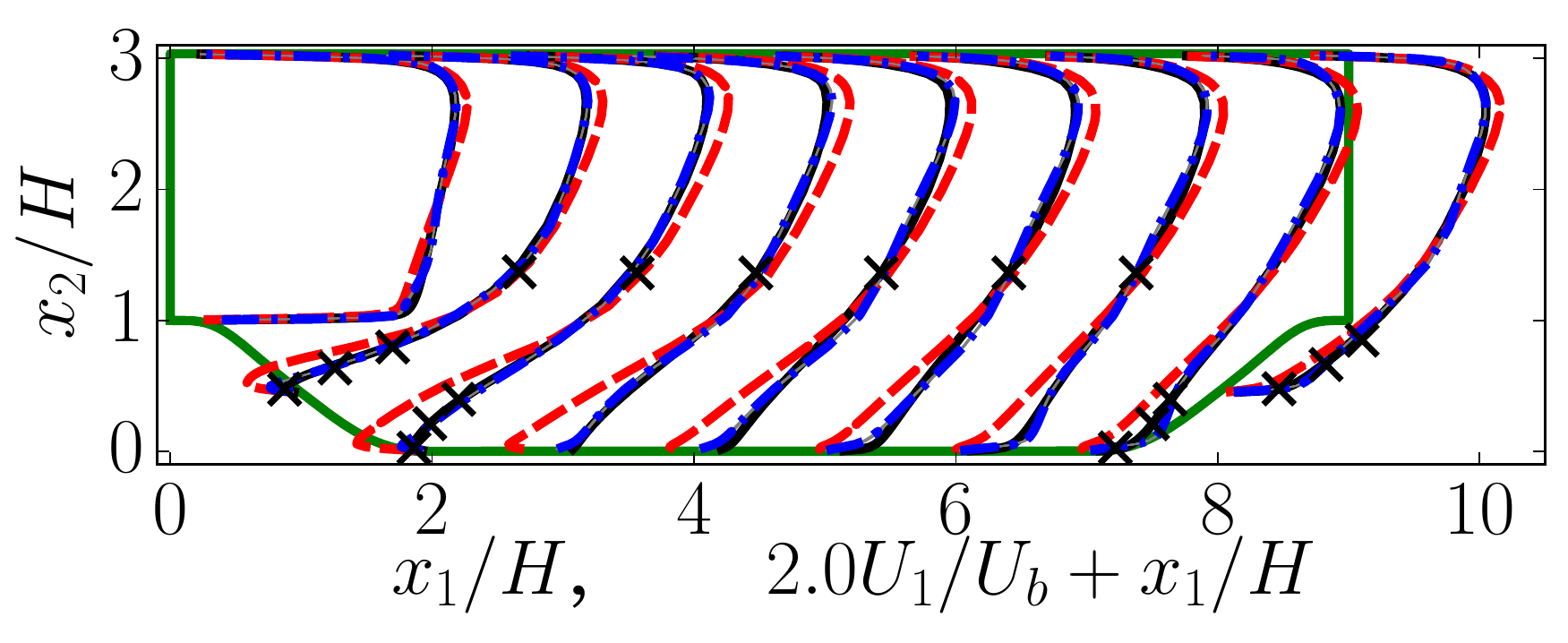}
              \caption{regularized, streamwise velocity}
          \end{subfigure}
          \caption{Inferred eddy viscosity field and propagated streamwise velocity field for the baseline and regularized methods using $200$ modes.}
          \label{fig:case3_results_profiles}
      \end{figure}

\begin{figure}
          \centering
          \includegraphics[width=0.3\textwidth]{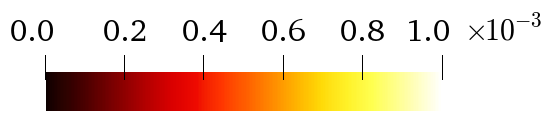}\\
          \begin{subfigure}[b]{0.4\linewidth}
              \includegraphics[width=\linewidth]{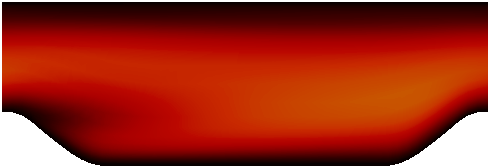}
              \caption{prior mean}
          \end{subfigure}
          \begin{subfigure}[b]{0.4\linewidth}
              \includegraphics[width=\linewidth]{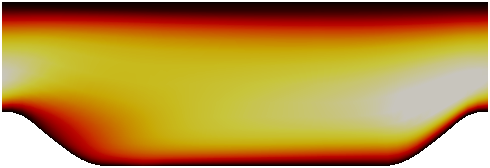}
              \caption{truth}
          \end{subfigure}
          \begin{subfigure}[b]{0.4\linewidth}
              \includegraphics[width=\linewidth]{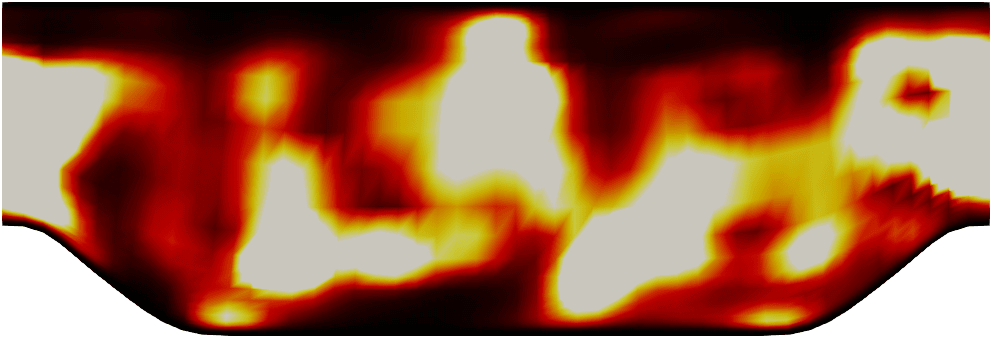}
              \caption{posterior mean, baseline method}
          \end{subfigure}
          \begin{subfigure}[b]{0.4\linewidth}
              \includegraphics[width=\linewidth]{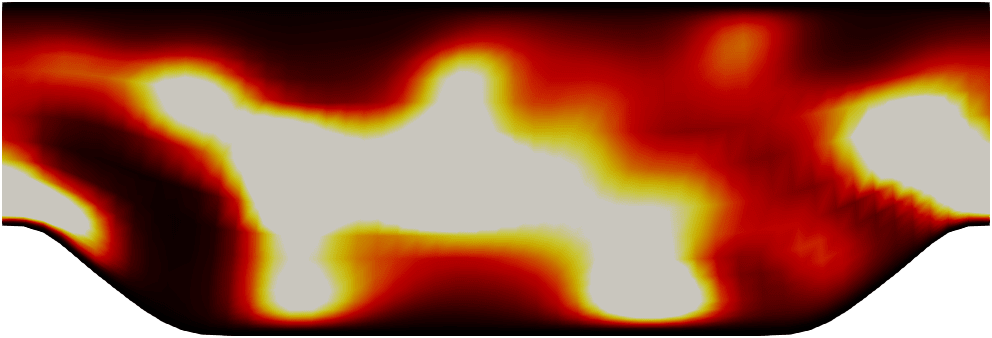}
              \caption{Posterior mean, regularized method}
          \end{subfigure}
          \caption{Inferred (posterior) eddy viscosity $\nu_t$ using the baseline and regularized methods and $200$ modes.}
          \label{fig:contour_nut}
      \end{figure}

      The case with~$200$ modes is used to show the performance of the proposed regularized method. 
      Profiles of the inferred eddy viscosity fields, as well as the propagated stream-wise velocity fields, are shown in Fig.~\ref{fig:case3_results_profiles} using both the baseline and regularized methods. 
      The baseline method is able to improve the velocity profiles in most of the domain. 
      The regularized method is similarly able to improve the velocity field.
      Although the baseline method improves the predicted velocity, the inferred eddy viscosity field is much farther from the true field than the prior mean. 
      The inferred eddy viscosity field in Fig.~\ref{fig:EnKF_nut} has magnitudes many times larger than the truth and exhibit much more oscillations. 
      Embedding the additional information into the inversion using the regularized method can result in improved results. 
      The inferred field in Fig.~\ref{fig:REnKF_nut} is still worse than the prior, but many of the problems in the inferred field with the baseline method have been significantly reduced. 
      Specifically, the inferred field is smoother and has smaller magnitudes. 
      The entire inferred fields are shown in Fig.~\ref{fig:contour_nut}, which more clearly shows the qualitative difference between the true field and the inferred field using either method. 
      The field inferred with the regularized method can be seen to reduce the magnitude and number of the oscillations compared to the field inferred with the baseline ensemble Kalman method. 
      To further improve the inferred eddy viscosity, more information, such as smoothness, could be embedded as constraints in the regularized ensemble Kalman method. 
      
\begin{figure}[!htb]
        \centering
        \includegraphics[width=0.6\linewidth]{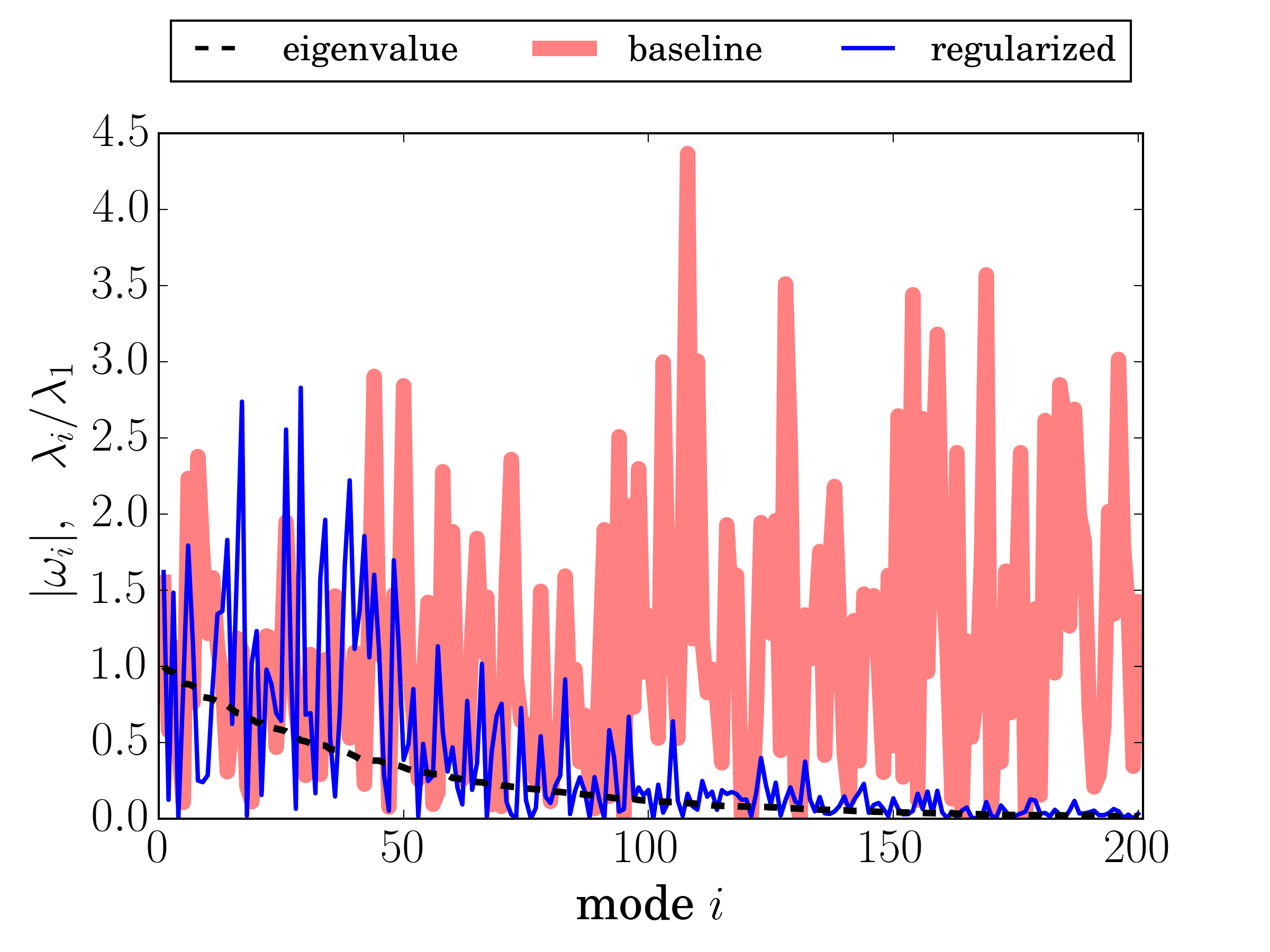}
        \caption{Magnitudes of the inferred KL coefficients for the periodic hills case using the baseline and regularized methods and 200 modes.}
        \label{fig:case3_inferred_omega}
      \end{figure}

      The magnitudes of the inferred coefficients for both methods are shown in Fig.~\ref{fig:case3_inferred_omega}. %
      The baseline ensemble Kalman method uses the modes indiscriminately, and the KL coefficients for the higher modes are large. %
      By contrast, the regularized method uses less of the higher modes, successfully enforcing our preference. %
      Furthermore, the trend of the decay of the magnitudes of the inferred coefficients is proportional to the eigenvalues as expected. 
      This is due to the specified weight matrix in Eq.~\eqref{eq:cov_RANS} penalizing each mode by the reciprocal~$1/\lambda_i$ of its respective eigenvalue. 
      
\begin{figure}[!htb]
        \centering
        \includegraphics[width=0.5\linewidth]{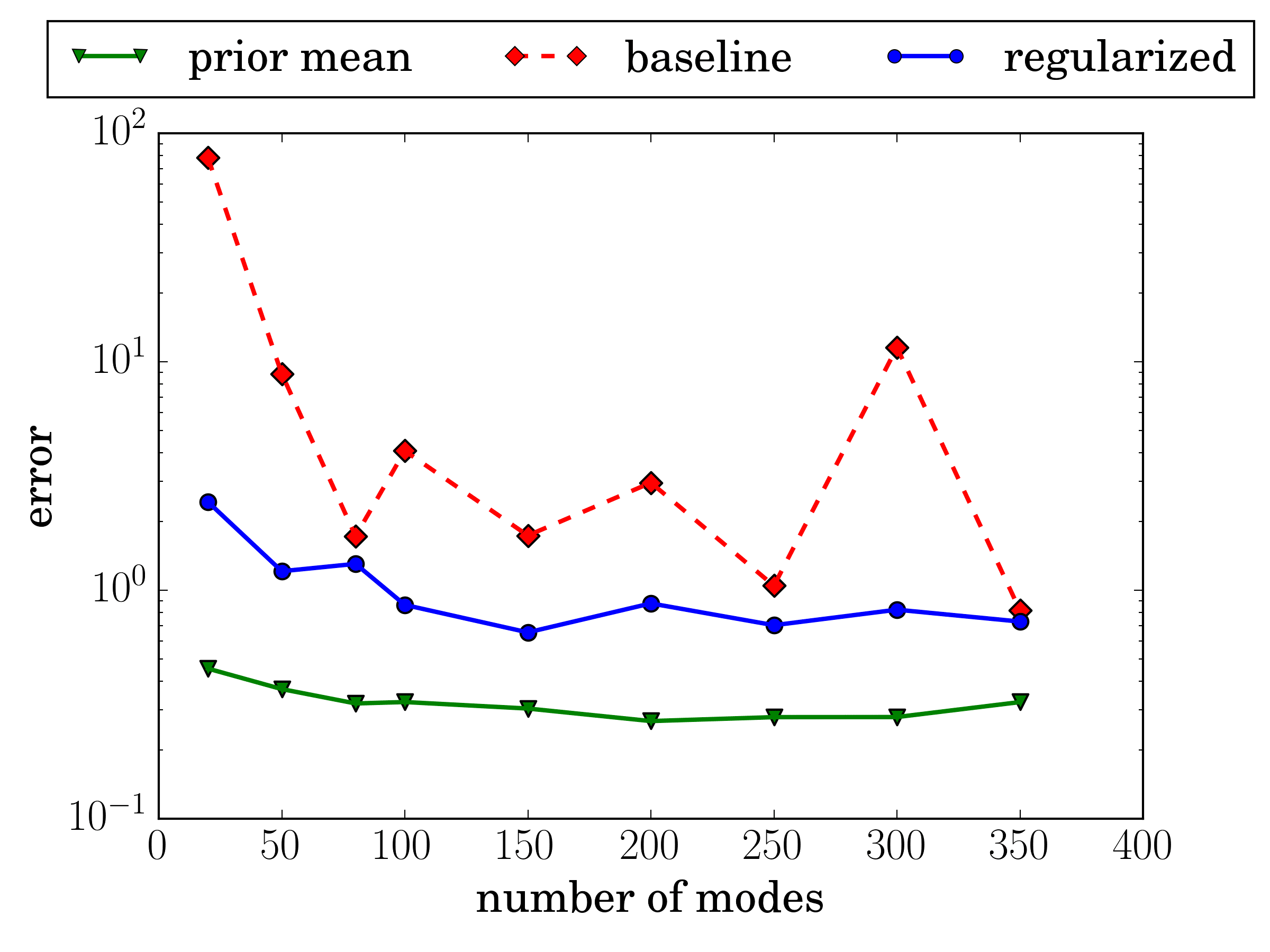}
        \caption{Error in the inferred eddy viscosity in the \textbf{periodic hill case} using the baseline and regularized methods as a function of the number of modes used in the field representation. Note that a logarithmic scale is used for the errors.}
        \label{fig:case3_error}
      \end{figure}

      The EnKF and regularized method both provide similar and satisfactory data fit in velocity regardless of the number of modes, and hence the plot for the error in velocity is omitted for brevity. The error in the inferred eddy viscosity is calculated using Eq.~\eqref{eq:error_def}. 
      The errors for the different methods are shown in Fig.~\ref{fig:case3_error} as a function of the number of the modes used in the representation. 
      The inference using the regularized method has a lower error for all cases tested. 
      It should be noted that this measure of error accounts for the entire field not only observation points. 
      With too few modes, the error is large because the number of modes is insufficient to represent the field. Consequently, in order to fit the observations well, the inversion scheme drives the field in the unobserved areas to depart significantly from the truth.
      However, the error tends to flatten out as the number of modes is increased. 
      
\section{Conclusion}
\label{sec:conclude}
    Inverse problems are common and important in many applications in computational physics.
    They consist of inferring causal parameters in the model from observations of model output. 
    The parameters can be scalar model parameters or physical fields, and the observations are typically sparse point observations of some, possibly different, physical fields. 
    The most straightforward way to solve such problems is minimizing a cost function that penalizes the  discrepancy of the inferred results with the observations. %
    This cost function is minimized using gradient-based methods with the gradients computed from the adjoint of the model. 
    However, many physical models used in practice do not have a readily available adjoint capability, and development of this capability requires significant effort. 
    This has prompted the development of ensemble-based models, such as the ensemble Kalman methods, which are widely used in practice. 
    Ensemble methods use the sample covariance rather than the gradient to guide the optimization. 
    The ensemble Kalman methods implicitly solve the same minimization problem and both ensemble- and gradient-based methods are equivalent under mild assumptions. 
    
    A problem with inverse problems is that they are generally ill-posed, with many possible solutions of the parameters leading to satisfactory results in the observation space. 
    This is typically solved by regularizing the problem by adding some additional constraint to the cost function. 
    For instance, smoothness can be enforced by penalizing the magnitude of the gradient of the field. 
    When directly optimizing the cost function with adjoint methods, this is straightforward to implement; however, it is not straightforward to implement such constraints in ensemble-based methods. 
    Here we propose a regularized ensemble Kalman update capable of embedding such additional knowledge into ensemble Kalman methods. 
    Additional constraints are added into the Bayesian formulation, and a derivative-free update scheme is derived from an optimization perspective. 
    This effectively bridges the gap between the ability to regularize the problem in both classes of methods and allows for general constraints to be enforced implicitly in the data assimilation problem. 
    
    Here we presented three different cases of increasing complexity, including inferring scalar parameters as well as one- and two-dimensional fields. 
    For the final test case we used the method to infer the closure field in the Reynolds-averaged Navier--Stokes equations, a case of significant practical importance in computational fluid dynamics. 
    Compared to using a traditional ensemble Kalman method, the proposed method performs just as well in the observation space, but by incorporating additional knowledge as regularization, the inference in the parameter space is greatly improved. 
    The results demonstrate that the proposed method correctly embeds the additional constraints. 
    The application of the proposed method to enforce other constraints, e.g., physical conservation laws, will be investigated in future studies.

\section*{Acknowledgment}
CMS is supported by the U.S. Air Force under agreement number FA865019-2-2204.
The U.S. Government is authorized to reproduce and distribute reprints for Governmental purposes notwithstanding any copyright notation thereon.
The authors would like to thank the reviewers for their constructive and valuable comments, which helped improve the quality and clarity of this manuscript.
\appendix

\section{Detailed derivation of regularized ensemble Kalman method}
\label{sec:derivation_REnKF}
The detailed derivation for the update scheme in the regularized ensemble Kalman method is presented here. The cost function with prior and general regularization terms can be formulated as
\begin{equation}
    J[\sx_j^\text{a}] = (\sx_j^\text{a}-\sx_j^\text{f})^\top \mathsf{P}^{-1}(\sx_j^\text{a}-\sx_j^\text{f})+
    (\mathcal{H}[\sx_j^\text{a}]-\oy_j)^\top \mathsf{R}^{-1} (\mathcal{H}[\sx_j^\text{a}]-\oy_j) +	(\mathcal{G}[\sx_j^\text{a}]^\top \mathsf{Q}^{-1} \mathcal{G}[\sx_j^\text{a}]) \text{.}
    \label{eq:costfunction_REnKF_A}
\end{equation}
The gradient of the cost function is then
\begin{equation}
    \frac{\partial J[\sx_j^\text{a}]}{\partial \sx_j^\text{a}} = \mathsf{P}^{-1}(\sx_j^\text{a}-\sx_j^\text{f})+(\mathcal{H'}[\sx_j^\text{a}])^\top \mathsf{R}^{-1}(\mathcal{H}[\sx_j^\text{a}]-\oy_j) + (\mathcal{G'}[\sx_j^\text{a}])^\top \mathsf{Q}^{-1} \mathcal{G}[\sx_j^\text{a}]  \text{.}
    \label{eq:grad_REnKF_A}
\end{equation}
To minimize the cost function, the gradient~\eqref{eq:grad_REnKF_A} is set to zero:
\begin{equation}
\label{eq:zero_grad1}
    \mathsf{P}^{-1}(\sx_j^\text{a}-\sx_j^\text{f}) + (\mathcal{H'}[\sx_j^\text{a}])^\top \mathsf{R}^{-1}(\mathcal{H} [\sx_j^\text{a}]-\oy_j) + (\mathcal{G'}[\sx_j^\text{f}])^\top \mathsf{Q}^{-1} \mathcal{G}[\sx_j^\text{f}] = 0  \text{.}
\end{equation}
The unknown terms $\mathcal{H}[\sx_j^\text{a}]$ and $\mathcal{H'}[\sx_j^\text{a}]$ are linearized as
\begin{subequations}
\begin{align}
    \mathcal{H}[\sx_j^\text{a}] &\approx \mathcal{H}[\sx_j^\text{f}] + \mathcal{H'}[\sx_j^\text{f}](\sx_j^\text{a}-\sx_j^\text{f}),
    \label{eq:linear_assumption1}
    \\
    \mathcal{H'}[\sx_j^\text{a}] &\approx \mathcal{H'}[\sx_j^\text{f}],
    \label{eq:linear_assumption2}
\end{align}
\end{subequations}
respectively.
With this linearization Equation \eqref{eq:zero_grad1} becomes
\begin{subequations}
\begin{align}
\label{eq:zero_grad2}
    &\mathsf{P}^{-1}(\sx_j^\text{a} -\sx_j^\text{f})= - (\mathcal{H'}[\sx_j^\text{f}])^\top \mathsf{R}^{-1}(\mathcal{H}[\sx_j^\text{f}]+\mathcal{H'}[\sx_j^\text{f}](\sx_j^\text{a} - \sx_j^\text{f})-\oy_j) -  (\mathcal{G'}[\sx_j^\text{f}])^\top \mathsf{Q}^{-1} \mathcal{G}[\sx_j^\text{f}]  \text{.}
\end{align}
\end{subequations}
Similarly, the constraint term is linearized as
\begin{equation}
    \mathcal{G}[\sx_j^\text{f}] \approx \mathcal{G}[\sx_j^\text{a}]  \qquad \text{and} \qquad \mathcal{G'}[\sx^\text{f}] \approx \mathcal{G'}[\sx^\text{a}].
\end{equation}
Note that a convergence condition is assumed for $\mathcal{G}[\sx_j^\text{a}]$ (i.e., the first derivate term is ignored) to simplify the derivation. 
Furthermore, the tangent linear operator $\mathsf{H}$ is used as an estimate of the observation operator $\mathcal{H}$, giving  $\mathcal{H}[\sx]=\mathsf{H}\sx$ and $\mathcal{H'}[\sx] = \mathsf{H}$. 
With these simplifications \eqref{eq:zero_grad2} can be written as
\begin{equation}
     \mathsf{P}^{-1} (\sx_j^\text{a} -\sx_j^\text{f}) + \mathsf{H}^\top \mathsf{R}^{-1}(\mathsf{H} \sx_j^\text{f}+\mathsf{H}(\sx_j^\text{a} - \sx_j^\text{f})-\oy_j) + \mathcal{G'}^\top \mathsf{Q}^{-1} \mathcal{G} = 0 \text{,}
     \label{eq:zero_grad3a}
\end{equation}
where the argument $\sx_j^\text{f}$ of the function $\mathcal{G}$ and its derivative $\mathcal{G}'$ are omitted for brevity of notation.
Equation~\eqref{eq:zero_grad3a} can be rearranged to be in the form of an update scheme as
\begin{equation}
     \sx_j^\mathsf{a} = \sx_j^\text{f} + \mathsf{P}(I + \mathsf{H}^\top \mathsf{R}^{-1}\mathsf{H}\mathsf{P})^{-1}\mathsf{H}^\top \mathsf{R}^{-1}(\oy_j - \mathsf{H} \sx_j^\text{f})- \mathsf{P}(I + \mathsf{H}^\top \mathsf{R}^{-1}\mathsf{H}\mathsf{P})^{-1}  \mathcal{G'}^\top \mathsf{Q}^{-1} \mathcal{G} \text{.}
    \label{eq:zero_grad3}
\end{equation}
The term $(I + \mathsf{H}^\top \mathsf{R}^{-1}\mathsf{H}\mathsf{P})^{-1}\mathsf{H}^\top$ can be written as 
\begin{equation}
    (I + \mathsf{H}^\top \mathsf{R}^{-1}\mathsf{H}\mathsf{P})^{-1}\mathsf{H}^\top = \mathsf{H}^\top(I + \mathsf{R}^{-1}\mathsf{H}\mathsf{P}\mathsf{H}^\top)^{-1} \text{,}
    \label{eq:transform formula}
\end{equation}
by starting with 
\begin{equation}
    \mathsf{H}^\top (I + \mathsf{R}^{-1}\mathsf{H}\mathsf{P} \mathsf{H}^\top) = (I + \mathsf{H}^\top \mathsf{R}^{-1}\mathsf{H}\mathsf{P})\mathsf{H}^\top \text{,}
\end{equation}
and 
taking the left multiplication $(I + \mathsf{H}^\top \mathsf{R}^{-1}\mathsf{H}\mathsf{P})^{-1}$ and right multiplication $(I + \mathsf{R}^{-1}\mathsf{H}\mathsf{P}\mathsf{H}^\top)^{-1}$ for both sides.
By substituting $(I + \mathsf{H}^\top \mathsf{R}^{-1}\mathsf{H}\mathsf{P})^{-1}\mathsf{H}^\top$ in \eqref{eq:zero_grad3} with $\mathsf{H}^\top(I + \mathsf{R}^{-1}\mathsf{H}\mathsf{P}\mathsf{H}^\top)^{-1}$ based on Eq.~\eqref{eq:transform formula}, 
Equation~\eqref{eq:zero_grad3} is written as
\begin{subequations}
\begin{align}
     \sx_j^\text{a} &=
     \sx_j^\text{f} + \mathsf{P}(I + \mathsf{H}^\top \mathsf{R}^{-1}\mathsf{H}\mathsf{P})^{-1}\mathsf{H}^\top \mathsf{R}^{-1}(\oy_j - \mathsf{H} \sx_j^\text{f})- \mathsf{P}(I + \mathsf{H}^\top \mathsf{R}^{-1}\mathsf{H}\mathsf{P})^{-1}  \mathcal{G'}^\top \mathsf{Q}^{-1} \mathcal{G}\\
     &= \sx_j^\text{f} + \mathsf{P}
     \mathsf{H}^\top(I + \mathsf{R}^{-1}\mathsf{H}\mathsf{P}\mathsf{H}^\top)^{-1}
    \mathsf{R}^{-1}(\oy_j - \mathsf{H} \sx_j^\text{f})- \mathsf{P}(I + \mathsf{H}^\top \mathsf{R}^{-1}\mathsf{H}\mathsf{P})^{-1}  \mathcal{G'}^\top \mathsf{Q}^{-1} \mathcal{G} \\
     &=\sx_j^\text{f}+ \mathsf{P}\mathsf{H}^\top ( \mathsf{R} + \mathsf{H} \mathsf{P} \mathsf{H}^\top)^{-1}(\oy_j-\mathsf{H} \sx_j^\text{f})-  \mathsf{P}(I + \mathsf{H}^\top \mathsf{R}^{-1}\mathsf{H}\mathsf{P})^{-1} \mathcal{G'}^\top \mathsf{Q}^{-1} \mathcal{G}.
     \label{eq:zero_grad4}
\end{align}
\end{subequations}
The quantity $(I + \mathsf{H}^\top \mathsf{R}^{-1}\mathsf{H}\mathsf{P})^{-1}$ in the last term of Eq.\eqref{eq:zero_grad4} can be expanded using the Woodbury formula~\cite{hager1989updating} as
\begin{equation}
    (I + \mathsf{H}^\top \mathsf{R}^{-1}\mathsf{H}\mathsf{P})^{-1} = I - \mathsf{H}^\top (\mathsf{R} + \mathsf{H} \mathsf{P} \mathsf{H}^\top)^{-1}\mathsf{H}\mathsf{P}.
    \label{eq:above}
\end{equation}
By substituting Eq.~\eqref{eq:above} into Eq.~\eqref{eq:zero_grad4}, Equation~\eqref{eq:zero_grad4} is written as
\begin{equation}
\begin{aligned}
    \sx_j^\text{a} = \sx_j^\text{f} + \mathsf{P}\mathsf{H}^\top ( \mathsf{R} + \mathsf{H} \mathsf{P} \mathsf{H}^\top)^{-1}(\oy_j-\mathsf{H} \sx_j^\text{f}) -  \mathsf{P}\mathcal{G'}^\top \mathsf{Q}^{-1} \mathcal{G}  + \mathsf{P}\mathsf{H}^\top (\mathsf{R} + \mathsf{H} \mathsf{P} \mathsf{H}^\top)^{-1}\mathsf{H}\mathsf{P}\mathcal{G'}^\top \mathsf{Q}^{-1} \mathcal{G}.
    \label{eq: zero_grad5}
\end{aligned}
\end{equation}
By combining the second and last terms in the right hand side of Eq.~\eqref{eq: zero_grad5}, the regularized analysis scheme  can be written as
\begin{equation}
    \sx_j^\text{a} = \sx_j^\text{f}+ \mathsf{P}\mathsf{H}^\top ( \mathsf{R} + \mathsf{H} \mathsf{P} \mathsf{H}^\top)^{-1}(\oy_j-\mathsf{H} \sx_j^\text{f} + \mathsf{H}\mathsf{P}\mathcal{G'}^\top \mathsf{Q}^{-1} \mathcal{G})- \mathsf{P}\mathcal{G'}^\top \mathsf{Q}^{-1} \mathcal{G}.
    \label{eq: general_update}
\end{equation}
By defining the Kalman gain matrix $\mathsf{K}$ and the correction $\delta$ as 
\begin{align}
    \mathsf{K} &=  \mathsf{P}\mathsf{H}^\top ( \mathsf{R} + \mathsf{H} \mathsf{P} \mathsf{H}^\top)^{-1}, \\
    \delta &= - \mathsf{P}\mathcal{G'}^\top \mathsf{Q}^{-1} \mathcal{G},
\end{align}
the final analysis scheme for the regularized ensemble Kalman method becomes
\begin{equation}
     \sx_j^\text{a} = \sx_j^\text{f} + \delta + \mathsf{K} (\oy_j-\mathsf{H} (\sx_j^\text{f} + \delta)) \text{.}
     \label{eq:REnKF_update}
\end{equation}
The formulation can be rewritten as a pre-processing step prior to a standard Kalman filter as
\begin{subequations}
    \begin{align}
        \tilde{\sx}_j^\text{f} & = \sx_j^\text{f} + \delta ,  \\
        \sx_j^\text{a} & = \tilde{\sx}_j^\text{f} + \mathsf{K} (\oy_j - \mathsf{H} \tilde{\sx}_j^\text{f}) \text{.}
        \label{eq:pre-correction-App}
    \end{align}
\end{subequations}
Alternatively, the regularization update can also be reformulated to be a post-processing step after a standard Kalman filter as
\begin{subequations}
    \begin{align}
        \hat{\sx}_j^\text{f} & =  \sx_j^\text{f} + \mathsf{K} (\oy_j - \mathsf{H} \sx^\text{f}_j) \\
        \sx_j^\text{a} & = \hat{\sx}_j^\text{f} + \mathsf{KHP} \mathcal{G'}^\top \mathsf{Q}^{-1} \mathcal{G} -  \mathsf{P} \mathcal{G'}^\top \mathsf{Q}^{-1} \mathcal{G} \\
        & = \hat{\sx}_j^\text{f} + (\mathsf{KH} - I) \mathsf{P} \mathcal{G'}^\top \mathsf{Q}^{-1} \mathcal{G} \text{,}
        \label{eq:post-correction}
    \end{align}
\end{subequations}
based on Eq.~\eqref{eq: general_update}. In the formula above, $\hat{\sx}$ indicates the updated state with standard Kalman filter to be distinguished from the pre-corrected state~$\tilde{\sx}$ in Eq.~\eqref{eq:pre-correction-App}.
Note that the updated state covariance $\hat{\mathsf{P}}$ with the Kalman analysis is formulated as $\hat{\mathsf{P}} =  (I - \mathsf{KH})\mathsf{P}$.
Hence, we can express the update scheme by applying a post-correction step
\begin{equation}
    \sx_j^\text{a} = \hat{\sx}_j^\text{f} + \hat{\delta}  
    \label{eq:post-correction-app}
\end{equation}
with the post-correction $\hat{\delta}$ defined as
\[
\hat{\delta} =  -\hat{\mathsf{P}}\mathcal{G'}^\top \mathsf{Q}^{-1} \mathcal{G} \text{.}
\]

\section{Comparison to the projection method and the observation augmentation method}
\label{app:projection}
The projection method is an extensively used approach for imposing general state constraints on Kalman filter methods~\cite{simon2002kalman,amor2018constrained,simon2010kalman}. The outline of deriving the update scheme based on the projection method is presented here to provide a clear comparison to our method. 
After the Kalman filter step, the projection method projects the updated state onto a constrained surface by solving a constrained optimization problem.
We assume that the updated state and model error covariance with the standard Kalman filter is $\hat{\sx}^\text{f}$ and $\hat{\mathsf{P}}$, respectively.
Further, the constrained estimate can be written as~\cite{simon2002kalman}
\begin{equation}
\sx^\text{a} = \arg \min (\sx^\text{a}  - \hat{\sx}^\text{f})^\top \hat{\mathsf{P}}^{-1} (\sx^\text{a} - \hat{\sx}^\text{f}),
\end{equation}
subject to the constraint
\begin{equation}
\mathsf{G} \hat{\sx}^\text{a} = \mathsf{z} ,
\end{equation}
where $\mathsf{G}$ is a known matrix.
The Lagrange multiplier method is used to solve this problem.
The Lagrangian can be formulated as
\begin{equation}
L = (\sx^\text{a} - \hat{\sx}^\text{f})^\top \hat{\mathsf{P}}^{-1} (\sx^\text{a} - \hat{\sx}^\text{f}) + 2 \boldsymbol{\lambda}^\top (\mathsf{G} \hat{\sx}^\text{a} - \mathsf{z}) \text{,}
\end{equation}
where $\boldsymbol{\lambda}$ denotes the vector of Lagrange multipliers.
To find the minimum, the first-order derivatives are taken with respect to the state~$\hat{\sx}^\text{f}$ and the Lagrange multiplier $\boldsymbol{\lambda}$ is taken to be zero, i.e.,
\begin{subequations}
\label{eq:lagrange-dev}
\begin{align}
\frac{\partial L}{\partial \sx^\text{a}} &= \hat{\mathsf{P}}^{-1} (\sx^\text{a} - \hat{\sx}^\text{f}) + \mathsf{G}^\top \boldsymbol{\lambda} = 0 \text{,} \\
\frac{\partial L}{\partial \boldsymbol{\lambda}} &= \mathsf{G} \hat{\sx}^\text{a} - \mathsf{z} = 0 \text{.}
\end{align}
\end{subequations}
Solving Eq.~\eqref{eq:lagrange-dev} yields the following:
\begin{subequations}
\begin{align}
\boldsymbol{\lambda} &= (\mathsf{G} \hat{\mathsf{P}} \mathsf{G}^\top)^{-1} (\mathsf{G} \hat{\sx}^\text{f} - \mathsf{z}) \\
\text{and} \quad
\sx^\text{a} & = \hat{\sx}^\text{f} - \hat{\mathsf{P}} \mathsf{G}^\top (\mathsf{G} \hat{\mathsf{P}} \mathsf{G}^\top)^{-1}(\mathsf{G} \hat{\sx}^\text{f} - \mathsf{z}) \text{.}
\label{eq: projection_update}
\end{align}
\end{subequations}
Equation.~\eqref{eq: projection_update} is thus the update scheme of the projection step to impose constraints. 
This is in contrast to our formulation in the post-correction form in Eq.~\eqref{eq:post-correction-app}:
\begin{equation*}
    \sx_j^\text{a} = \hat{\sx}_j^\text{f} - \hat{\mathsf{P}}\mathcal{G'}^\top \mathsf{Q}^{-1} \mathcal{G} \text{,} 
\end{equation*}
which requires computing the derivative~$\mathcal{G'}$ of the constraint function with respect to the state and uses the weight matrix $\mathsf{Q}^{-1}$ to account for the precision of the constraints.
Therefore, the difference between the two methods is evident in terms of the motivation, the derivation procedure, and the final update scheme.

Another commonly used constrained Kalman filter method is the observation augmentation method, also known as the perfect measurement method. This method consists of augmenting the observation model with the constraints on the state with zero variance as
\begin{equation}
    \left[
  \begin{matrix}
   \oy\\
   \mathsf{z}
  \end{matrix}
  \right]
  = \left[
  \begin{matrix}
   \mathsf{H}\\
   \mathsf{G}
  \end{matrix}
  \right] \mathsf{x}
  + \left[
  \begin{matrix}
   \epsilon \\
   0
  \end{matrix}
  \right]
\end{equation}
The augmented observation error covariance can be written as
\begin{equation}
\tilde{\mathsf{R}} = 
\left[
    \begin{matrix}
    \mathsf{R} & 0 \\
    0 & 0 
    \end{matrix}
\right] .
\end{equation}
One can also extend this method to enforce soft constraints by adding noise in the constraint function and modify the error covariance matrix $\tilde{\mathsf{R}}$ accordingly. 
The observation augmentation method does not require modifications to the standard Kalman update scheme as in our method, but they have to compute and store the augmented Kalman gain matrix, which may increase computational costs, particularly when the constraint space is large.  For example, enforcing smooth solution (i.e., small gradient at each cell) on a mesh of $10^5$ cells require augmenting the observation matrix by $10^5 \times 10^5$, making it extremely expensive for matrix inversion when calculating the Kalman gain matrix.

\section{Sensitivity study of algorithmic parameters in regularization}
\label{sec:tunable_parameter}
    As implemented here the regularization term $\delta$ has three hyper-parameters ($\chi_0$, $S$, and $d$) in Eqs.~\eqref{eq:lambda_collapse} and~\eqref{eq:regularization_function}.
    We take $\chi_0 = 0.1$,~$S = 5$, and~$d = 2$ as reference and investigate the effects of different~$\chi_0$,~$S$, and~$d$ for the parameter estimation problem in Section~\ref{sec:parameter_estimation}. The inferred parameter $\wv$ with different tunable parameters~($\chi_0$, $S$, and $d$) for an equality constraint (case C1 in Section~\ref{sec:parameter_estimation}) are shown in Table~\ref{tab:summary_diff_lambda_eq}. It can be seen that with an equality constraint, the proposed method is robust, and there is a large range of hyper-parameters that result in good inference.
      \begin{table}[!htb]
        \caption{Summary of the sensitivity study for an equality constraint (case C1) in the parameter estimation problem in Section~\ref{sec:parameter_estimation}. The inferred parameters $\wv$ are shown for different values of $\chi_0$, $S$, and $d$. The values in bold indicate the reference values.}
        \label{tab:summary_diff_lambda_eq}
        \begin{tabular}{l l | l l l}
        \hline
        \hline
        parameter & value & prior $(-2, -2)$ & prior $(0,0)$ & prior $(2, 2)$ \\
        \hline
        $\chi_0$ & $0.01$ & $(0.86, 1.07)$ & $(0.81,1.14)$ & $(0.93, 1.04)$\\
        & $\bm{0.1}$ & $(1.06, 0.93)$ & $(1.06, 0.93)$ & $(1.02, 0.98)$\\
        & $0.5$ & $(0.94, 1.05)$ & $(0.94, 1.05)$ & $(0.98, 1.00)$\\
        & $1.0$ & $(1.05, 0.94)$ & $(0.94, 1.06)$ & $(1.02, 0.98)$\\
        \hline
        $S$ & $1$ & $(0.92, 1.08)$ & $(1.07,0.92)$ & $(0.98, 1.02)$\\
        & $\bm{5}$ & $(1.06, 0.93)$ & $(1.06, 0.93)$ & $(1.02, 0.98)$\\
        & $20$ & $(1.09, 0.90)$ & $(1.07, 0.92)$ & $(1.04, 0.94)$\\
        & $50$ & $(0.86, 1.13)$ & $(0.91, 1.08)$ & $(1.05, 0.94)$\\
        \hline
        $d$ & $0.1$ & $(1.08, 0.91)$ & $(1.07,0.92)$ & $(1.01, 0.98)$\\
        & $\bm{2}$ & $(1.06, 0.93)$ & $(1.06, 0.93)$ & $(1.02, 0.98)$\\
        & $10$ & $(1.09, 0.90)$ & $(1.08, 0.91)$ & $(1.00, 0.98)$\\
        & $50$ & $(1.10, 0.89)$ & $(0.91, 1.08)$ & $(1.02, 0.98)$\\
        \hline
        \end{tabular}
      \end{table}
      
However, with inequality constraints the method is not as robust as with equality constraints. The result of using an inequality constraint (case C2 in Section~\ref{sec:parameter_estimation}) are shown in Table~\ref{tab:summary_diff_lambda_ineq}. With an inequality constraint, the method is more sensitive to these hyper-parameters. 
If the inequality constraint overcorrects the inferred parameters and then turns off, this can lead to the inference diverging. 

      \begin{table}[!htb]
        \caption{Summary of the sensitivity study for an inequality constraint (case C2) in the parameter estimation problem in Section~\ref{sec:parameter_estimation}. The inferred parameters $\wv$ are shown for different values of $\chi_0$, $S$, and $d$. The values in bold indicate the reference values.}
        \label{tab:summary_diff_lambda_ineq}
        \begin{tabular}{l l | l l l}
        \hline
        \hline
        parameter & value & prior $(-2, -2)$ & prior $(0,0)$ & prior $(2, 2)$ \\
        \hline
        $\chi_0$ & $0.01$ & $(0.07, 0.07)$ & $(0.91,0.96)$ & $(0.95, 0.96)$\\
        & $\bm{0.1}$ & $(1.07, 1.03)$ & $(0.96, 0.98)$ & $(0.94, 0.96)$\\
        & $0.5$ & $(0.83, 0.98)$ & $(1.07, 0.94)$ & $(0.94, 0.94)$\\
        & $1.0$ & Diverge & Diverge & Diverge\\
        \hline
        $S$ & $1$ & Diverge & $(1.07,0.92)$ & $(0.95, 0.96)$\\
        & $\bm{5}$ & $(1.07, 1.03)$ & $(0.96, 0.98)$ & $(0.94, 0.96)$\\
        & $20$ & $(0.31, 0.31)$ & $(0.87, 1.07)$ & $(0.95, 0.93)$\\
        & $50$ & $(0.29, 0.29)$ & $(1.00, 0.96)$ & $(0.95, 0.96)$\\
        \hline
        $d$ & $0.1$ & $(0.19, 0.19)$ & Diverge & $(0.94, 0.95)$\\
        & $\bm{2}$ & $(1.07, 1.03)$ & $(0.96, 0.98)$ & $(0.94, 0.96)$\\
        & $10$ & Diverge & $(0.99, 0.93)$ & $(0.97, 0.95)$\\
        & $50$ & $(1.01, 0.87)$ & $(0.96, 1.04)$ & $(0.96, 0.94)$\\
        \hline
        \end{tabular}
      \end{table}

\afterpage{\clearpage}

\end{document}